\documentclass[aps,prb,twocolumn,showpacs,floatfix]{revtex4}  
\usepackage{amsmath,amssymb,amstext,graphicx,bm}

\begin{document}
\title{Spin relaxation at the singlet-triplet crossing in a quantum dot}
\author{Vitaly N. Golovach,\footnote{
E-mail: {\sf vitaly.golovach@physik.lmu.de}}$^{1,2}$
Alexander Khaetskii,$^{1,3}$ and Daniel Loss$^1$}
\affiliation{$^1$Department of Physics, University of Basel,
 Klingelbergstrasse 82, CH-4056 Basel, Switzerland}
\affiliation{$^2$
Physics Department, Arnold Sommerfeld Center for Theoretical Physics, and Center 
for NanoScience, Ludwig-Maximilians-Universit\"at M\"unchen, 80333 M\"unchen, Germany}
\affiliation{$^3$Institute of Microelectronics Technology,
Russian Acedemy of Sciences,
142432 Chernogolovka,
Moscow District,
Russia}

\date{\today}


\begin{abstract}
We study spin relaxation in a two-electron quantum dot
in the vicinity of the singlet-triplet crossing.
The spin relaxation occurs due to a
combined effect of the spin-orbit, Zeeman, and 
electron-phonon interactions.
The singlet-triplet relaxation rates exhibit strong
variations as a function of the singlet-triplet splitting.
We show that the Coulomb interaction between the electrons
has two competing effects on the singlet-triplet spin relaxation.
One effect is to enhance the relative strength of spin-orbit
coupling in the quantum dot, resulting in larger spin-orbit
splittings and thus in a stronger coupling of spin to charge.
The other effect is to make the charge density profiles 
of the singlet and triplet look similar to each other,
thus diminishing the ability of charge environments 
to discriminate between singlet and triplet states.
We thus find essentially different channels of
singlet-triplet relaxation for the case of strong and weak
Coulomb interaction.
Finally, for the linear in momentum Dresselhaus and Rashba 
spin-orbit interactions, we calculate the singlet-triplet
relaxation rates to leading order in the spin-orbit interaction, 
and find that they are proportional to the second power of
the Zeeman energy, in agreement with recent experiments on
triplet-to-singlet relaxation in quantum dots.
\end{abstract}
\maketitle

                     \section{Introduction}                      %
\label{IntroSec}
The interest to the electron spin in semiconductors
has revived in recent years due to potential
applications in spintronics\cite{Spintronics,Wolf} 
and quantum computing.\cite{Loss97,Cerletti}
On the way to building spin-based qubits, 
a number of interesting spin-related phenomena have been
studied experimentally in semiconductor quantum dots. 
For instance, single-shot read-out of an individual 
electron spin in a quantum dot
has been demonstrated,\cite{SingleShotNature,AmashaZumbuhl} 
optically programmable electron spin memory using 
arrays of self-assembled quantum dots
has been implemented,\cite{Kroutvar}
entanglement of a two-electron correlated state has been 
predicted\cite{Golovach04} and accessed in transport measurements,\cite{Zumbuel}
mixing of singlet and triplet states in double quantum dots 
(due to hyperfine interaction with nuclear spins) 
has been studied,\cite{PettaT2Science,JohnsonPettaT2Nature,KopensFolkScience,Laird06,Coish05}
direct access to the spin-orbit interaction in nanowires
has been demonstrated.\cite{fasth}
The use of electron spin in quantum information
requires long spin coherence times.
Identifying the mechanisms that govern spin decay in 
nanostructures is very important. 
Knowledge about the dominant mechanism can help one build
structures with the least coupling of spin to the environment 
and therefore increase the
spin lifetimes by orders of magnitude.

In semiconductor quantum dots,\cite{KAT}
the electron spin interacts with the environment 
via a number of interactions.  
The spin-orbit, hyperfine, and Zeeman interactions
are extremely relevant for the coherent spin dynamics.
The decoherence time $T_2$ is limited by two kinds of
processes: (i) spin-flip processes
and (ii) dephasing processes.
The relaxation time $T_1$ is determined only by spin-flip 
processes.
The characteristic relaxation time $T_1$ is typically very long in semiconductors.
Recently, $T_1$ was measured in electrostatically defined lateral 
GaAs quantum dots,\cite{SingleShotNature,AmashaZumbuhl}
observing $T_1$ from  $100\,\mu{\rm s}$ 
to $1\,{\rm s}$, depending on the strength of the external magnetic field.
In self-assembled InGaAs quantum dots\cite{Kroutvar}
$T_1$ varies from $100\,\mu{\rm s}$ to $20\,{\rm ms}$.
Dephasing of spin occurs due to environmental changes
that affect the phase of a coherent superposition of ``spin up'' and 
``spin down'' states.
In a Zeeman field, the process (i) is associated with exchanging energy with the environment,
whereas the process (ii) is purely elastic and results from adiabatic
(with respect to Zeeman energy) dynamics of the environment.

There are two major mechanisms that govern the electron spin 
lifetimes in quantum dots.
The spin-orbit interaction is
responsible for the spin relaxation ($1/T_1$), 
limiting the spin coherence time 
to $T_2\leq 2T_1$.
The hyperfine interaction is
responsible for the spin dephasing 
and gives the main contribution to the
decoherence rate ($1/T_2$).
The   decoherence time $T_2^*$  (averaged over many runs) has
recently been accessed in experiment,\cite{PettaT2Science}
finding $T_2^*\simeq 10\,{\rm ns}$ and 
$T_2\simeq 1\,\mu{\rm s}$
for a GaAs quantum (double-) dot, in agreement
with theory.\cite{KhaetskiiLossGlazman,MerculovEfrosRosen}
The hyperfine interaction has recently been investigated experimentally in 
Refs.~\onlinecite{PettaT2Science,JohnsonPettaT2Nature,KopensFolkScience}.

In GaAs quantum dots, the electrons are strongly confined
in one direction and can be described by a two-dimensional Hamiltonian.
The spin-orbit interaction is related to the absence of inversion symmetry, 
either in the elementary crystal cell or at the heterointerface,
and is described by the spin-orbit terms in the electron Hamiltonian\cite{Dyak,Rashba} 
that are linear in the two-dimensional electron momentum. 
These terms can be removed in the case of localized electrons 
by a spin-dependent unitary transformation\cite{Khaetskii,Halperin,Aleiner,GKL}
(for the case of delocalized electrons see Ref.~\onlinecite{Khaetski1}).
As a result, the contribution to the spin-flip rate proportional 
to the second power of the spin-orbit coupling constant 
appears only if one takes into account the Zeeman 
splitting in the electron spectrum. 
This results in unusually low spin-flip rates.
For a quantum dot defined in a two-dimensional electron layer,
the linear in momentum terms in the spin-orbit interaction are dominant,
provided the dot lateral size $\lambda$ is much larger than the
layer thickness $d$ ($\lambda\gg d$).

The spin-orbit interaction in quantum dots is a weak perturbation
on top of the dot confining potential.
The spin-orbit length $\lambda_{SO}$---the distance a bulk electron travels until
its spin precesses around in the spin-orbit field---is typically much larger
than the quantum dot lateral size $\lambda$.
The spin-orbit interaction introduces small corrections ($\sim\lambda/\lambda_{SO}$) to the spin states
of the quantum dot and, more importantly, mediates a coupling
between the spin and the electron orbital bath.
Thus, for example, the electron spin couples efficiently to
acoustic phonons.\cite{Khaetskii,GKL}
Coupling to a functioning QPC has also been considered.\cite{BGL}
Recent experiments\cite{SingleShotNature,Kroutvar,AmashaZumbuhl} suggest that 
the spin-phonon coupling dominates the spin relaxation
in GaAs quantum dots in strong magnetic fields ($B\gtrsim 1\,{\rm T}$).
In contrast to spin relaxation, spin dephasing due to 
spin-orbit interaction is inefficient in GaAs quantum dots,\cite{GKL}
leading to $T_2\simeq 2T_1$ for a realistic system with 
$\lambda/\lambda_{SO}\ll 1$.

The spin-orbit interaction can also be used as a means of control over
the electron spin in quantum dots. An oscillating electric field
couples to the electron spin via the spin-orbit interaction and
produces an electric-dipole-induced spin resonance in the quantum dot.\cite{EDSR}
The coupling of spin to electric fields can be envisioned as
being due to a spin-electric moment of electron.\cite{EDSR}
A dipole-dipole coupling between such moments is usually much stronger than the
magnetic dipole-dipole coupling and occurs between distant
spin-$1/2$ quantum dots coupled with each other only through 
the Coulomb interaction.\cite{TGL}

Two-electron quantum dots provide a unique possibility to verify the
dominant mechanisms of spin relaxation in quantum dots.
The two-particle ground state is a singlet at zero magnetic field.
The triplet level is separated from the singlet
by a sizable energy ($E_{TS}\sim 1\,{\rm meV}$ in GaAs quantum dots).
In an orbital magnetic field $B$,
the singlet and triplet levels can be brought to intersection
at $B\sim 1\,{\rm T}$ in GaAs quantum dots. 
The singlet-triplet splitting, $E_{TS}=E_{T}-E_{S}$,
was studied as a function of magnetic field in a number of experiments
(see {\em e.g.} Refs.~\onlinecite{FujisawaT1Nature,Kyriakidis,Zumbuel,HansonST2el}).
Furthermore, by applying an in-plane component of magnetic field, one can 
control the Zeeman energy $E_Z$ independently of $E_{TS}$.
The triplet-to-singlet relaxation was studied
in Refs.~\onlinecite{FujisawaT1Nature,HansonST2el,Sasaki05,Meunier}
with the help of a pulsed relaxation measurement technique.\cite{Fujisawa}
In Ref.~\onlinecite{FujisawaT1Nature}, the spin lifetime reached up to
$200\,\mu{\rm s}$ and was limited by cotunneling processes.
In Ref.~\onlinecite{HansonST2el}, the spin lifetime reached
$2.6\,{\rm ms}$ and showed a clear $E_Z^{-2}$ dependence.

In the present paper, we consider a two-electron quantum dot
and calculate the triplet-to-singlet relaxation rates, applying
the method used in Refs.~\onlinecite{Khaetskii,GKL} to the
two-electron case.
We emphasize the necessity of three ingredients for the spin relaxation,
\begin{displaymath}
\mbox{spin relaxation}=\mbox{spin-orbit}\times\mbox{Zeeman}\times\mbox{orbital bath},
\end{displaymath}
where ``orbital bath'' stands for any environmental bath that couples 
to the electron charge.
As an example of orbital bath, in this paper, we consider 
the electron-phonon interaction, which governs the spin relaxation
in single-electron GaAs quantum dots.\cite{Khaetskii,GKL,SingleShotNature,Kroutvar,AmashaZumbuhl}
As in the single-electron problem, the effect of the spin-orbit interaction 
cancels at the lowest order in the absence of the Zeeman field.
As a result, the lowest-order singlet-triplet relaxation rates are 
proportional to $E_Z^{2}$, as observed in experiment.

In the two-electron case studied here, the
Coulomb interaction between electrons plays an important role
and the relaxation rates depend on the scale $E_{TS}$,
which can be varied by magnetic fields.
In the case of acoustic phonons, an enhancement of
spin relaxation is expected at a characteristic scale
$E_{TS}\sim \hbar s/\lambda$, where $s$ is the speed of sound and
$\lambda$ is a lateral size of the quantum dot.
For the simplicity of discussion, we do not resolve the Zeeman splitting
of the triplet level until the end of this section.
In the regime $E_{Z}\ll E_{TS}\ll \hbar s/\lambda$, we obtain a relaxation rate 
$\propto E_{Z}^2E_{TS}^3$ in contrast to the dependence $\propto E_{Z}^5$
found in the single-electron quantum dots.\cite{Khaetskii,GKL,Kroutvar,AmashaZumbuhl}
Note that the experimental technique used in Ref.~\onlinecite{HansonST2el}
allows one to study spin relaxation across the singlet-triplet crossing
and thus to probe a wide range of energy scales,\cite{Meunier}
$0\leq |E_{TS}|\lesssim 1\,{\rm meV}$.

The spin-orbit interaction gives rise to avoided crossings of the singlet and triplet 
levels in a magnetic field.
For the linear in momentum spin-orbit interaction, we find that the singlet undergoes
avoided crossings with the triplets $|T_\pm\rangle$, and the energy gap which determines 
the anti-crossing is given by $\Delta\simeq E_Z (r_{12}/\lambda_{SO})$, where 
$r_{12}$ is the average distance between the electrons.
$r_{12}$ is an increasing function of the Coulomb interaction strength 
$\lambda/a_B^*$ (ratio of dot confinement size to effective Bohr radius),
and therefore, the energy gap $\Delta$ is enhanced by Coulomb interaction.
At an avoided crossing, the spin selection rules are maximally violated
and one would expect a strong coupling of spins to the orbital bath.
The anti-crossing energy $\Delta$ is, however, very small compared
to the optimal phonon-emission energy $\hbar s/\lambda$, and 
the spin-flip rate is, therefore, extremely small at the avoided crossing.
[The suppression occurs here mainly due to a small density of states
of acoustic phonons at the phonon emission energy $\hbar\omega_q=\Delta$.]
As a result, the maximum of spin relaxation takes place
away from the point of avoided crossing, at an energy
$E_{TS}$ satisfying $|E_{TS}|\sim \hbar s/\lambda$.
The spin relaxation rate has, therefore, a non-monotonous
behavior as function of the magnetic field around the singlet-triplet
transition.
In the case considered here (small Zeeman energy $E_Z\ll\hbar s/\lambda$),
a maxima in the spin relaxation rate occurs on each side of the
singlet-triplet transition, {\em i.e.} at energies
$E_{TS}\simeq \pm \hbar s/\lambda$.

The Coulomb interaction has two competing effects on the singlet-triplet relaxation.
On the one hand, it enhances the mixing between the singlet and $|T_\pm\rangle$-triplet states.
This enhancement happens, as mentioned above, through the renormalization of the
energy gap $\Delta$.
On the other hand, the Coulomb interaction reduces the difference between the
singlet and triplet charge densities, $|\Psi_S({\bm r})|^2$ and $|\Psi_T({\bm r})|^2$.
This effect is easy to understand on the example of a double dot.
With increasing the strength of the Coulomb interaction the double occupancy
of the electrons on one and the same quantum dot decreases and the
charge distributions of the singlet and triplet states become alike.
The parameter that controls the double occupancy in the double dot is
the ratio of interdot tunnel coupling to the onsite Coulomb repulsion.
In the case of the single dot with two electrons, the role of this
parameter is played by the ratio $a_B^*/\lambda$.
Thus, in the limit $a_B^*/\lambda\ll 1$ the charge distributions
of singlet and triplet coincide.
Therefore, the sensitivity of the singlet and triplet states to environmental
fluctuations that couple to the electron charge is decreased by the Coulomb 
interaction.
The mechanism of spin relaxation in the limit of strong Coulomb interaction
is governed by virtual transitions to states outside the singlet-triplet
sector.

Spin relaxation in a two-electron quantum dot has been studied
theoretically in Refs.~\onlinecite{Dickmann,Florescu,Chaney,Climente}.
In Refs.~\onlinecite{Dickmann,Florescu} the spin-orbit interaction Hamiltonian was
truncated to the subspace of the lowest lying singlet and triplet
states under consideration, despite the fact that the matrix elements
of the spin-orbit interaction, being proportional to the electron
momentum and hence to the transition energy 
[${\bm p}_{nm}=(im^*/\hbar)(E_n-E_m){\bm r}_{nm}$], 
couple increasingly strongly to states higher in energy.
As a result, an important contribution to the spin transition amplitudes
has been overlooked in Refs.~\onlinecite{Dickmann,Florescu}.
This contribution cancels out the triplet-to-singlet relaxation amplitudes
calculated in Refs.~\onlinecite{Dickmann,Florescu} at the first-order of spin-orbit 
interaction and in the absence of a Zeeman splitting.
Furthermore, Ref.~\onlinecite{Florescu} reports
a magnitude of the singlet-triplet avoided-crossing splitting
on the order of the matrix element of the spin-orbit interaction
($\Delta\sim \hbar\beta/\lambda$), whereas the magnitude of the
splitting is much smaller and given by the product of the matrix elements
of the Zeeman and spin-orbit interactions divided by the orbital
level spacing ($\Delta\sim E_Zm^*\beta\lambda/\hbar$).
Here, $\beta$ is the coupling constant of the linear in momentum
Dresselhaus spin-orbit interaction 
(neglecting the Rashba interaction for simplicity) 
and we have assumed that 
$r_{12}\simeq\lambda$, which corresponds to a weak or moderate
Coulomb interaction strength.
In Refs.~\onlinecite{Chaney,Climente}, the spin-relaxation rates
have been computed numerically and the spin-orbit interaction
has therefore been accounted for to high orders of perturbation theory
in a sufficiently large subspace of states.
Our results on the relaxation rates agree with those reported in
Refs.~\onlinecite{Chaney,Climente} in the regime where the 
second and higher order in spin-orbit interaction contributions
to relaxation amplitudes can be neglected.
Parametrically, this regime corresponds to
Zeeman splittings much larger than the spin-orbit interaction
matrix elements, $E_Z\gg \hbar\beta/\lambda$.

The paper is organized as follows.
In Sec.~\ref{ModHamSec}, we introduce the model Hamiltonian, which we analyze
in great detail throughout the paper.
In Sec.~\ref{EnWaveFuncs}, 
we find the energy spectrum and the wave functions
of the two interacting electrons
in the absence of spin-orbit interaction
using a variational method.
In Sec.~\ref{MESOSec}, we evaluate the matrix elements of the spin-orbit interaction
with the wave functions found in Sec.~\ref{EnWaveFuncs}.
In Sec.~\ref{ESPSO}, we revisit the problem of energy spectrum and wave functions
of two electrons,
but this time in the presence of the spin-orbit interaction.
In Sec.~\ref{ESPSOgeq0},
we show that without a Zeeman splitting 
no spin relaxation occurs inside any singlet-triplet
subspace at the first order in spin-orbit interaction
(i.e. second order for the rates) for coupling of spin to any orbital bath.
In Sec.~\ref{ESPSOgfinite}, we consider a finite Zeeman splitting
and evaluate the energy spectrum and wave functions of the
two interacting electrons in the presence of the spin-orbit interaction.
Here, we obtain the avoided crossings between the singlet and triplet
levels as mentioned above.
Because of the avoided crossings, the total spin is not a conserved quantity
and spin relaxation takes place between the quantum dot eigenstates
under the action of potential fluctuations (e.g. electron-phonon interaction).
In Sec.~\ref{ssecWeakME}, we consider the admixture of excited
states to the singlet and triplet under consideration
and show that together with the phonon potential these virtual processes
give rise to an additional channel of spin relaxation inside the 
singlet-triplet subspace.
In Sec.~\ref{spinrelaxationphonon}, we introduce the quantum dot relaxation
rates, which we evaluate in subsequent subsections.
In Sec.~\ref{RelaxRates}, we evaluate the singlet-triplet
relaxation rates and discuss the effect of the Coulomb 
interaction on the relaxation rates in a variety of regimes.
In Sec.~\ref{RelaxRatesTT}, we evaluate the triplet-triplet
relaxation rates and find that,
for typical GaAs quantum dot with moderate strength of Coulomb interaction,
the rates are comparable to the 
spin-flip rates in the one-electron regime and at the same
Zeeman splitting.
The paper contains also a number of appendices which complement the
main text.
In Appendix~\ref{appA}, we study the variational
parameters introduced in Sec.~\ref{EnWaveFuncs} and used
throughout the paper.
In Appendix~\ref{appdeltaFormula}, we give
crossover formulae
for important parameters governing the singlet-triplet
crossing.
In Appendix~\ref{appFidelity}, we estimate
the fidelity of our variational method and show
that it is exact in the limiting cases of strong and weak Coulomb 
interaction. 
In Appendix~\ref{appMEofV}, we calculate the matrix elements
of the remainder of our variational treatment.
These matrix elements allow us to estimate the fidelity
in Appendix~\ref{appFidelity} and can be used to improve
our variational treatment.

\section{Hamiltonian}
\label{ModHamSec}
We consider two interacting electrons in a quantum dot described
 by the Hamiltonian
\begin{eqnarray}\label{H0}
H_0&=&H_d+H_{SO}+H_Z, \\
\label{Hd}
H_d&=&\sum_{i=1,2}\left(\frac{{p}_i^2}{2m^*}+\frac{m^*\omega_0^2}{2}r_i^2\right)+
\frac{e^2}{\kappa|{\bf r}_1-{\bf r}_2|},\\\label{HSO}
H_{SO}&=&\sum_{i=1,2}\left[\beta(-p_i^x\sigma_i^x+p_i^y\sigma_i^y)
+\alpha(p_i^x\sigma_i^y-p_i^y\sigma_i^x)\right],\;\;\;\;\\
H_Z&=&\frac{1}{2}g\mu_B{\bf B}\cdot(\mbox{\boldmath $\sigma$}_1
+\mbox{\boldmath $\sigma$}_2),
\label{HZ}
\end{eqnarray}
 where
${\bf r}_j=(x_j,y_j)$ is the $j$-th electron radius-vector and
${\bf p}_j=-i\hbar\partial/\partial {\bf r}_j +(e/c){\bf A}({\bf r}_j)$ 
is the kinetic momentum, with  
${\bf A}(x_j,y_j)=(-y_j,x_j)B_z/2$ being the vector-potential;
$\mbox{\boldmath $\sigma$}_j=(\sigma_j^x,\sigma_j^y,\sigma_j^z)$ 
are Pauli matrices.
The spin-orbit coupling is given by the Hamiltonian (\ref{HSO}),
where the term proportional to $\beta$ originates from the
bulk Dresselhaus spin-orbit coupling, which is due to
the absence of inversion symmetry in the GaAs lattice;
the term proportional to $\alpha$ represents the Rashba spin-orbit coupling,
which can be present in quantum wells if the confining potential 
(in our case along the z-axis) is asymmetric.
The axes $x$ and $y$ point along the main crystallographic
directions in the $(001)$ plane of GaAs.
The magnetic field 
${\bf B}=B(\cos\varphi\sin\theta,\sin\varphi\sin\theta,\cos\theta)$ 
determines the spin quantization axis and the magnitude
of the Zeeman splitting, $E_Z=g\mu_BB$, where $g$ is the electron
$g$-factor in GaAs.
The orbital effect of ${\bf B}$ is given by the
component $B_z=B\cos\theta$ entering in ${\bf A}({\bf r})$,
and it allows one to control the singlet-triplet splitting
$E_{TS}=E_{TS}(B_z)$.
In this work, we use a parabolic confining potential for the quantum dot, 
which is not necessarily the case in experiment.
The use of the parabolic confinement allows us, however, to give numerical
estimates for the spin relaxation rates in the presence of the Coulomb interaction
between the electrons.
We discuss the generality of our results and the validity
of the parabolic approximation for the dot confinement in Sec.~\ref{secConclusions}.

We consider the acoustic phonons as a major source of orbital fluctuations
in the quantum dot.
The potential of acoustic phonons reads
\begin{eqnarray}  
U_{\rm ph}({\bf r}_1,{\bf r}_2)&=&
\sum_{{\bf q}j}{F(q_z)
\over 
\sqrt{2\rho_c\omega_{{q}j}/\hbar} } 
(e\beta_{{\bf q}j}-iq\Xi_{{\bf q}j})\nonumber\\  
&&\times
\left(e^{i{\bf q}_\parallel \cdot{\bf r}_1}+e^{i{\bf q}_\parallel \cdot{\bf r}_2}\right)
(b_{-{\bf q}j}^\dag+b_{{\bf q}j}),\quad
\label{Uph}  
\end{eqnarray}
 where $b_{{\bf q}j}^\dag$ creates an acoustic phonon with wave vector  
 ${\bf q}=({\bf q}_\parallel,q_z)$, branch index $j$, and dispersion   
 $\omega_{{q}j}$; $\rho_c$ is the sample density  
 [volume is set to unity in (\ref{Uph})].  
The factor $F(q_z)$ equals unity for  
 $|q_z|\ll d^{-1}$ and vanishes for $|q_z|\gg d^{-1}$, where $d$ is the width
 of the 2D layer .  
We take into account both the piezoelectric ($\beta_{{\bf q}j}$)   
 and deformation potential ($\Xi_{{\bf q}j}$)   
 kinds of electron-phonon interaction.\cite{GantmakherLevinson}

Apart form coupling to the acoustic phonons, the electron spin couples
also to a variety of other degrees of freedom in the quantum dot
environment, e.g. spins of nuclei, excitations on
the Fermi surface (in surrounding metallic gates and 2DEGs), 
shot noise of a QPC,
charge switches of impurities, fluctuations in the electrostatic 
confinement (noisy gates), $B$-field fluctuations,
magnetic impurities, etc.  
Different sources of spin decay can be relevant in different regimes 
and also depending on the type of the experimental setup.
In the quantum dots, in which the spin relaxation has been measured 
so far,\cite{SingleShotNature,Kroutvar,AmashaZumbuhl,HansonST2el,Meunier} 
good agreement is obtained with the theoretical predictions
of Refs.~\onlinecite{Khaetskii,GKL}, which are
derived from the combined effect of the electron-phonon,
spin-orbit, and Zeeman interactions.
On the other hand, the relaxation rates obtained from other
mechanisms, such as coupling to a functioning QPC,\cite{BGL}
nuclear spins alone\cite{KhaetskiiLossGlazman,MerculovEfrosRosen}
and in combination with electron-phonon 
interaction,\cite{ErlingssonNazarovFalko,ErlingssonNazarov}
etc., show that these mechanisms can be relevant only in special regimes,
which are usually harder to achieve in experiment.
We, therefore, consider the phonons as the orbital bath in this paper.
We note, however, that
our method is quite general and many other fluctuating potentials
can also be considered as long as they are Markovian
with respect to the spin.

\section{Energy spectrum and wave functions}                      %
\label{EnWaveFuncs}

We consider here the Schr\"odinger equation 
 $H_d|{\bf n}\rangle=E_{\bf n}|{\bf n}\rangle$, with
 $H_d$ given in Eq.~(\ref{Hd}),
 and find the energies and wave functions
 of two interacting electrons.
We use the separation of variables in the
Schr\"odinger equation
 in terms of the center of mass, 
${\bf R}= ({\bf r}_1 +{\bf r}_2)/2$,
 and the relative motion,  
${\bf r}={\bf r}_1 -{\bf r}_2$, coordinates.
The wave functions can then be written as 
$|{\bf n}\rangle=|NM\rangle|nm\rangle$, with
\begin{eqnarray}\label{NM}
&&|NM\rangle={\rm R}_{NM}(R)e^{iM\varphi_R}/\sqrt{2\pi},\;\;\;M=0,\pm 1,\pm 2,...,\;\;\;\;\;\\
&&|nm\rangle={\cal R}_{nm}(r)e^{im\varphi_r}/\sqrt{2\pi},\;\;\;m=0,\pm 1,\pm 2,...,\;\;\;\;\;
\label{nm}
\end{eqnarray}
where $\varphi_R$ and $\varphi_r$ are the polar angles
of ${\bf R}$ and ${\bf r}$, {\em resp}.
Note that ${\bf r}_1\leftrightarrow{\bf r}_2$ corresponds to
$\varphi_r\to\varphi_r+\pi$.
Thus, $m=0,\pm 2,\pm 4,...$ refers to singlet states
(symmetric orbital wave function),
and $m=\pm 1,\pm 3,\pm 5,...$, to triplet states
(antisymmetric orbital wave function).
The radial component in Eq.~(\ref{NM}) reads\cite{Jacak}
\begin{equation}\label{funRNM}
{\rm R}_{NM}(R)=\sqrt{\frac{2N!}{(N+|M|)!}}
\frac{R^{|M|}}{\Lambda^{|M|+1}}L_N^{|M|}\left(R^2/\Lambda^2\right)
e^{-R^2/2\Lambda^2},
\end{equation}
where $L_N^{|M|}$ is a Laguerre polynomial ($N=0,1,2,...$) and 
$\Lambda=\sqrt{\hbar/2m^*\omega}$, with
$\omega=\sqrt{\omega_0^2+\omega_c^2/4}$
and the cyclotron frequency $\omega_c=eB_z/m^*c$.
The radial component in Eq.~(\ref{nm}) is given by
${\cal R}_{nm}(r)=f_{nm}(r)/\sqrt{r}$, where
the function $f_{nm}(r)$ obeys the
Schr\"{o}dinger equation
${\cal H}_{m}f_{nm}(r)=\varepsilon_{nm}f_{nm}(r)$, with
the Hamiltonian
\begin{equation}\label{Hm}
{\cal H}_{m}=\frac{\hbar^2}{m^*}
\left(-\frac{\partial^2}{\partial r^2}+
\frac{m^2-1/4}{r^2}+\frac{1}{a_B^*r}\right)+\frac{m^*\omega^2}{4}r^2,
\end{equation}
where
 $a_B^*=\hbar^2\kappa/m^*e^2$ is the Bohr radius and 
$r=\left|{\bf r}_1-{\bf r}_2\right|$.
Another length scale in the problem is $\lambda=\sqrt{2\hbar/m^*\omega}$,
which gives the characteristic extension of $f_{nm}(r)$ in
the absence of interaction ($a_B^*\gg\lambda$).
The functions 
$f_{nm}(r)$ obey the boundary conditions $f_{nm}(0)=f_{nm}(\infty)=0$,
and are normalized as follows
\begin{equation}
\int_0^{\infty}f_{nm}(r)f_{n'm}(r)dr=\delta_{nn'}.
\end{equation}
The energy spectrum of the Hamiltonian (\ref{Hd})
is given by
\begin{equation}\label{ENMnm}
E_{NMnm}=
\hbar\omega\left(2N+|M|+1\right)+\frac{\hbar\omega_c}{2}\left(M+m\right)+
\varepsilon_{nm}, 
\end{equation}
 where $\varepsilon_{nm}$ are the eigenvalues of
 ${\cal H}_{m}$ in Eq.~(\ref{Hm}).
Note that the quantum number $m$ enters in ${\cal H}_{m}$ as a parameter.
We apply a variational method to find the ground state
of ${\cal H}_{m}$ for each $m$.
For this, we consider first a Hamiltonian $\tilde{\cal H}_{m}$,
 similar to Eq.~(\ref{Hm}), but with 
 $1/a_B^*r\to\gamma/r^2$ and with
 $\omega\to\tilde\omega$.
From the solution of
 the Schr\"{o}dinger equation for $\tilde{\cal H}_{m}$,
 we obtain a complete set of functions
\begin{eqnarray}
\tilde f_{nm}(r)&=&\sqrt{\frac{2\Gamma(n+1+\sqrt{m^2+\gamma})}
{n!\Gamma^2(1+\sqrt{m^2+\gamma})}}
\frac{r^{1/2+\sqrt{m^2+\gamma}}}{\tilde\lambda^{1+\sqrt{m^2+\gamma}}}
\nonumber\\
&&
\times
e^{-r^2/2\tilde\lambda^2}
{}_1F_1\left(-n,1+\sqrt{m^2+\gamma};r^2/
\tilde\lambda^2\right),
\;\;\;\;\;\;\;\;
\label{fgnm}
\end{eqnarray}
 where $n=0,1,2,...$,
 $\tilde\lambda=\sqrt{2\hbar/m^*\tilde\omega}$, 
 $\Gamma(z)$ is the gamma function,
 and ${}_1F_1(a,b;z)$ is the confluent hypergeometric function of the
 first kind; $\gamma\geq 0$ and $\tilde\omega$ are variational parameters. 
Next, we evaluate the ground state energy of the Hamiltonian (\ref{Hm}),
 using $\tilde f_{0m}(r)$ as a trial wave function, {\em i.e.}
 we calculate 
$\varepsilon_{0m}=
 \langle\tilde f_{0m}|{\cal H}_m|\tilde f_{0m}\rangle$ and obtain
\begin{eqnarray}
\varepsilon_{0m}&=&\hbar\tilde\omega
\left(1+
\frac{m^2+\gamma/2}{\sqrt{m^2+\gamma}}
+\frac{\tilde\lambda}{2a_B^*}
\frac{\Gamma(1/2+\sqrt{m^2+\gamma})}{\Gamma(1+\sqrt{m^2+\gamma})}\right.
\nonumber\\
&&
\left.
+\left(1+\sqrt{m^2+\gamma}\right)
\frac{\omega^2-\tilde\omega^2}{2\tilde\omega^2}
\right).
\;\;\;
\label{eg0m}
\end{eqnarray}
Minimizing $\varepsilon_{0m}$ 
 in Eq.~(\ref{eg0m}) with respect to 
 $\gamma$ and $\tilde\omega$, we obtain the ground state energy
 and wave function of ${\cal H}_m$ within the variational method.
A detailed analysis of the dependence of $\tilde\omega$ and $\gamma$
 on the Coulomb interaction strength $\lambda/a_B^*$ is given in
 Appendix~\ref{appA}.
In the limit of vanishing Coulomb interaction ($\lambda/a_B^*\to 0$),
we obtained $\gamma=0$ and $\tilde\omega=\omega$, as expected.
In the limit of strong Coulomb interaction ($\lambda/a_B^*\gg 1$),
the variational parameter $\gamma$ increases like 
$\gamma=(3/4)(\lambda/2a_B^*)^{4/3}$, whereas
$\tilde\omega$ tends to a constant $\tilde\omega=\sqrt{3}\omega/2$.

Now we focus on the singlet-triplet crossing, between
the singlet state $|\psi_S\rangle=|0000\rangle$ and 
the triplet state $|\psi_T\rangle=|000,-1\rangle$, which occurs 
with applying an orbital magnetic field $B_z$
in the presence of the Coulomb interaction 
(the Zeeman interaction is set to zero here). 
Using Eq.~(\ref{ENMnm}), we obtain for the singlet-triplet splitting 
($E_{TS}\equiv E_T-E_S$)
\begin{equation}
E_{TS}=
\frac{\hbar}{2}\delta
\sqrt{\omega_c^2+4\omega_0^2}-\frac{\hbar\omega_c}{2},
\label{EST}
\end{equation}
where $\omega_c=eB_z/m^*c$ and 
$\delta=(\varepsilon_{01}-\varepsilon_{00})/\hbar\omega$.
The Coulomb interaction enters in $\delta=\delta(\lambda/a_B^*)$
and, for vanishing strength of the Coulomb interaction, we have
$\delta(0)=1$.
Furthermore, in leading order, we have
$\delta=1-\sqrt{\pi}\lambda/4a_B^*$, for $\lambda/a_B^*\ll1$,
and $\delta={1\over2}(2a_B^*/\lambda)^{2/3}$, for $\lambda/a_B^*\gg1$.
A crossover formula and a plot for $\delta(\lambda/a_B^*)$
are given in Appendix~\ref{appdeltaFormula}.

Singlet-triplet splitting in two-electron quantum dots
has been studied experimentally down to zero
magnetic fields.\cite{Zumbuel,HansonST2el}
In relation to such measurements, we would like to point out
the following observation.
The derivative of $E_{TS}$ (considering only the ground singlet and first excited triplet states) 
with respect to $\omega_c$ in the limit
$\omega_c\to 0$ has a universal value
\begin{equation}
\left.\frac{\partial E_{TS}}{\partial \omega_c}\right|_{\omega_c\to0}=-\frac{\hbar}{2}
\label{dESTdB}
\end{equation}
for all quantum dots with circular symmetry.
This result follows from the identity\cite{LandauLifshitz}
$\partial E_{\bf n}/\partial\omega_c=(\partial H/\partial\omega_c)_{\bf nn}$,
which is valid in general for any Hamiltonian $H$ dependent on a parameter $\omega_c$.
For the two-electron case considered here we have $H=H_d$
\begin{equation}
\left.\frac{\partial H_d}{\partial \omega_c}\right|_{\omega_c\to0}=
\frac{1}{2}\sum_{i=1,2}[\mbox{\boldmath $r$}_{i}\times\mbox{\boldmath $p$}_i]_z,
\end{equation}
where we used the cylindrical gauge.
Taking into account conservation of the $z$-component of the orbital angular momentum
for circularly symmetric quantum dots, we have that $E_{TS}$
of the ground singlet and first excited triplet obeys Eq.~(\ref{dESTdB}).

According to Eq.~(\ref{EST}), 
if $\delta=1$, the singlet-triplet crossing does not occur
even for arbitrarily large $B_z$,
i.e. there is no singlet-triplet crossing 
without Coulomb interaction.
The singlet-triplet crossing takes place at $B_z=B_z^*$,
due to the orbital effect of {\bf B}, in the presence of Coulomb interaction.
For strong Coulomb interaction, we have $\delta\ll1$
and the singlet-triplet crossing occurs at a small value of $B_z$, 
$B_z^*\ll \omega_0m^*c/e$, and thus, one can neglect the magnetic 
field dependence in $\lambda$ and $\delta$.
In this case, the degeneracy field is given by
$B_z^*=2\delta\omega_0m^*c/e$, or equivalently by
$\omega_c^*=\omega_0(2a_B^*/\lambda_0)^{2/3}$, where $\omega_c^*=eB_z^*/m^*c$
and $\lambda_0=\sqrt{2\hbar/m^*\omega_0}$. 
For weak Coulomb interaction, 
the singlet-triplet degeneracy occurs at 
$\omega_c^*=2\pi^{-1/3}\omega_0(2a_B^*/\lambda_0)^{2/3}$.
A crossover formula and a plot for $\omega_c^*$ are given in 
Appendix~\ref{appdeltaFormula}.

The accuracy of our variational method is discussed in
Appendix~\ref{appFidelity}. We remark that our
variational method is exact in the limiting cases
of weak and strong Coulomb interaction.
In the crossover regime ($\lambda/a_B^*\sim 1$),
the overlap of the ground state wave function $\tilde{f}_{0m}(r)$
with the exact one is close to unity
(see Appendix~\ref{appFidelity}).

Finally, the remaining part of ${\cal H}_m$, which was
not captured by the variational method, is studied analytically 
in Appendix~\ref{appMEofV}.
With the matrix elements calculated there, one can, in principle,
find also the wave functions $f_{nm}(r)$ of the excited states ($n>0$).
However, in the crossover regime ($\lambda/a_B^*\sim 1$),
no expansion parameter is present and an exact diagonalization of
${\cal H}_m=\tilde{\cal H}_m+V$ is required.
In this paper, we will not need knowledge of the excited states.
 
\section{Matrix elements of $H_{SO}$} %
\label{MESOSec}
The spin-orbit Hamiltonian $H_{SO}$ in Eq.~(\ref{HSO}) 
connects different spin components and orbital wave functions.
The matrix elements of $H_{SO}$  read
\begin{equation}\label{HSOnm}
\langle{\bf ns}|H_{SO}|{\bf n}'{\bf s}'\rangle=
i\left(E_{\bf n}-E_{\bf n'}\right)\sum_{j=1,2}
\langle{\bf n}|\mbox{\boldmath $\xi$}_j|{\bf n'}\rangle\cdot
\langle {\bf s}|\mbox{\boldmath $\sigma$}_j|{\bf s}'\rangle,
\end{equation}
 where $E_{\bf n}$ and $\left|{\bf n}\right\rangle=\left|NMnm\right\rangle$
 are, respectively, the eigenenergies and eigenfunctions of $H_d$ 
 (see Sec.~\ref{EnWaveFuncs}),
 and ${\bf s}$ stands for the spin quantum numbers of two electrons.
The vector $\mbox{\boldmath $\xi$}_j$
 has a simple form in the coordinate frame: $x'=(x+y)/\sqrt{2}$,
 $y'=-(x-y)/\sqrt{2}$, $z'=z$, namely
\begin{equation}
\mbox{\boldmath $\xi$}_j=(y'_j/\lambda_-,x'_j/\lambda_+,0),
\;\;\;\;\;\;\;\; j=1,2, 
\end{equation}
where
 $\lambda_{\pm}=\hbar/m^*(\beta\pm\alpha)$ are the spin-orbit lengths.
To obtain Eq.~(\ref{HSOnm}), 
we used the definition of momentum,
${\bf p}_j=(im^*/\hbar)\left[H_d,{\bf r}_j\right]$, 
with $H_d$ given in Eq.~(\ref{Hd}).
Note that, due to the linear in $p$ form of 
$H_{SO}$, the matrix elements (\ref{HSOnm})
vanish at the point of singlet-triplet degeneracy.\cite{note2}

The spin states of two electrons $|{\bf s}\rangle$ 
are the singlet ($S$) and triplet ($T$) states:
\begin{eqnarray}\label{Sstate}
|S\rangle&=&\frac{1}{\sqrt{2}}\left(|\uparrow\downarrow\rangle 
-|\downarrow\uparrow\rangle\right),
\\
|T_+\rangle&=&|\uparrow\uparrow\rangle,
\;\;\;\;\;\;\;\;
|T_-\rangle=|\downarrow\downarrow\rangle,\label{Tpmstate}
\\
|T_0\rangle&=&\frac{1}{\sqrt{2}}
\left(|\uparrow\downarrow\rangle
 +|\downarrow\uparrow\rangle\right),
\label{Tzerostate}
\end{eqnarray}
where arrows in the first (second) place denote the spin state of the
 first (second) electron.
Using the representation in terms of 
${\bf R}= ({\bf r}_1 +{\bf r}_2)/2$
and 
${\bf r}={\bf r}_1 -{\bf r}_2$,
we write
\begin{equation}\label{sumviaSands}
\sum_{j=1,2}\mbox{\boldmath $\xi$}_j\cdot\mbox{\boldmath $\sigma$}_j=
\mbox{\boldmath $\xi$}^R\cdot\mbox{\boldmath $\Sigma$}+
\frac{1}{2}
\mbox{\boldmath $\xi$}^r\cdot\mbox{\boldmath $\sigma$},
\end{equation}
where 
$\mbox{\boldmath $\Sigma$}=
\mbox{\boldmath $\sigma$}_1+
\mbox{\boldmath $\sigma$}_2$
and
$\mbox{\boldmath $\sigma$}=
\mbox{\boldmath $\sigma$}_1-
\mbox{\boldmath $\sigma$}_2$,
and also
$\mbox{\boldmath $\xi$}^R=(
\mbox{\boldmath $\xi$}_1+
\mbox{\boldmath $\xi$}_2)/2$
and
$\mbox{\boldmath $\xi$}^r=
\mbox{\boldmath $\xi$}_1-
\mbox{\boldmath $\xi$}_2$.
Then, it is easy to see that the following matrix elements
vanish
\begin{equation}
\langle S|H_{SO}|S\rangle=
\langle T_0|H_{SO}|T_0\rangle=
\langle T_+|H_{SO}|T_-\rangle=0.
\end{equation}
Next, since the operator $\mbox{\boldmath $\Sigma$}$ is nonzero
only in the triplet subspace and 
$\mbox{\boldmath $\sigma$}$ only between the singlet and triplets, 
all matrix elements of $H_{SO}$ in the triplet subspace 
are proportional to $\mbox{\boldmath $\xi$}^R$, and those
connecting singlet and triplet are proportional to
$\mbox{\boldmath $\xi$}^r$.
For a magnetic field along $\mbox{\boldmath $l$}={\bf B}/B$, we thus have
\begin{eqnarray}
{\langle{\bf n}T_\pm|H_{SO}|{\bf n}'T_\pm\rangle}&=&\pm2i
\left(E_{\bf n}-E_{{\bf n}'}\right)
\langle{\bf n}|\mbox{\boldmath $\xi$}^R|{\bf n}'\rangle
\cdot\mbox{\boldmath $l$},\;\;\;\;\;\;\;
\label{belowconv0}
\\
{\langle{\bf n} T_\pm|H_{SO}|{\bf n}'T_0\rangle}&=&i\sqrt{2}
\left(E_{\bf n}-E_{{\bf n}'}\right)\nonumber\\
&&\times\langle{\bf n}|\mbox{\boldmath $\xi$}^R|{{\bf n}'}\rangle
\cdot\left(
\mbox{\boldmath $X$}\mp i\mbox{\boldmath $Y$}\right),
\;\;\;\;\;\;\;
\label{belowconv}
\end{eqnarray}
where
$\mbox{\boldmath $X$}$, $\mbox{\boldmath $Y$}$ and
$\mbox{\boldmath $Z$}\equiv\mbox{\boldmath $l$}$ are
unit vectors of a coordinate frame with 
$\mbox{\boldmath $Z$}$ along the spin quantization axis.
One possible choice of $\mbox{\boldmath $X$}$ and 
$\mbox{\boldmath $Y$}$ is:
$\mbox{\boldmath $X$}=(\sin\varphi',-\cos\varphi',0)$ and
$\mbox{\boldmath $Y$}=
(\cos\varphi'\cos\theta,\sin\varphi'\cos\theta,-\sin\theta)$,
where $\varphi'=\varphi-\pi/4$ and we wrote vectors
in the frame $(x',y',z)$.
Note that for the vector $\mbox{\boldmath $l$}$ in this frame 
we have 
$\mbox{\boldmath $l$}=
(\cos\varphi'\sin\theta,\sin\varphi'\sin\theta,\cos\theta)$.
The matrix elements of $\mbox{\boldmath $\xi$}^R$ can
be easily evaluated using the 
wave functions in Eqs.~(\ref{NM}) and (\ref{funRNM}).
To avoid extra phase factors due to the $(\pi/4)$-rotation 
in going from $(x,y,z)$ to $(x',y',z)$, 
we replace $\varphi_R\to\varphi_{R'}$
and $\varphi_r\to\varphi_{r'}$ in Eqs.~(\ref{NM}) and (\ref{nm}), 
respectively.
Thus, e.g., we have $R_\pm\equiv R_{x'}\pm iR_{y'}=Re^{\pm i\varphi_{R'}}$ and 
$|NM\rangle={\rm R}_{nm}(R)e^{iM\varphi_{R'}}/\sqrt{2\pi}$.
Then, the matrix elements 
$\langle NM|\mbox{\boldmath $\xi$}^R|N'M'\rangle$ 
of the vector
\begin{equation}
\mbox{\boldmath $\xi$}^R=\left(i\frac{R_--R_+}{2\lambda_-},
\frac{R_-+R_+}{2\lambda_+},0\right)
\end{equation}
are given in terms of the following matrix elements,
\widetext
\begin{eqnarray}\label{RNMpmWide}
\langle NM|R_\pm|N'M'\rangle&=&
\Lambda\delta_{M,M'\pm1}\left(\delta_{N,N'}\sqrt{N+\frac{|M|+|M'|+1}{2}}-
\delta_{N+|M|,N'+|M'|}\sqrt{\frac{N+N'+1}{2}}\right).
\end{eqnarray}
\endwidetext\noindent
In particular, for the ground state $N'=M'=0$, we have
only $\langle 0,\pm1|R_\pm|00\rangle=\Lambda$ non-zero, which yields
\begin{equation}
\langle NM|
\mbox{\boldmath $\xi$}^R
|00\rangle=
\sum_\pm
\frac{\Lambda}{2}
\delta_{N,0}\delta_{M,\pm1}
\left(\mp i\lambda_-^{-1},
\lambda_+^{-1},0\right).
\end{equation}

Next, we calculate the remaining three matrix elements 
of $H_{SO}$ in Eq.~(\ref{HSOnm}). 
They connect the singlet and triplet states, and
therefore, the second term in Eq.~(\ref{sumviaSands})
contributes here.
We obtain
\begin{eqnarray}
{\langle{\bf n}T_0|H_{SO}|{\bf n}'S\rangle}
&=&i\left(E_{\bf n}-E_{{\bf n}'}\right)
\langle{\bf n}|\mbox{\boldmath $\xi$}^r|{\bf n}'\rangle
\cdot\mbox{\boldmath $l$},\;\;\;\;\;\;\;
\label{belowconvST0}\\
{\langle{\bf n} T_\pm|H_{SO}|{\bf n}'S\rangle}
&=&\mp\frac{i}{\sqrt{2}}\left(E_{\bf n}-E_{{\bf n}'}\right)\nonumber\\
&&\times
\langle{\bf n}|\mbox{\boldmath $\xi$}^r|{{\bf n}'}\rangle
\cdot\left(
\mbox{\boldmath $X$}\mp i\mbox{\boldmath $Y$}\right),
\;\;\;\;\;\;\;
\label{belowconvST}
\end{eqnarray}
where $\mbox{\boldmath $X$}$ and $\mbox{\boldmath $Y$}$ have
been defined above.
Analogously to the previous case, we
express $\mbox{\boldmath $\xi$}^r$ in terms of 
$r_\pm=r_{x'}\pm ir_{y'}$,
\begin{equation}
\mbox{\boldmath $\xi$}^r=\left(i\frac{r_--r_+}{2\lambda_-},
\frac{r_-+r_+}{2\lambda_+},0\right),
\label{xirrprm}
\end{equation}
which allows us to evaluate 
$\langle nm|\mbox{\boldmath $\xi$}^r|n'm'\rangle$, 
provided we know  
$\langle nm|r_\pm|n'm'\rangle$.
Using the wave function in Eq.~(\ref{nm}), 
we obtain
\begin{equation}\label{rpmnmintegral}
\langle nm|r_\pm|n'm'\rangle=
\delta_{m,m'\pm1}\int_0^\infty r\tilde f_{nm}(r)\tilde f_{n'm'}(r)dr,
\end{equation}
where we recalled that
${\cal R}_{nm}(r)=\tilde f_{nm}(r)/\sqrt{r}$, with
$\tilde f_{nm}(r)$ 
given in Eq.~(\ref{fgnm}).
The function $\tilde f_{nm}(r)$ contains the variational parameters
$\gamma$ and $\tilde\omega$, which are determined from the minimum
of the ground state energy of ${\cal H}_m$ in Eq.~(\ref{Hm}),
for each $|m|$ independently. 
Therefore, apart from the explicit dependence on $|m|$
in Eq.~(\ref{fgnm}), the function $\tilde f_{nm}(r)$ depends
on $|m|$ also via $\gamma$ and 
$\tilde\lambda\sim\sqrt{1/\tilde\omega}$.
In what follows, we provide $\gamma$ and 
$\tilde\lambda\sim\sqrt{1/\tilde\omega}$ 
with an index $m$ .
We thus rewrite Eq.~(\ref{fgnm}) in the following form
\widetext
\begin{eqnarray}
\tilde f_{nm}(r)=
\sqrt{\frac{2\Gamma(n+t_m)}{n!}}
\frac{r^{t_m-1/2}}{\Gamma(t_m)\tilde\lambda_m^{t_m}}
\exp{\left(\frac{-r^2}{2\tilde\lambda_m^2}\right)}
{}_1F_1\left(-n,t_m;r^2/
\tilde\lambda_m^2\right),
\;\;\;\;\;\;\;\;
\label{fgnmnew}
\end{eqnarray}
where $t_m=1+\sqrt{m^2+\gamma_m}$ and
$\tilde\lambda_m=\sqrt{2\hbar/m^*\tilde\omega_m}$.
Then, evaluating the integral in Eq.~(\ref{rpmnmintegral}),
we obtain\cite{GradsteinRyzhik}
\begin{equation}\label{rnmF2lam}
\langle nm|r_\pm|n'm'\rangle=
\lambda_{mm'}
\delta_{m,m'\pm1}
\sqrt{\frac{\Gamma(n+t_m)\Gamma(n'+t_{m'})}{n!n'!\Gamma(t_m)\Gamma(t_{m'})}}
F_2\left(
\frac{1+t_m+t_{m'}}{2}
;-n,-n';t_m,t_{m'};
\frac{2\tilde\lambda_{m'}^2}{\tilde\lambda_m^2+
\tilde\lambda_{m'}^2},
\frac{2\tilde\lambda_m^2}{\tilde\lambda_{m'}^2+
\tilde\lambda_m^2}\right),
\end{equation}
where $F_2(\alpha;\beta,\beta';\gamma,\gamma';x,y)$ 
is a hypergeometric function of two variables, see below. 
In Eq.~(\ref{rnmF2lam}), we used the notation
\begin{eqnarray}
\lambda_{mm'}=
\frac{
\tilde\lambda_m^{1+t_{m'}}\tilde\lambda_{m'}^{1+t_m}}
{\sqrt{\Gamma(t_m)\Gamma(t_{m'})}}
\left(\frac{2}{\tilde\lambda_m^2+
\tilde\lambda_{m'}^2}\right)^{
\frac{1+t_m+t_{m'}}{2}}
\Gamma\left(\frac{1+t_m+t_{m'}}{2}\right).
\label{lambdamm}
\end{eqnarray}
\endwidetext
\noindent
We note that the diagonal part of $\lambda_{mm'}$ 
gives the average distance between the electrons in a
state with $n=0$,
\begin{equation}
{\langle r\rangle}_m\equiv\lambda_{mm}
=\tilde\lambda_m\frac{\Gamma(3/2+\sqrt{m^2+\gamma_m})}
{\Gamma(1+\sqrt{m^2+\gamma_m})}.
\label{rmnew}
\end{equation}
In particular, for strong Coulomb interaction, 
we have ${\langle r\rangle}_m=\tilde\lambda_m\gamma_m^{1/4}=
\lambda(\lambda/2a_B^*)^{1/3}$, in the leading asymptotic order of
$\lambda/a_B^*\gg 1$.

The function
$F_2(\alpha;\beta,\beta';\gamma,\gamma';x,y)$ used in Eq.~(\ref{rnmF2lam})
is defined  by the series\cite{GradsteinRyzhik}
\begin{equation}\label{F2Hyper}
F_2(\alpha;\beta,\beta';\gamma,\gamma';x,y)=
\sum_{k,l=0}^\infty\frac{(\alpha)_{k+l}
(\beta)_k(\beta')_l}{(\gamma)_k(\gamma')_lk!l!}x^ky^l,
\end{equation}
where $(x)_k=\Gamma(x+k)/\Gamma(x)$ is the Pochhammer symbol,
and the variables $x$ and $y$ should obey the condition
$|x|+|y|<1$ to guarantee convergence.
In our case, however, we have
$x=2\tilde\lambda_{m'}^2/(\tilde\lambda_m^2+
\tilde\lambda_{m'}^2)$ and
$y=2\tilde\lambda_m^2/(\tilde\lambda_{m'}^2+
\tilde\lambda_m^2)$, which gives $|x|+|y|=2>1$.
Nevertheless, the series in 
Eq.~(\ref{F2Hyper}) contains a finite number of terms, 
and thus converges.
Indeed, since the second and third arguments of 
$F_2(\alpha;\beta,\beta';\gamma,\gamma';x,y)$ 
in Eq.~(\ref{rnmF2lam}) are
negative integers, $\beta=-n$ and $\beta'=-n'$, 
the Pochhammer symbols $(\beta)_k$ and $(\beta')_l$
in Eq.~(\ref{F2Hyper})
are identically zero for $k>n$ and $l>n'$,
and thus, the summation in Eq.~(\ref{F2Hyper})
ends at $k=n$ and $l=n'$.

Finally, we note that, in particular, for $n=n'=0$, 
Eq.~(\ref{rnmF2lam}) gives
\begin{equation}
\langle 0m|r_\pm|0m'\rangle=
\lambda_{mm'}
\delta_{m,m'\pm1}.
\label{rmp00mm}
\end{equation}
In the absence of Coulomb interaction,
Eq.~(\ref{rnmF2lam}) reduces to a simpler expression, 
as given in Eq.~(\ref{RNMpmWide}), but with capital letters 
replaced by lower-case ones.

\section{Electron states with Spin-orbit interaction} %
\label{ESPSO}
Using the results of  Sections~\ref{EnWaveFuncs} 
and~\ref{MESOSec}, we proceed now to construct
a perturbation theory in the vicinity of the 
singlet-triplet crossing.
We assume that the spin-orbit interaction in the
quantum dot is weak.
As we show below, the small parameter 
of our perturbation theory
is given by the average distance between the electrons
in the dot divided by the spin-orbit length,
${\langle r\rangle}_m/\lambda_{SO}\ll 1$, where 
${\langle r\rangle}_m$ is given in Eq.~(\ref{rmnew}) and
$\lambda_{SO}$ is the smallest absolute value of 
$\lambda_{\pm}=\hbar/m^*(\beta\pm\alpha)$.
Next, we consider separately the case of zero and finite Zeeman
interaction.
We show that, in the absence of the Zeeman interaction, 
the effect of the spin-orbit interaction
cancels out at the first order of perturbation theory,
similarly to the one-electron quantum dots.\cite{Khaetskii,Halperin,Aleiner,GKL}
In the presence of a Zeeman splitting, we
derive an effective spin-orbit interaction and
analyze first its matrix elements 
between the states of the lowest in energy singlet-triplet
subspace. These matrix elements lead to avoided
crossings inside the singlet-triplet
subspace and allow for the potential fluctuations
(e.g. electron-phonon interaction) to cause
direct transitions between the quantum dot eigenstates.
Finally, we analyze the matrix elements, which involve
states outside of the considered singlet-triplet subspace.
These matrix elements allow for virtual transition to occur,
during which the spin state may be changed.
We show that the triplet-triplet relaxation is fully 
governed by such virtual transitions.

\subsection{Case of zero Zeeman interaction ($g=0$)}
\label{ESPSOgeq0}
We consider the Hamiltonian $H=H_d+H_{SO}$.
In this case, the matrix elements of the spin-orbit
interaction vanish at the singlet-triplet degeneracy point,
see Eq.~(\ref{HSOnm}), and we can use perturbation theory
for the non-degenerate case\cite{LandauLifshitz} 
up to a close vicinity of the singlet-triplet degeneracy point.
We have to make sure only that, within each degenerate multiplet, 
such as a triplet, the basis states are chosen correctly.
For this, we write down the correction to the dot Hamiltonian 
in the second order of $H_{SO}$,
\begin{eqnarray}\label{SWtrtdHZn}
\Delta H^{(2)}&=&
\frac{1}{2}\left[\left(\hat{L}_d^{-1}H_{SO}\right),H_{SO}\right]
\nonumber\\
&=&\frac{\hbar}{m^*\lambda_-\lambda_+}
\sum_j\left(x_jp_{jy}-y_jp_{jx}\right)\sigma_{jz},
\label{corr2order}
\end{eqnarray}
where $\hat{L}_dA=[H_{d},A]$ for $\forall A$, and
the $z$-axis is perpendicular to the plane of the 2DEG.
Going to the relative coordinates in 
Eq.~(\ref{corr2order}) and neglecting transitions between levels
which never intersect, we obtain
\begin{eqnarray}
\Delta H^{(2)}\simeq
\frac{\hbar^2}{2m^*\lambda_-\lambda_+}
\left[\ell^R_z+\ell^r_z
+\frac{eB_z}{\hbar c}\left(R^2+\frac{r^2}{4}\right)
\right]\Sigma_z,
\;\;\;\;\;\;
\label{corr2ordernew}
\end{eqnarray}
where $\ell^R_z=-i{\partial}/{\partial\varphi_R}$ and
$\ell^r_z=-i{\partial}/{\partial\varphi_r}$ and we have used
the cylindrical gauge 
${\bf A}({\bf r})={1\over2}[{\bf B}\times{\bf r}]$.
Equation (\ref{corr2ordernew}) gives a spin-orbit induced
spin-splitting of the dot energy levels. 
This splitting can be viewed as an effective Zeeman 
energy which sets the quantization axis along 
the [001] crystallographic direction.
The terms which we neglected in Eq.~(\ref{corr2ordernew})
are proportional to $\sigma_z=\sigma_{1z}-\sigma_{2z}$
and thus violate the symmetry of a Zeeman interaction.
However, since these terms are purely off-diagonal, their
contribution goes to higher orders. 

The leading order correction to the 
two-electron wave function can be easily found 
using the matrix elements of $H_{SO}$ calculated in Section~\ref{MESOSec}.
We choose the unit vectors 
${\bf X}$, ${\bf Y}$ and ${\bf Z}\equiv\mbox{\boldmath $l$}$ 
to point along the axes $x'$, $y'$ and $z$, 
respectively (this fixes our spin measurement frame).
The singlet and triplet states in the first order of $H_{SO}$ then become,
\widetext
\begin{eqnarray}
|SNMnm\rangle'&=&|SNMnm\rangle 
+|NM\rangle
\frac{m^*}{\hbar\sqrt{2}}
\sum_{n'}\left\{
|n',m-1\rangle 
\left(|T_+\rangle
\alpha
+
|T_-\rangle
\beta
\right)
\langle n',m-1|r_{-}|nm\rangle
\right.
\nonumber\\
&&
\left.
+|n',m+1\rangle 
\left(|T_-\rangle
\alpha+
|T_+\rangle
\beta
\right)
\langle n',m+1|r_{+}|nm\rangle
\right\},\label{SEZ0}\\
|T_\pm NMnm\rangle'&=&|T_\pm NMnm\rangle 
-|SNM\rangle \frac{m^*}{\hbar\sqrt{2}}
\sum_{n'}\left\{
|n',m\pm1\rangle \alpha \langle n',m\pm1|r_\pm|nm\rangle +
|n',m\mp1\rangle \beta \langle n',m\mp1|r_\mp|nm\rangle
\right\}
\nonumber\\
&&
\pm |T_0nm\rangle \frac{\sqrt{2}m^*}{\hbar}
\sum_{N'}\left\{
|N',M\pm1\rangle \alpha \langle N',M\pm1|R_\pm|NM\rangle +
|N',M\mp1\rangle \beta \langle N',M\mp1|R_\mp|NM\rangle
\right\},\label{TpmEZ0}
\;\;\;\;\;\;\;\;\;\\
|T_0 NMnm\rangle'&=&|T_0 NMnm\rangle 
+|nm\rangle \frac{\sqrt{2}m^*}{\hbar}
\sum_{N'}\left\{
|N'M+1\rangle
\left(|T_-\rangle\alpha-|T_+\rangle\beta\right)
\langle N',M+1|R_+|NM\rangle\right.\nonumber\\
&&\left.-|N'M-1\rangle
\left(|T_+\rangle\alpha-|T_-\rangle\beta\right)
\langle N',M-1|R_-|NM\rangle
\right\}.\label{T0EZ0}
\end{eqnarray}
\endwidetext
\noindent
We note that Eqs.~(\ref{SEZ0}), (\ref{TpmEZ0}) and (\ref{T0EZ0})
can be rewritten in the following general form
\begin{eqnarray}
|{\bf ns}\rangle'&=&\left(1-S
\right)|{\bf ns}\rangle,
\label{transfS}\\
S&=&i\mbox{\boldmath $\xi$}^R\cdot\mbox{\boldmath $\Sigma$}+
{i\over2}
\mbox{\boldmath $\xi$}^r\cdot\mbox{\boldmath $\sigma$}=i
\sum_{j=1,2}\mbox{\boldmath $\xi$}_j\cdot\mbox{\boldmath $\sigma$}_j,
\label{Sformula}
\end{eqnarray}
which can be viewed as a spin-dependent gauge transformation
$|{\bf ns}\rangle'=\exp(-S)|{\bf ns}\rangle$
in leading order of $S$.
We note that this transformation is identical to that
used for a single-electron quantum 
dot,\cite{Khaetskii,Halperin,Aleiner,GKL} with the only difference
that we sum over electrons here.
Using the states in Eq.~(\ref{transfS}),
it is easy to see that the matrix elements of any scalar
potential, such as, e.g., the electron-phonon interaction
in Eq.~(\ref{Uph}), are diagonal in the spin index,
\begin{equation}
\langle {\bf ns}|e^SU_{\rm ph}({\bf r}_1,{\bf r}_2)e^{-S}|{\bf n's'}\rangle
=\left(U_{\rm ph}\right)_{\bf nn'}\delta_{ss'},
\label{decoupling}
\end{equation}
because $S$ in Eq.~(\ref{Sformula}), being a function of coordinates only,
commutes with scalar potentials.
Thus, the spin degrees of freedom in the quantum dot 
decouple from all scalar potential fluctuations in the first
order of $H_{SO}$. 
Next, we discuss the validity of this statement for 
quantum dots of arbitrary shape.

We arrived at Eqs.~(\ref{transfS}) and (\ref{Sformula}) 
by considering the harmonic confining potential. 
For this confining potential and any other one
possessing a center of inversion in the $(x,y)$-plane,
the diagonal in orbit part
\begin{equation}
S_{{\bf ns},{\bf ns}'}=i\sum_{j}
\left(\mbox{\boldmath $\xi$}_j\right)_{\bf nn}\cdot
\left(\mbox{\boldmath $\sigma$}_j\right)_{{\bf ss}'}
\label{diagofSpart}
\end{equation}
is identically zero for all orbital states $|{\bf n}\rangle$, 
or it can be made so by shifting the origin of coordinates.
This allowed us to choose the quantization axis for all orbital
levels equally (along $z$), 
as required by the interaction (\ref{corr2ordernew}).
We now consider a dot confining potential of an arbitrary shape,
for which $S_{{\bf ns},{\bf ns}'}$ in Eq.~(\ref{diagofSpart})
is not necessarily zero.
We also allow for an arbitrary number of electrons in the quantum dot.
As we show below, the transformation (\ref{transfS})
takes then the form
\begin{eqnarray}
|{\bf ns}\rangle'&=&\left(1-S_{\cal Q}
\right)e^{-S_{\cal P}}|{\bf ns}\rangle,
\label{transfSnew}
\end{eqnarray}
where  $S_{\cal P}$ and $S_{\cal Q}$ are some operators, respectively, 
diagonal and off-diagonal in the Hilbert space of $H_d$,
i.e. ${\cal P}S_{\cal P}=S_{\cal P}$ and 
${\cal Q}S_{\cal Q}=S_{\cal Q}$, with ${\cal Q}=1-{\cal P}$ and
${\cal P}A=\sum_{\bf n}A_{\bf nn}|{\bf n}\rangle\langle{\bf n}|$ 
for $\forall A$.
Here, we assume for simplicity that the Hamiltonian $H_d$ has no
orbital degeneracies (otherwise one had to regroup the sets 
${\bf n}$ and ${\bf s}$ such that $E_{\bf n}$ were non-degenerate).
The order of the two transformations in Eq.~(\ref{transfSnew})
corresponds to applying first the non-degenerate perturbation
theory (transformation $1-S_{\cal Q}$) and then the
degenerate one (transformation $e^{-S_{\cal P}}$).

We start with writing down the following formally exact equality 
of the Schrieffer-Wolff transformation\cite{BirPikus,EDSR}
\begin{equation}
e^S\left(H_d+H_{SO}\right)e^{-S}=H_d+\Delta H,
\label{SchWffexact}
\end{equation}
where $S=-S^\dagger$ is chosen such that ${\cal P}\Delta H=\Delta H$.
The Hamiltonian $H'=H_d+\Delta H$ is diagonal in the basis of
$H_d$ and has the same energy spectrum as the Hamiltonian 
$H=H_d+H_{SO}$; 
the latter is diagonal in a basis related by 
$\exp(-S)$ to the basis of $H_d$. 
Since in our case ${\cal P}H_{SO}=0$ and thus 
$\Delta H={\cal O}(H_{SO}^2)$, it follows from Eq.~(\ref{SchWffexact})
that $\left[H_d,S\right]=H_{SO}$ in leading order of $H_{SO}$. 
In matrix form, the latter equation reads
\begin{equation}
\left(E_{\bf n}-E_{{\bf n}'}\right)S_{{\bf ns},{\bf n}'{\bf s}'}=
i\left(E_{\bf n}-E_{\bf n'}\right)\sum_{j}
\left(\mbox{\boldmath $\xi$}_j\right)_{{\bf nn}'}\cdot
\left(\mbox{\boldmath $\sigma$}_j\right)_{{\bf ss}'},
\label{SchWeq}
\end{equation}
where we used Eq.~(\ref{HSOnm}) for matrix elements of $H_{SO}$.
Clearly, for $E_{\bf n}\neq E_{\bf n}'$, we obtain 
from Eq.~(\ref{SchWeq}) that
$S_{{\bf ns},{\bf n}'{\bf s}'}=i\sum_{j}
\left(\mbox{\boldmath $\xi$}_j\right)_{{\bf nn}'}\cdot
\left(\mbox{\boldmath $\sigma$}_j\right)_{{\bf ss}'}$,
which coincides with what one obtains from Eq.~(\ref{Sformula}).
Further, we note that the fully diagonal part $S_{{\bf ns},{\bf ns}}$
gives rise to phase factors, which can be absorbed into
the definition of basis states. 
These phase factors, obviously, do not affect the 
spin relaxation, and can be chosen arbitrarily.
The diagonal in orbit part $S_{{\bf ns},{\bf ns}'}$
remains undefined by Eq.~(\ref{SchWeq}), because
the perturbation $H_{SO}$ does not lift the spin
degeneracy in the first order (${\cal P}H_{SO}=0$).
Next, we split the transformation (\ref{SchWffexact}) 
into a product,
\begin{equation}
e^{-S}\approx (1-S_{\cal Q})e^{-S_{\cal P}},
\label{expS1mSQexpSP} 
\end{equation}
where we retain only the leading order of $S_{\cal Q}$.
For $S_{\cal Q}$, we again have
$\left[H_d,S_{\cal Q}\right]=H_{SO}$,
which now gives
\begin{equation}
S_{\cal Q}=(1-{\cal P})
\sum_{j}i\mbox{\boldmath $\xi$}_j\cdot\mbox{\boldmath $\sigma$}_j,
\label{SQform}
\end{equation}
where the factor $(1-{\cal P})$ zeroes out the diagonal in orbit
part.
Note that $S_{\cal Q}$ does not, in general, commute with scalar
potentials and, unless $S_{\cal P}$ can be chosen such as 
in Eq.~(\ref{diagofSpart}), 
there can be, in principle, spin relaxation in the first order of $H_{SO}$.

As we have seen above, the spin splitting in the dot 
is given by $\Delta H$ in Eq.~(\ref{SchWffexact}) and
occurs in the second or higher orders of $H_{SO}$.
From Eq.~(\ref{SchWffexact}), one obtains at the leading order
\begin{equation}
\Delta H=\frac{\cal P}{2}\left[S_{\cal Q},H_{SO}\right] + {\cal O}(H_{SO}^3).
\end{equation}
The operator $S_{\cal P}$ is chosen such that $\Delta H$ is diagonal also in 
the spin subspace of each orbital level.
Therefore, $S_{\cal P}$ depends on the details of $H_d$, 
i.e. on the confining potential and the number of electrons
in the quantum dot. 
Next we briefly discuss $\Delta H$ for the two-electron case considered in 
this paper.

The fine structure of a two-electron quantum dot in the absence of
orbital degeneracy 
(such as singlet-triplet degeneracy)
is described by a Hamiltonian of the
following form
\begin{equation}
\Delta H={\bf A}\cdot\mbox{\boldmath $\Sigma$}
+\sum_{\mu\nu}B_{\mu\nu}\Sigma_\mu\Sigma_\nu,
\label{finestruct}
\end{equation}
where $A_\mu$ and $B_{\mu\nu}=B_{\nu\mu}$ 
are operators diagonal in the orbital space of $H_d$.
In Eq.~(\ref{finestruct}), we did not include energy shifts of the singlet
levels, since they are negligible on the scale of the dot level spacing.
On the same reason, we also neglect the triplet shifts by setting 
$\sum_\mu B_{\mu\mu}=0$.
The term ${\bf A}\cdot\mbox{\boldmath $\Sigma$}$ in Eq.~(\ref{finestruct})
can be viewed as a spin-orbit induced Zeeman interaction.
In leading order, ${\bf A}$ is perpendicular to the 2DEG and 
can be found from Eq.~(\ref{corr2ordernew}), which remains valid 
also in this case.
On average ${\bf A}$ has the following magnitude
\begin{equation}
\langle{A}\rangle=
\frac{\hbar\omega_c}{4\lambda_-\lambda_+}\sum_{j=1,2}
\langle{r}_j^2\rangle,
\label{finestrA}
\end{equation}
where $\langle\dots\rangle$ denotes averaging over many orbital levels.
Note that ${\bf A}$ vanishes at zero magnetic field  
due to time reversal symmetry.
The remaining terms (second order in $\Sigma$) in Eq.~(\ref{finestruct})
describe on-site spin anisotropy.
For a given triplet level, we can choose a coordinate frame
in which $\langle{\bf n}|B_{\mu\nu}|{\bf n}\rangle\sim\delta_{\mu\nu}$,
and since $\sum_\mu B_{\mu\mu}=0$, there remain only two independent 
components of $B_{\mu\nu}$.
We note that $B_{\mu\nu}$ is of higher order in $H_{SO}$ than 
${\bf A}$; however, since the spin anisotropy interaction is
time-reversal invariant, $B_{\mu\mu}\Sigma_\mu^2$ are the only terms left 
at zero magnetic field.

We now discuss the relevance of the fine structure (\ref{finestruct}) 
to spin relaxation in the quantum dot.
We have seen above that, for quantum dots without
center of inversion in the $(x,y)$-plane,
the spin-orbit interaction can, in principle,
lead to spin coupling to potential fluctuations
at the first order of $H_{SO}$.
First, however, we mention several cases when
this coupling results in 
zero relaxation rates for two electrons and at $g=0$.
We stress that the statements below are formulated for
the case of a linear in the electron momentum $H_{SO}$.

(i) There can be no relaxation between singlet and triplet states
due to spin-bath coupling at the first order of $H_{SO}$ 
and in the absence of Zeeman splitting.
Indeed, $S_{{\bf ns},{\bf ns}'}$ in Eq.~(\ref{diagofSpart})
has no matrix elements between any singlet and triplet
states of the quantum dot.
The right-hand side of Eq.~(\ref{diagofSpart})
can be rewritten as follows
\begin{equation}
{\cal P}\sum_{j=1,2}i\mbox{\boldmath $\xi$}_j\cdot\mbox{\boldmath $\sigma$}_j
=i{\cal P}\mbox{\boldmath $\xi$}^R\cdot\mbox{\boldmath $\Sigma$},
\label{PSdiag}
\end{equation}
using the symmetry with respect to exchange of electrons.
As a result, the singlet state is block-diagonal 
in coupling to scalar potential fluctuations [cf. Eq.~(\ref{decoupling})],
i.e. a singlet can relax only to a singlet. 

(ii) There can be no relaxation between spin states belonging to one
and the same orbital level in the first order of $H_{SO}$ without Zeeman 
interaction. This is obvious since ${\cal P}[S_{\cal Q},U_{\rm ph}]=0$,
see Eq.~(\ref{decoupling}) with $S\to S_{\cal Q}$.

After (i) and (ii) we are now left only with transitions between
triplet states of different orbital levels.
Let the triplet states be fine-split with some characteristic
energy $\Delta_{\rm fs}$. This fine splitting can be
calculated from the Hamiltonian (\ref{finestruct}), using
coupling constants obtained from the perturbation theory expansion of 
Eq.~(\ref{SchWffexact}).
However, for the sake of making a simple argument here,
we assume $\Delta_{\rm fs}$ to be an independent parameter.
Clearly, if $\Delta_{\rm fs}=0$, we are free to construct any 
linear combination of states within a triplet; so we can
recover Eqs.~(\ref{transfS}), (\ref{Sformula}) and 
(\ref{decoupling}), which yield no spin relaxation.
To be more rigorous, we formulate an additional condition when there is
no spin decay in the first order of $H_{SO}$.

(iii) If the orbital relaxation time $\tau$ is
much shorter than $\hbar/\Delta_{\rm fs}$, then 
the spin relaxation between triplet states
belonging to different orbital levels is suppressed.
As mentioned above, note that, for $\Delta_{\rm fs}\ll\hbar/\tau$,
one can choose to work with the states in Eqs.~(\ref{transfS}) 
and (\ref{Sformula}) instead of the true eigenstates 
(\ref{transfSnew}), because the internal evolution 
of the states (\ref{transfS}), occurring due to the fine structure
splitting $\Delta_{\rm fs}$, is slower than the lifetime $\tau$.
Thus, the spins can be considered as decoupled from the potential
fluctuations during the time $\sim\hbar/\Delta_{\rm fs}$.
As a result, the spin relaxation occurs due to coupling at the 
second or higher orders of $H_{SO}$.

Finally, we conclude that spin relaxation due to spin-bath 
couplings at the first order of $H_{SO}$ and in the absence of 
the Zeeman interaction is possible only
between spin triplet states belonging to different orbital levels and only 
if $\Delta_{\rm fs}>\hbar/\tau$.
The interaction which causes spin relaxation is proportional to
$(\mbox{\boldmath $\xi$}^R_{\bf nn}
-\mbox{\boldmath $\xi$}^R_{\bf mm})
\cdot\mbox{\boldmath $\Sigma$}$, where ${\bf n}$ and ${\bf m}$
denote the orbital states involved in the relaxation process.
We do not consider this mechanism in further detail here, 
since it refers to an asymmetric quantum dot, 
which goes beyond our model.

\subsection{Case of finite Zeeman interaction ($g\neq0$):
Degenerate perturbation theory}
\label{ESPSOgfinite}

We consider now the Hamiltonian $H=H_d+H_Z+H_{SO}$
in the vicinity of a singlet-triplet crossing.
We assume the Zeeman spitting $E_Z=g\mu_BB$ to be large 
compared to both the fine structure splitting $\Delta_{\rm fs}$ 
and the level broadening $\hbar/\tau$.
This allows us to set the spin quantization axis along
the applied magnetic field ${\bf B}$.
In Section~\ref{ESPSOgeq0}, we found the wave functions 
of two electrons in the quantum dot for the case without 
Zeeman interaction.
It is convenient now to use these functions as basis
states for studying the effect of Zeeman interaction.
We perform the following unitary 
transformation,\cite{Khaetskii,Halperin,Aleiner,GKL}
$\tilde{H}=e^{S_{\cal Q}}(H_d+H_Z+H_{SO})e^{-S_{\cal Q}}$, with 
$S_{\cal Q}$ given in leading order in Eq.~(\ref{SQform}).
After this transformation, a basis state $|{\bf ns}\rangle$,
associated with the Hamiltonian $\tilde{H}$,
will correspond to the basis state 
$|{\bf ns}\rangle'=(1-S_{\cal Q})|{\bf ns}\rangle$, 
associated with the Hamiltonian $H=H_d+H_Z+H_{SO}$.
In first order of 
$H_{SO}$, it is straightforward to obtain
\begin{eqnarray}
\label{HHdHZHZSO}
\tilde{H}&=&H_d+H_Z+H_{Z}^{SO},\\
H_Z^{SO}&=&E_{Z}\sum_{j}
\left[\mbox{\boldmath $l$}\times{\cal Q}\mbox{\boldmath $\xi$}_j\right]\cdot
\mbox{\boldmath $\sigma$}_j,
\label{DeltaHZ}
\end{eqnarray}
where $\mbox{\boldmath $l$}={\bf B}/B$ is the spin quantization direction, and $E_Z=g\mu_BB$
is the Zeeman energy.
The projector ${\cal Q}$ zeroes out the diagonal part of  $\mbox{\boldmath $\xi$}_j$.
Below, we consider the harmonic confining potential, for
which we achieve ${\cal Q}\mbox{\boldmath $\xi$}_j=\mbox{\boldmath $\xi$}_j$
by choosing the origin of coordinates in the dot center.
In Eq.~(\ref{DeltaHZ}), we can rewrite 
the sum over two electrons in terms of the relative coordinates,
\begin{equation}
\sum_{j}
\left[\mbox{\boldmath $l$}\times\mbox{\boldmath $\xi$}_j\right]\cdot
\mbox{\boldmath $\sigma$}_j=
[\mbox{\boldmath $l$}\times\mbox{\boldmath $\xi$}^R]\cdot
\mbox{\boldmath $\Sigma$}+
\frac{1}{2}
\left[\mbox{\boldmath $l$}\times\mbox{\boldmath $\xi$}^r\right]\cdot
\mbox{\boldmath $\sigma$}.
\label{ZeemanSOrelco}
\end{equation}
The first term on the right-hand side of Eq.~(\ref{ZeemanSOrelco})
admixes only triplet states with different orbital wave functions and is, therefore, responsible for triplet-to-triplet
relaxation.
This term can be taken into account
by means of perturbation 
theory for the non-degenerate case.
In contrast, the second term on the right-hand side of 
Eq.~(\ref{ZeemanSOrelco}) is {\em strong} 
at the singlet-triplet
crossing, since it connects singlet and 
triplet states which are degenerate.
For this term, we use perturbation theory for the 
degenerate case.
The matrix elements of $H_Z^{SO}$ in Eq.~(\ref{DeltaHZ}),
therefore, read
\begin{eqnarray}
\label{DHZTpmT0}
{\langle{\bf n} T_\pm|H_Z^{SO}|{\bf n}'T_0\rangle}&=&\mp i\sqrt2E_Z
\langle {\bf n}|\mbox{\boldmath $\xi$}^R|{\bf n}'\rangle
\cdot\left(
\mbox{\boldmath $X$}\mp i\mbox{\boldmath $Y$}\right),
\;\;\;\;\;\;\;\;\\
{\langle{\bf n} T_\pm|H_Z^{SO}|{\bf n}'S\rangle}&=&\frac{i}{\sqrt{2}}E_Z
\langle {\bf n}|\mbox{\boldmath $\xi$}^r|{\bf n}'\rangle
\cdot\left(
\mbox{\boldmath $X$}\mp i\mbox{\boldmath $Y$}\right),
\;\;\;\;\;\;\;\;
\label{DHZTS}\\
{\langle{\bf n} T_0|H_Z^{SO}|{\bf n}'S\rangle}&=&{\langle{\bf n} T_-|H_Z^{SO}|{\bf n}'T_+\rangle}=0\,,
\label{HZSOTSTTzero}
\end{eqnarray}
where $\mbox{\boldmath $X$}$ and $\mbox{\boldmath $Y$}$ are unit vectors
in the plane perpendicular to the applied magnetic field ${\bf B}$
(see Sec.~\ref{MESOSec}). 
Note that, according to Eq.~(\ref{HZSOTSTTzero}), there is no admixture between
any singlet $|S\rangle$ and triplet $|T_0\rangle$ states, as well as between
any triplet states: $|T_+\rangle$ and $|T_-\rangle$.
This will have an effect on the spin relaxation rates in Sec.~\ref{RelaxRates},
where the corresponding rates for the transitions
$S\leftrightarrow T_0$ and $T_-\leftrightarrow T_+$ are found to be
zero in leading order in spin-orbit interaction.
Next, we use Eqs.~(\ref{DHZTpmT0}) 
and (\ref{DHZTS}) to find the two-electron 
wave functions for the lowest in energy 
singlet-triplet crossing in the quantum dot.

We consider first a close vicinity of the singlet-triplet crossing, 
where the contribution of {\em strong} matrix elements is dominant.
Substituting Eq.~(\ref{xirrprm}) into Eqs.~(\ref{DHZTpmT0})$-$(\ref{HZSOTSTTzero}) and using
Eq.~(\ref{rmp00mm}), we obtain for
$|\Psi_S\rangle=|0000S\rangle$ and
$|\Psi_{T_{0,\pm}}\rangle=|000,-1,T_{0,\pm}\rangle$ 
the following expression
\begin{eqnarray}
\langle\Psi_{T_\pm}|H_Z^{SO}|\Psi_S\rangle&=&
\frac{iE_Z\lambda_{1,0}}{2\sqrt2}
\left[
\left(
\pm\frac{\cos\theta}{\lambda_-}-\frac{1}{\lambda_+}
\right)\cos{\varphi'}
\right.\nonumber\\
&&\left.+i
\left(
\frac{1}{\lambda_-}\mp\frac{\cos\theta}{\lambda_+}
\right)\sin{\varphi'}
\right],
\label{TDHZS}\\
\langle\Psi_{S,T_\pm}|H_Z^{SO}|\Psi_{T_0}\rangle&=&\langle\Psi_{T_+}|H_Z^{SO}|\Psi_{T_-}\rangle=0,
\label{TSTTzeroes}
\end{eqnarray}
where we used the convention for 
$\mbox{\boldmath $X$}$ and $\mbox{\boldmath $Y$}$ given below 
Eq.~(\ref{belowconv}).
Next, introducing 
$\Delta_\pm\equiv\left|\langle\Psi_{T_\pm}|H_Z^{SO}|\Psi_S\rangle\right|$,
we obtain
\begin{eqnarray}
\Delta_\pm&=&
\frac{|E_Z|\lambda_{1,0}}{2\sqrt2}
\sqrt{\frac{1-l_{x'}^2}{\lambda_-^2}+\frac{1-l_{y'}^2}{\lambda_+^2}
\mp\frac{2l_z}{\lambda_-\lambda_+}},
\;\;\;\;\;
\label{gapsD}
\end{eqnarray}
where
$\mbox{\boldmath $l$}=
(\cos\varphi'\sin\theta,\sin\varphi'\sin\theta,\cos\theta)$
gives the dependence of $\Delta_\pm$ on the direction of the
magnetic field.
Note that, since we are considering a particular singlet and triplet, 
namely
$|\Psi_S\rangle=|0000S\rangle$ and
$|\Psi_{T_{0,\pm}}\rangle=|000,-1,T_{0,\pm}\rangle$,
the directions of ${\bm l}$, for which these levels intersect
each other in the absence of Zeeman and spin-orbit interactions,
correspond to ${\bm l}$ lying in the northern hemisphere
($\theta < \pi/2$).
For ${\bm l}$ in the southern hemisphere ($\theta >\pi/2$),
a different triplet, $|000,1,T_{0,\pm}\rangle$,
lies lower in energy than the considered one.
Nevertheless, we shall follow the triplet 
$|\Psi_{T_{0,\pm}}\rangle=|000,-1,T_{0,\pm}\rangle$
to all values of $B$ and directions of ${\bm l}$,
since it makes the following consideration simpler.
Note that the results for the other triplet, $|000,1,T_{0,\pm}\rangle$,
are obtained by replacing $B_z\to -B_z$, which, in particular,
corresponds to replacing $\Delta_{\pm}\to \Delta_{\mp}$.

The quantities $\Delta_\pm$ in Eq.~(\ref{TDHZS}) give the strength 
of the singlet-triplet mixing.
The Coulomb interaction enters in Eq.~(\ref{TDHZS})
through the length scale $\lambda_{1,0}$.
Note that $\lambda_{1,0}\sim \lambda_{0,0}, \lambda_{1,1}$ for
all values of $\lambda/a_B^*$. 
One can interpret $\lambda_{1,0}$ as an
average distance between the electrons, $\lambda_{1,0}\simeq \langle r\rangle$.
In the limit of strong Coulomb repulsion,
all three quantities $\lambda_{0,0}$, $\lambda_{1,0}$, and $\lambda_{1,1}$,
converge to one another and become equal to 
${\langle r\rangle}=\lambda(\lambda/2a_B^*)^{1/3}$.
The average distance between the electrons is increased
by the presence of the Coulomb interaction (see Eq.~(\ref{rmnew})).
Therefore, we expect that the Coulomb interaction generally
enhances the effects related to the spin-orbit interaction.
In particular, this enhancement is expected to be strong  
in quantum dots with large $\lambda/a_B^*$.

\begin{figure}
 \begin{center}
  \includegraphics[angle=0,width=0.3\textwidth]{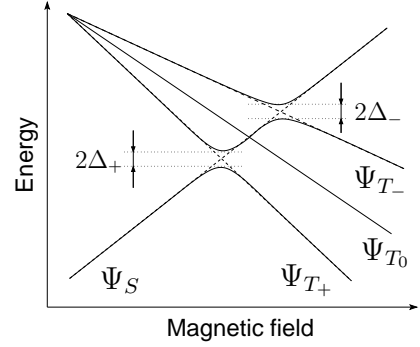}
    \caption{\small
Sketch of the singlet-triplet crossing 
in a quantum dot with spin-orbit interaction.
The triplet level spilts in three due to the Zeeman
interaction. The singlet $|\Psi_S\rangle$ undergoes 
avoided crossing with the triplets 
$|\Psi_{T_\pm}\rangle$, with the splittings $\Delta_{\pm}$
given in Eq.~(\ref{gapsD}).
At the same time, the degeneracy of 
$|\Psi_S\rangle$ and $|\Psi_{T_0}\rangle$ at the crossing
point is not lifted.
}
    \label{STtranSchetch}
 \end{center}
\end{figure}

The singlet-triplet crossing is shown schematically 
in Fig.~\ref{STtranSchetch}. 
The triplet splits into three levels,
$E_{T_\pm}=E_T\mp |E_Z|$ and $E_{T_0}=E_T$, 
due to the Zeeman interaction.
We assume a negative $g$-factor, {\em i.e.} $E_Z=-|E_Z|$, which is the case for GaAs.
The avoided crossing of the singlet state $|\Psi_S\rangle$
with the triplets $|\Psi_{T_\pm}\rangle$ occurs
due to the interaction $H_Z^{SO}$, with
the matrix elements given in Eq.~(\ref{TDHZS}).
The splitting energies 
are the doubled $\Delta_\pm$ in Eq.~(\ref{gapsD}).
Note that the degeneracy of the singlet state with the 
triplet $|\Psi_{T_0}\rangle$ at the crossing point
is not lifted.
This degeneracy, however, can be lifted in higher orders 
of $H_{SO}$.
Next, we consider the interaction between the states:
$|\Psi_S\rangle,|\Psi_{T_+}\rangle$, and $|\Psi_{T_-}\rangle$.
And since the state $|\Psi_{T_0}\rangle$ couples to neither
of these states, see Eq.~(\ref{ZeemanSOrelco}) and the discussion 
thereafter, we disregard it here.
Writing the wave function in the form
\begin{equation}
|\Psi\rangle=a|\Psi_S\rangle+b|\Psi_{T_+}\rangle+
c|\Psi_{T_-}\rangle,
\label{Psiabc}
\end{equation}
we obtain the following set of equations
\begin{eqnarray}
\left(
\begin{array}{ccc}
E_S-E & 
W_{{ST}_+} &
W_{ST_-}\\
W_{T_+S} &
E_{T_+}-E & 0\\
W_{T_-S}&
0& E_{T_-}-E
\end{array}
\right)
\left(
\begin{array}{c}
a\\b\\c
\end{array}
\right)=0,
\end{eqnarray}
where $W_{ST_\pm}$ stands for $\langle \Psi_S|H_Z^{SO}|\Psi_{T_\pm}\rangle$.
The characteristic equation reads
\begin{equation}
E_S-E-
\frac{\Delta_+^2}{E_{T_+}-E}-\frac{\Delta_-^2}{E_{T_-}-E}=0,
\label{seceqST}
\end{equation}
where $E_{T_\pm}=E_T\mp |E_Z|$, and $E_S$ and $E_T$ are the energies of
singlet $|\Psi_S\rangle$ and triplet state $|\Psi_{T_0}\rangle$, respectively.
Since $\Delta_\pm\ll E_Z$, we can solve Eq.~(\ref{seceqST}) 
in the secular approximation.
Setting in turn $\Delta_-$ and $\Delta_+$ to zero
in Eq.~(\ref{seceqST}), we find 
the following expressions, for the lower ($+$) and upper ($-$) solid curves in Fig.~\ref{STtranSchetch},
\begin{eqnarray}
E_\pm=\frac{1}{2}
\left[
E_{T_\pm}+E_S\mp
\sqrt{\left(E_{T_\pm}-E_S\right)^2+
4\Delta_\pm^2}
\right].
\label{Epmonehalf}
\end{eqnarray}
Next, multiplying Eq.~(\ref{seceqST}) by $(E_{T_+}-E)(E_{T_-}-E)$ and dividing 
the obtained polynomial  by $E^2-(E_++E_-)E+E_+E_-$, we obtain 
in leading order the solution for the S-shaped curve 
in Fig.~\ref{STtranSchetch},
\begin{eqnarray}
E_0&=&E_T+\frac{1}{2}\sqrt{(E_{T_+}-E_S)^2+4\Delta_+^2}
\nonumber\\
&&-\frac{1}{2}\sqrt{(E_{T_-}-E_S)^2+4\Delta_-^2}.
\label{E0ETpm0}
\end{eqnarray}
We note that, in the case of positive $g$-factor, {\em i.e.} $E_Z=|E_Z|$, one should
modify Eqs.~(\ref{Epmonehalf}) and (\ref{E0ETpm0}) as follows: $E_{T_\pm}\to E_{T_\mp}$ and $\Delta_\pm\to\Delta_\mp$. 
Finally, the wave function of a level with the eigenenergy 
$E=E_\pm,E_0$
is given by Eq.~(\ref{Psiabc}), with the coefficients
\begin{eqnarray}
b&=&\frac{\langle \Psi_{T_+}|H_Z^{SO}|\Psi_S\rangle}{E-E_{T_+}}a,
\;\;\;\;
c=\frac{\langle \Psi_{T_-}|H_Z^{SO}|\Psi_S\rangle}{E-E_{T_-}}a,\nonumber
\;\;\;\;\;\;\;\\
a&=&\left[1+\frac{\Delta_+^2}{(E-E_{T_+})^2}+
\frac{\Delta_-^2}{(E-E_{T_-})^2}\right]^{-1/2}.
\label{aaacoeff}
\end{eqnarray}
Equation~(\ref{Psiabc}) together with 
Eqs.~(\ref{Epmonehalf})$-$(\ref{aaacoeff}) allows one to calculate
transition rates between quantum dot levels under the action of a perturbation.
As a perturbation, in the next Section,
we consider the electron-phonon interaction.

\begin{figure}
 \begin{center}
  \includegraphics[angle=0,width=0.4\textwidth]{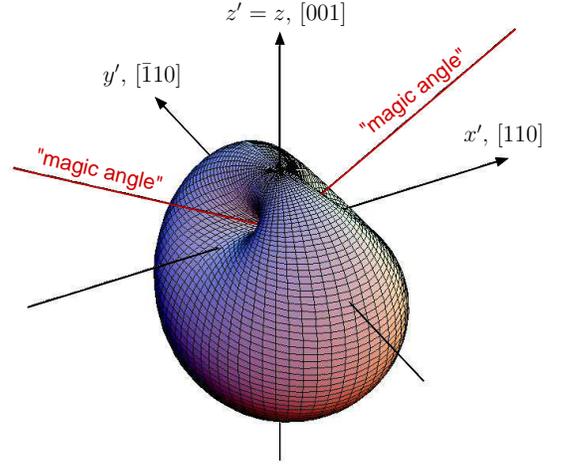}
    \caption{\small
(Color online) Angular dependence of the splitting $\Delta_+$ as given by Eq.~(\ref{gapsD}), 
evaluated for $\lambda_+/\lambda_-=2$.
The direction in space corresponds to the spin-quantization axis, 
given by $\mbox{\boldmath $l$}={\bf B}/B$.
At the ``magic angles'', the value of $\Delta_+$ equals zero.
The angular dependence of the splitting $\Delta_-$ in Eq.~(\ref{gapsD}) is obtained
from this figure by reflecting it in the $(x',y')$-plane.
Note that the circular symmetry of the quantum dot was essential for
obtaining this angular dependence.
}
    \label{DeltaPlus1}
 \end{center}
\end{figure}

Now, we return to Eq.~(\ref{gapsD}) and consider the particular case
when the magnetic field ${\bf B}$ is perpendicular to the 2DEG plane 
($\theta=0$). 
Then, according to Eq.~(\ref{gapsD}), 
we have $\Delta_+=|E_Z|\lambda_{1,0}m^*|\alpha|/\hbar\sqrt{2}$ 
and $\Delta_-=|E_Z|\lambda_{1,0}m^*|\beta|/\hbar\sqrt{2}$. 
These expressions, in principle, allow one to access the Rashba 
(Dresselhaus) coupling constant $\alpha$ ($\beta$) separately 
in a measurement of 
$\Delta_+$ ($\Delta_-$).
Accessing $\alpha$ and $\beta$ separately was, so far,
possible only by analyzing the weak (anti)localization
data in diffusive 2DEGs.\cite{Miller}
The splittings $2\Delta_{\pm}$ can, in principle,
be deduced from transport spectroscopy of the quantum dot.
GaAs quantum dots have, however, a fairly small splitting.
To give an estimate, we take typical GaAs values:
$\lambda_{1,0}=100\,{\rm nm}$ for the inter-electron distance,
$\hbar/m^*\beta=8\,\mu{\rm m}$ for the spin-orbit length,\cite{Miller} 
and a magnetic field $B=10\,{\rm T}$ (with $g=-0.44$ ); 
as a result, we obtain 
$2\Delta_-\approx 20\,\mu{\rm eV}$.
A different possibility to access $\Delta_\pm$
is to measure the spin relaxation times 
and to deduce the strength of the spin-orbit interaction therefrom.
We also note that an analogous spin splitting (though of a much larger
value) has been observed by means of transport spectroscopy 
in a recent experiment on an elongated quantum dot defined electrostatically 
inside a InAs nanowire.\cite{Fasth}

Equation~(\ref{gapsD}) has a strong
dependence on the direction of $\mbox{\boldmath $l$}={\bf B}/B$.
For some angles --- ``magic angles'' --- 
the effect of the spin-orbit interaction vanishes at the leading order.
From Eq.~(\ref{gapsD}), we find that $\Delta_+=0$ for
$\cos\theta=\lambda_-/\lambda_+$ and $\varphi'=0,\pi$, provided
$|\lambda_+|>|\lambda_-|$. 
Alternatively, if $|\lambda_-|>|\lambda_+|$,
we find that $\Delta_+=0$ for $\cos\theta=\lambda_+/\lambda_-$ 
and $\varphi'=\pi/2,3\pi/2$.
Similarly for $\Delta_-$, we have the same expressions, but with 
$\cos\theta\to-\cos\theta$.
In Fig.~\ref{DeltaPlus1},
we plot the angular dependence of $\Delta_+$ given in Eq.~(\ref{gapsD}),
calculated for the ratio $\lambda_+/\lambda_-=2$.
A similar figure can be obtained also for $\Delta_-$ in Eq.~(\ref{gapsD}),
by reflecting Fig.~\ref{DeltaPlus1} in the $(x,y)$-plane.
Clearly, the ``magic angles'' can be used to determine the relative strength
between the Rashba ($\alpha$) and Dresselhaus ($\beta$) coupling constants
(recall that $1/\lambda_\pm=m^*(\beta\pm\alpha)/\hbar$).

\begin{figure}
 \begin{center}
  \includegraphics[angle=0,width=0.4\textwidth]{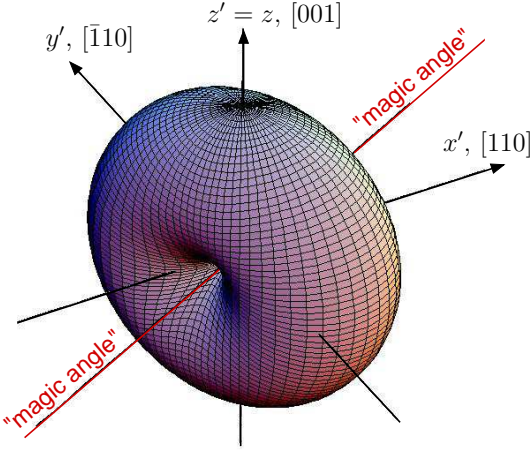}
    \caption{\small
(Color online) Opposite to Fig.~\ref{DeltaPlus1}, we plot
here the angular dependence of the splittings $\Delta_\pm$ 
(with $\Delta_+=\Delta_-$) 
as given by Eq.~(\ref{gapsDrealpsi}), which describes a quantum dot without
circular symmetry.
The degree of asymmetry needs to be large enough, such that in a magnetic
field causing the singlet-triplet crossing the imaginary part of 
$\bar{\bf r}=\langle\psi_T|{\bf r}|\psi_S\rangle$ to be negligible.
For the plot we assumed $\bar r_{y'}\lambda_+/\bar r_{x'}\lambda_-=2$.
}
    \label{DeltaSym1}
 \end{center}
\end{figure}

The angular dependence of the splittings $\Delta_\pm$ in Eq.~(\ref{gapsD})
is specific to the circular symmetry of the
harmonic confining potential used above. 
In the opposite extreme case, when the two-electron wave functions
can be chosen real (or their imaginary parts due to $B_z$ are negligible)
we obtain $\Delta_+=\Delta_-$ and
\begin{eqnarray}
\Delta_\pm&=&
\frac{|E_Z|}{\sqrt{2}}
\sqrt{\frac{\bar r_{y'}^2}{\lambda_-^2}+
\frac{\bar r_{x'}^2}{\lambda_+^2}
-\left(
\frac{\bar r_{y'}}{\lambda_-}l_{x'}
+\frac{\bar r_{x'}}{\lambda_+}l_{y'}
\right)^2},
\;\;\;\;\;\;\;\;
\label{gapsDrealpsi}
\end{eqnarray}
where $\bar{\bf r}=\langle\psi_T|{\bf r}|\psi_S\rangle$ is
the dipolar matrix element (not to be confused with $\langle r\rangle$), 
which we assume to be real.
We note that this corresponds to a quantum dot elongated
in one direction, or to a double quantum dot. 
The singlet-triplet crossing can occur in this case
at a value of $B_z$ much smaller
than the characteristic confining magnetic field.
Then, the imaginary part of  $\bar{\bf r}$ is negligible.
Note that Eq.~(\ref{gapsDrealpsi}) can be rewritten in the form
\begin{displaymath}
\Delta_\pm=\frac{|E_Z|}{\sqrt{2}}
\sqrt{\frac{\bar r_{y'}^2}{\lambda_-^2}+\frac{\bar r_{x'}^2}{\lambda_+^2}}
\sqrt{1-\cos^2(\varphi'-\varphi_0)\sin^2\theta}, 
\end{displaymath}
with $\varphi_0$
satisfying $\tan\varphi_0=\bar r_{x'}\lambda_-/\bar r_{y'}\lambda_+$.
Then, it is easy to see that Eq.~(\ref{gapsDrealpsi}) describes
a doughnut (or more precisely a {\em horn} torus) with the rotation
axis at $\varphi'=\varphi_0$ and $\theta=\pi/2$, see
Fig.~\ref{DeltaSym1}.
In this case,
the ``magic angle'' coincides with the rotation axis of the torus.
Knowing the rotation axis, one can find the relative strength of
$\alpha$ and $\beta$, provided the asymmetry of the 
quantum dot in the directions $x'$ and $y'$ is known.
In Fig.~\ref{DeltaSym1}, we plotted the angular dependence
of $\Delta_\pm$ assuming $\bar r_{y'}\lambda_+/\bar r_{x'}\lambda_-=2$.

\subsection{
Case of finite Zeeman interaction ($g\neq0$):
Non-degenerate perturbation theory
}
\label{ssecWeakME}
In Sec.~\ref{ESPSOgfinite}, we have
taken into account the matrix elements (\ref{TDHZS}) 
by means of the degenerate
perturbation theory and we obtained two avoided
crossings between the singlet and the triplet 
$|\Psi_{T_\pm}\rangle$.
We expect that accounting for these ({\em strong}) 
matrix elements
should be sufficient in the majority of cases and it
should give the dominant channel of spin relaxation 
in the vicinity of the singlet-triplet crossing.
To complete our analysis of the first order
in $H_{SO}$ contributions to the spin
relaxation, we must also consider the rest of the matrix 
elements in Eqs.~(\ref{DHZTpmT0}) and (\ref{DHZTS}).
We call these matrix elements {\em weak}.
The {\em weak} matrix elements,
despite the fact that they do not induce strong 
mixing between singlet and triplet states,
can, in principle, govern the spin relaxation in
special cases, when the {\em strong} matrix elements
become, on some reason, inefficient.
Examples of such cases include: positions
far away from the singlet-triplet crossing, where
$E_{TS}$ is comparable to the energy distance 
to other excited states; 
magnetic field ${\bf B}$ along a ``magic angle'';
and the limit of strong Coulomb interaction.

We use perturbation theory for the non-degenerate case
and account for the admixture of distant in energy 
states to the singlet-triplet subspace under consideration.
For these we first separate the weak matrix
elements from the Hamiltonian into a perturbation 
(cf. Eq.~(\ref{HHdHZHZSO})),
\begin{eqnarray}\label{tildeHnewV}
\tilde H&=&\tilde H_0+\tilde V_{\rm w},\\
\tilde H_0&=&H_d+H_Z+{\cal P}_{ST}H_{Z}^{SO},\\
\tilde V_{\rm w}&=&(1-{\cal P}_{ST})H_Z^{SO},
\end{eqnarray}
where ${\cal P}_{ST}$ projects onto the singlet-triplet subspace,
${\cal P}_{ST}A=\sum_{\mu\nu}A_{\mu\nu}|\mu\rangle\langle\nu|$,
with $\mu,\nu\in (\Psi_S,\Psi_{T_\pm},\Psi_{T_0})$ and $\forall A$.
Note that all the strong matrix elements are included
into $\tilde H_0$, whereas all the weak ones into $\tilde V_{\rm w}$.
Next, we transform
$\tilde H$ to ${\tilde H}'=(1+S_{\rm w})\tilde H(1-S_{\rm w})$,
with $S_{\rm w}\ll1$ obeying $[\tilde H_0,S_{\rm w}]=\tilde V_{\rm w}$.
After this transformation, we obtain
${\tilde H}'=\tilde H_0+{\cal O}(H_{SO}^2)$.
At the same time, by doing this transformation,
we correct the basis states to take into account the perturbation
$\tilde V_{\rm w}$ to leading order.

Considering now the phonon potential $U_{\rm ph}({\bf r}_1,{\bf r}_2,t)$,
we apply the same unitary transformations,
$\tilde U_{\rm ph}'=(1+S_{\rm w})\tilde U_{\rm ph}(1-S_{\rm w})$, 
and obtain
\begin{equation}\label{UphtoUphpSZ}
U_{\rm ph}\to U_{\rm ph}+[S_{\rm w},U_{\rm ph}].
\end{equation}
The transformation matrix $S_{\rm w}$ 
can, formally, be written as follows
\begin{equation}\label{Sw1mPST}
S_{\rm w}=\frac{1-{\cal P}_{ST}}{\hat L_d+\hat L_Z}H_Z^{SO}+{\cal O}(H_{SO}^2),
\end{equation}
where $\hat L_dA=[H_d,A]$ and $\hat L_ZA=[H_Z,A]$, 
for $\forall A$.
For simplicity, we assume small Zeeman splittings, 
$E_Z\ll \hbar\omega_0$,
which is usually the case for GaAs quantum dots. 
Neglecting then the Liouvillean $\hat L_Z$ in the denominator 
of Eq.~(\ref{Sw1mPST}) and going to relative coordinates, 
we obtain
\begin{equation}
S_{\rm w}=E_Z\left\{
\frac{1}{\hat L_d}
[\mbox{\boldmath $l$}\times
\mbox{\boldmath $\xi$}^R]\cdot
\mbox{\boldmath $\Sigma$}+
\frac{1-{\cal P}_{ST}}{2\hat L_d}
[\mbox{\boldmath $l$}\times
\mbox{\boldmath $\xi$}^r]\cdot
\mbox{\boldmath $\sigma$}
\right\}.
\label{SwHarmonic}
\end{equation}
The first term in Eq.~(\ref{SwHarmonic}) is easy to 
evaluate for the harmonic confinement and because the 
Coulomb interaction does not couple to the
coordinate ${\bf R}$.
It gives
\begin{eqnarray}
\frac{E_Z}{\hat L_d}
[\mbox{\boldmath $l$}\times
\mbox{\boldmath $\xi$}^R]\cdot
\mbox{\boldmath $\Sigma$}&=&
\frac{-E_Z}{2m^*\omega_0^2}
[\mbox{\boldmath $l$}\times
\mbox{\boldmath $D$}^R]\cdot
\mbox{\boldmath $\Sigma$}+f({\bf R}),
\label{term1SwR}
\\
\mbox{\boldmath $D$}^R&:=&\left(
\lambda_-^{-1}
\partial/\partial R_{y'},
\lambda_+^{-1}
\partial/\partial R_{x'},0
\right),
\label{defDR}
\;\;\;\;\;\;\;\;\;
\end{eqnarray}
where $f({\bf R})$ is a function of coordinates only,
which drops out upon substitution
of $S_{\rm w}$ into Eq.~(\ref{UphtoUphpSZ}).
The second term in Eq.~(\ref{SwHarmonic})
is more difficult to evaluate and
it is the subject of a separate publication.

We note here that the first term in Eq.~(\ref{SwHarmonic})
is responsible for the triplet-triplet relaxation.
In combination with the phonon potential (or any other orbital bath),
it results in finite relaxation rates for the
transitions $T_+\leftrightarrow T_0$ and $T_-\leftrightarrow T_0$.
The transition $T_+\leftrightarrow T_-$ is, however, forbidden
at the first order of $H_{SO}$, because it requires a change
of spin $\Delta S_z=2$.
We evaluate the triplet-triplet relaxation rates in Sec.~\ref{RelaxRatesTT}.

The second term in Eq.~(\ref{SwHarmonic}) is similar by nature
to the first term, however, it gives rise to singlet-triplet 
relaxation. The projector $1-{\cal P}_{ST}$, multiplying the second term, 
ensures that only distant in energy states are admixed to the
singlet-triplet subspace under consideration.
In the vicinity of the singlet-triplet (avoided) crossing,
there is a strong admixture between the singlet and triplet 
states inside the singlet-triplet subspace, see 
Sec.~\ref{ESPSOgfinite}.
Therefore, the second term can usually be neglected, except
for special cases mentioned above.
The second term contributes, due to its spin symmetry, to the 
transitions $T_+\leftrightarrow S$ and $T_-\leftrightarrow S$,
as also the {\em strong} matrix elements do.
In this paper, we neglect the contribution of the
second term to the spin relaxation.

\section{Phonon-induced spin relaxation} %
\label{spinrelaxationphonon}
The spin dynamics, decoherence, and relaxation are
described by the quantum dot density matrix $\rho(t)$.
Within the Bloch-Wangsness-Redfield theory,\cite{Slichter}
the density matrix obeys a set of linear
differential equations, 
\begin{equation}
\frac{d}{dt}\rho_{\alpha\alpha'}(t)=
-i\omega_{\alpha\alpha'}\rho_{\alpha\alpha'}(t)
+\sum_{\beta\beta'}R_{\alpha\alpha'\beta\beta'}\rho_{\beta\beta'}(t),
\label{RedfieldEq}
\end{equation}
where $R_{\alpha\alpha'\beta\beta'}=\left(R_{\alpha'\alpha\beta'\beta}\right)^*$ are time-independent constants,
which form the Redfield tensor, and
$\omega_{\alpha\alpha'}=(E_{\alpha}-E_{\alpha'})/\hbar$
are the transition frequencies,
with $E_\alpha$ being the energy of the level $\alpha$.
Equations (\ref{RedfieldEq}) are usually derived 
microscopically using the Born-Markov 
approximation.\cite{Slichter}
In this approximation, the system-bath coupling
is assumed to be sufficiently weak, so that the
next-to-leading order Born terms contribute at times
when the density matrix has already reached its
equilibrium value within the desired accuracy.
Additionally, the bath correlation time $\tau_c$
is assumed to be much shorter than the characteristic 
times occurring in Eq.~(\ref{RedfieldEq}).
As a matter-of-fact, the evolution of the 
density matrix
on times shorter or comparable to $\tau_c$ is
not described by Eq.~(\ref{RedfieldEq}) and is
rarely of practical interest.
For the phonon bath considered
in this paper, the correlation time
$\tau_c$ is extremely short.
It is the time it takes a phonon to 
traverse a distance equal to the quantum dot size,
i.e. $\tau_c=\langle r\rangle/s$,
where $\langle r\rangle$ is the average distance 
between the electrons and $s$ is the speed of 
sound in the sample.
For GaAs quantum dots, we estimate 
$\tau_c\simeq10\,{\rm ps}$.
At the same time, the characteristic
spin relaxation times in GaAs quantum dots
range from $100\,\mu{\rm s}$ to $1\,{\rm s}$,
as measured 
experimentally.\cite{SingleShotNature,AmashaZumbuhl,Kroutvar,FujisawaT1Nature,HansonST2el,Meunier} 

In the Born-Markov approximation, the Redfield tensor 
reads
\begin{eqnarray}
R_{\alpha\alpha'\beta\beta'}&=&
\Gamma_{\beta'\alpha'\alpha\beta}^{+}+\Gamma_{\beta'\alpha'\alpha\beta}^{-}\nonumber\\
&&-\delta_{\alpha'\beta'}\sum_{\gamma}\Gamma_{\alpha\gamma\gamma\beta}^{+}
-\delta_{\alpha\beta}\sum_{\gamma}\Gamma_{\beta'\gamma\gamma\alpha'}^{-},
\;\;\;\;\;\;\;\;\;
\label{RedfieldRtens}
\end{eqnarray}
where 
$(\Gamma_{\alpha\alpha'\beta\beta'}^{+})^*=
\Gamma_{\beta'\beta\alpha'\alpha}^{-}$ are given as follows
\begin{equation}
\Gamma_{\alpha\alpha'\beta\beta'}^{+}=\int_{0}^\infty \frac{dt}{\hbar^2}
e^{-i\omega_{\beta\beta'}t}
\overline{\langle\alpha|U_{\rm int}(t)|\alpha'\rangle\langle\beta|U_{\rm int}|\beta'\rangle},
\label{Gpaabb}
\end{equation} 
\begin{equation}
\Gamma_{\alpha\alpha'\beta\beta'}^{-}=\int_{0}^\infty \frac{dt}{\hbar^2}
e^{-i\omega_{\alpha\alpha'}t}
\overline{\langle\alpha|U_{\rm int}|\alpha'\rangle\langle\beta|U_{\rm int}(t)|\beta'\rangle}.
\label{Gmaabb}
\end{equation} 
Here, the bar denotes averaging over the bath degrees 
of freedom, $|\alpha\rangle$ is a state of the quantum dot,
and $U_{\rm int}$ is the interaction between the quantum dot
and the bath.
The time dependence of $U_{\rm int}(t)$ is taken 
with respect to the bath only,
$U_{\rm int}(t)=\exp(iH_Bt/\hbar)U_{\rm int}\exp(-iH_Bt/\hbar)$, where $H_B$ is the 
Hamiltonian of the bath.
In the expressions above it is assumed that $\overline{U_{\rm int}(t)}=0$.

In our case, the system-bath interaction is obtained from the transformation 
in Eq.~(\ref{UphtoUphpSZ})
\begin{equation}
U_{\rm int}(t)=U_{\rm ph}(t)+[S_{\rm w},U_{\rm ph}(t)],
\label{UintSB}
\end{equation}
with $S_{\rm w}$ given in Eq.~(\ref{SwHarmonic}).
The time dependence of the phonon potential $U_{\rm ph}(t)$ is governed by the Hamiltonian 
of the free phonons,
\begin{equation}
H_{B}\equiv H_{\rm ph}=\sum_{{\bf q}j}\hbar\omega_{{\bf q}j}
\left(b_{{\bf q}j}^\dag b_{{\bf q}j}+1/2\right),
\end{equation}
where the index $j$ runs over the three branches of acoustic phonons. 
The optical phonons can be neglected, since usually $|E_{TS}|,|E_Z|<\hbar\omega_{\rm opt}$, 
where $\hbar\omega_{\rm opt}$ is the energy of optical phonons.
The acoustic phonon potential can be rewritten in terms of relative coordinates
as follows
\begin{eqnarray}
U_{\rm ph}({\bf R},{\bf r},t)&=&
\sum_{{\bf q}j}
M_{{\bf q}j}
F(q_z)e^{i{\bf q}_\parallel \cdot{\bf R}}
\cos\left({\bf q}_\parallel \cdot{\bf r}/2\right)
\nonumber\\
&&\times
\left[b_{-{\bf q}j}^\dag(t)+b_{{\bf q}j}(t)\right],
\nonumber \\
M_{{\bf q}j}&=&
\sqrt{{2\hbar\over\rho_c\omega_{{q}j}}} 
(e\beta_{{\bf q}j}-iq\Xi_{{\bf q}j}),
\label{Uphnew}  
\end{eqnarray}
where $b_{{\bf q}j}(t)=b_{{\bf q}j}\exp(-i\omega_{{\bf q}j}t)$ gives the time
dependence mentioned above for $U_{\rm ph}(t)$.
The averaging over the bath (denoted by bar
in Eqs.~(\ref{Gpaabb}) and (\ref{Gmaabb}))
is evaluated as a thermodynamic mean using the free phonon statistical operator
\begin{equation}
\rho_{\rm ph}=Z_{\rm ph}^{-1}e^{-H_{\rm ph}/T},
\end{equation}  
where $T$ is the temperature in units of energy ($k_B=1$) 
and $Z_{\rm ph}$ is the phonon partition function.

In addition to the Born-Markov approximation,
a further important simplification arises in 
the case of the spin-phonon coupling considered here.
The elements of the Redfield tensor
$R_{\alpha\alpha'\beta\beta'}$
are much smaller than the corresponding transition
frequencies
$\omega_{\alpha\alpha'}$ for $\alpha\neq\alpha'$.
One can then expand in terms of $R_{\alpha\alpha'\beta\beta'}/\omega_{\alpha\alpha'}\ll 1$
in the characteristic equation of the master equation (\ref{RedfieldEq}).
At the zeroth order, the off-diagonal part of $\rho(t)$ decouples
from the master equation (\ref{RedfieldEq}) and can be considered separately.
The diagonal part, then, obeys the Pauli master equation
\begin{eqnarray}
\frac{d}{dt}\rho_{\alpha\alpha}(t)&=&\sum_{\beta}R_{\alpha\alpha\beta\beta}\rho_{\beta\beta}(t)\nonumber\\
&=&\sum_\beta\left[\Gamma_{\alpha\beta}\rho_{\beta\beta}-
\Gamma_{\beta\alpha}\rho_{\alpha\alpha}\right],
\label{PauliMasterEqs}
\end{eqnarray}
where $\Gamma_{\alpha\beta}$ is obtained from Eqs.~(\ref{Gpaabb}) and (\ref{Gmaabb}) as follows
\begin{equation}
\Gamma_{\alpha\beta}=\Gamma_{\beta\alpha\alpha\beta}^{+}+\Gamma_{\beta\alpha\alpha\beta}^{-}.
\end{equation}
In our notation,
$\Gamma_{\alpha\beta}$ is the probability per unit time to transit from state $\beta$ to state $\alpha$.
For the rates $\Gamma_{\alpha\beta}$ the time integral can be extended to $-\infty$ in the lower bound,
arriving at the Fermi ``golden rule'' rate
\begin{equation}
\Gamma_{\alpha\beta}=\int_{-\infty}^{+\infty} \frac{dt}{\hbar^2}
e^{-i\omega_{\alpha\beta}t}
\overline{\langle\beta|U_{\rm int}(t)|\alpha\rangle\langle\alpha|U_{\rm int}|\beta\rangle}.
\label{GabGR}
\end{equation}
Provided the bath is at thermal equilibrium (which is our assumption),
the rates $\Gamma_{\alpha\beta}$  obey the detailed balance equation
\begin{equation}\label{detailedbalance}
\Gamma_{\alpha\beta}=\Gamma_{\beta\alpha}\exp\left({\frac{E_\beta-E_\alpha}{T}}\right).
\end{equation}

\begin{figure}
 \begin{center}
  \includegraphics[angle=0,width=0.23\textwidth]{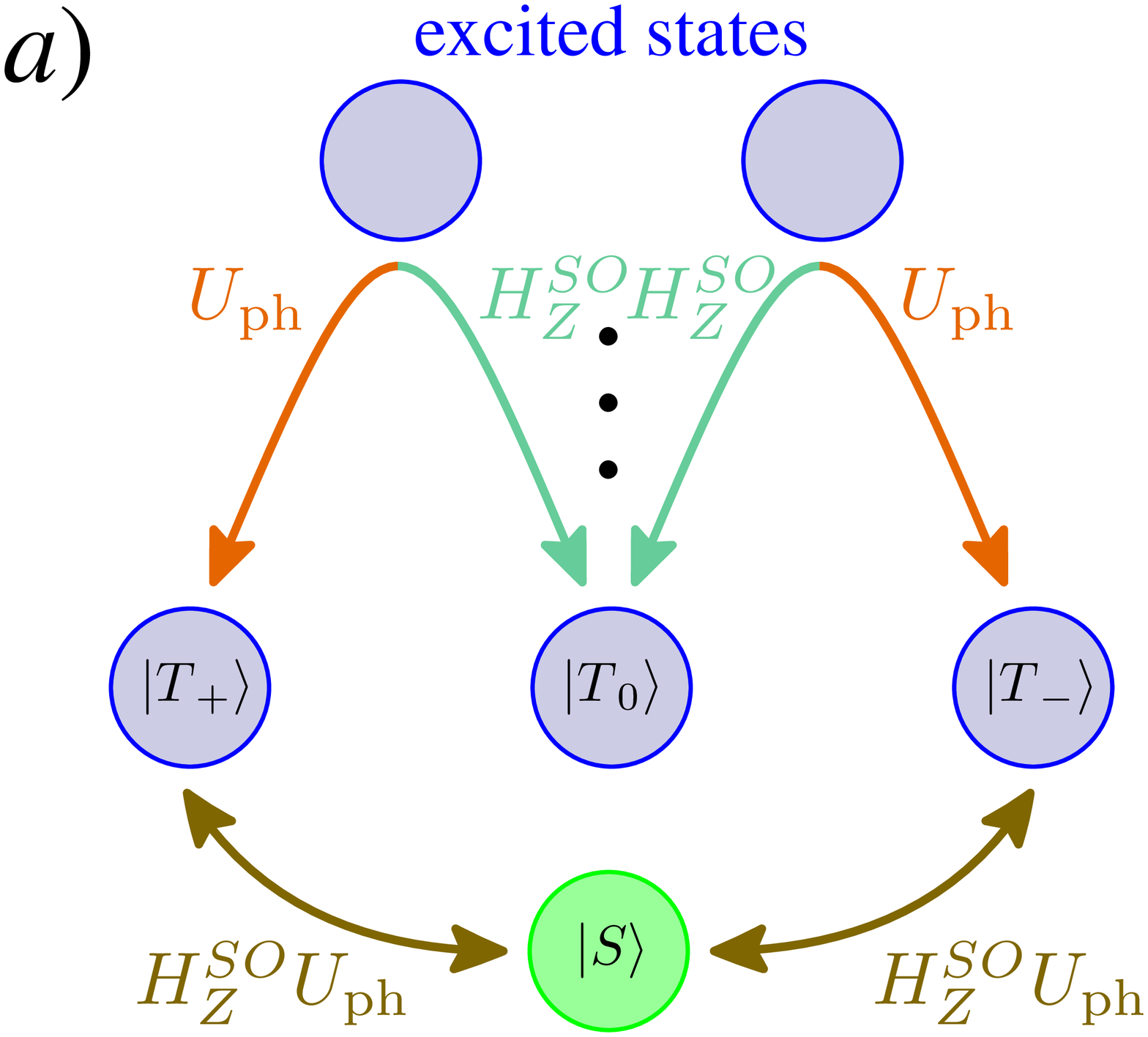}
\hfill
   \includegraphics[angle=0,width=0.23\textwidth]{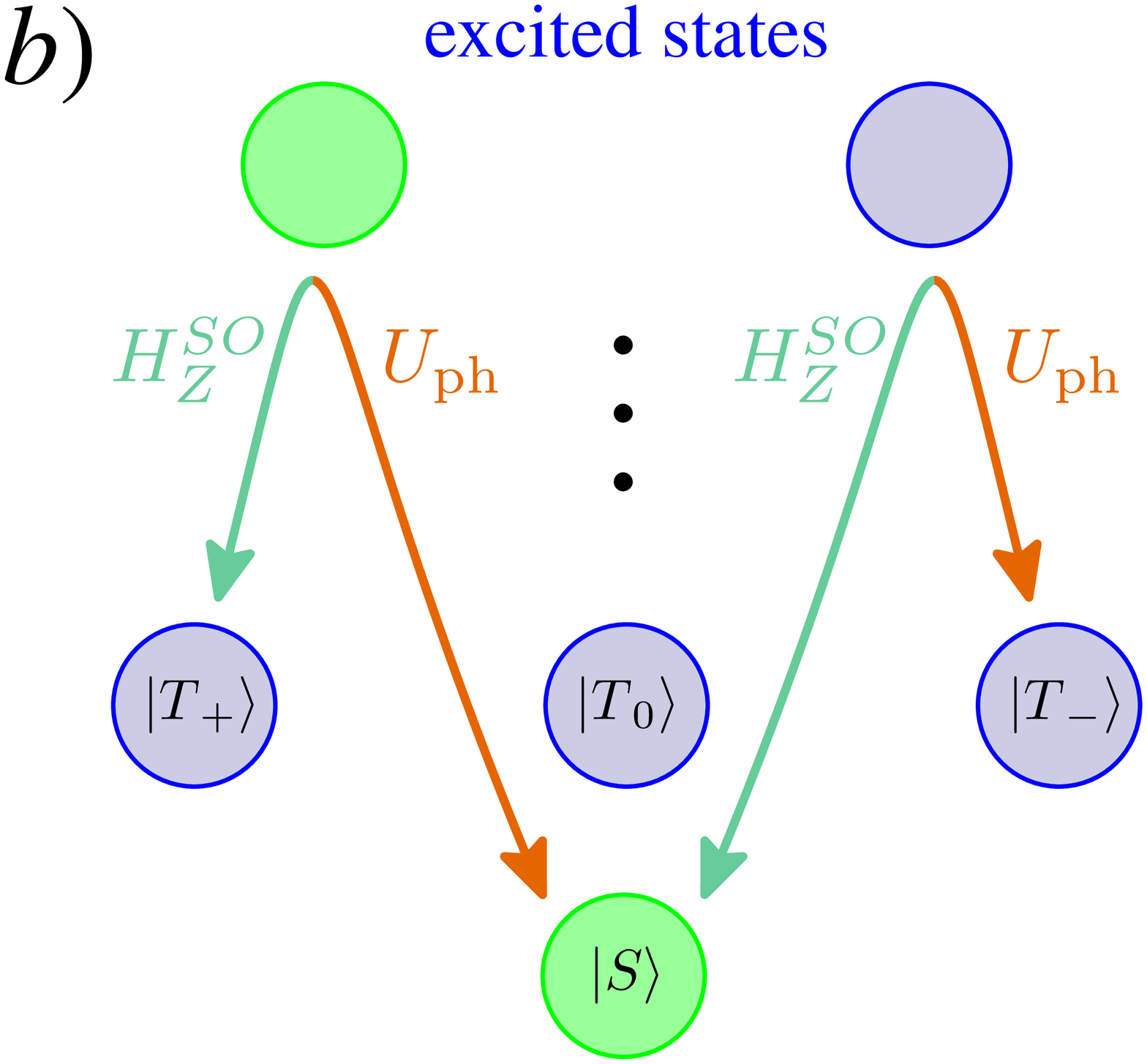}
    \caption{\small
(Color online) Spin relaxation processes accounted for (a) and neglected (b) in our treatment.
The arrows show transitions between states of the lowest in energy 
singlet-triplet subspace.
The excited states (upper row) are involved only in virtual 
transitions as intermediate states.
(a) We take into account the {\em strong} matrix elements,
which mix the singlet-triplet subspace inside itself,
allowing phonon-induced transitions to occur between 
the singlet and two triplet states.
We denote these transitions by the lower-most arrows, where the
product $H_{Z}^{SO}U_{\rm ph}$ next to the arrows denotes
that the transition occurs under the simultaneous action of 
$H_{Z}^{SO}$ and $U_{\rm ph}$.
Additionally, we take into account the 
{\em weak} matrix elements for the triplet-triplet transition.
In contrast to the singlet-triplet transition, 
a triplet-triplet transition requires
the participation of an excited state.
During the virtual process, $H_{Z}^{SO}$ and $U_{\rm ph}$ 
act separately on the two halves of the process 
(e.g. going up with $H_{Z}^{SO}$ and coming down with $U_{\rm ph}$).
We denote this difference by placing $H_{Z}^{SO}$ and $U_{\rm ph}$
on opposite sides of the arrow.
Note that $H_{Z}^{SO}$ connects states with different spin, 
whereas $U_{\rm ph}$ with the same.
(b) Transitions between the singlet and triplet states
due to the {\em weak} matrix elements are neglected.
These processes are relevant only in special cases mentioned
in the text (see Sec.~\ref{ssecWeakME}).
}
    \label{realVSvitrual}
 \end{center}
\end{figure}

In what follows, we calculate the relaxation rates 
for transitions between the quantum dot eigenstates, 
using the system-bath interaction (\ref{UintSB}) 
and the definition (\ref{GabGR}).
Before proceeding, we summarize our main results
from Sec.~\ref{ESPSO} regarding the admixture processes
contributing to spin relaxation.
The admixture processes are illustrated in Fig.~\ref{realVSvitrual},
where the arrows denote finite relaxation rates
occurring either due to the direct mixing inside the
singlet-triplet subspace (Sec.~\ref{ESPSOgfinite})
or due to virtual excitations to states outside
the singlet-triplet subspace (Sec.~\ref{ssecWeakME}).
In Fig.~\ref{realVSvitrual}(a), we show the processes
which we take into account in our treatment, 
whereas in Fig.~\ref{realVSvitrual}(b)
we show the processes which we neglect.
The neglected processes were discussed in Sec.~\ref{ssecWeakME}.
Finally, we remark that the following relaxation rates 
are identically zero at the leading order of spin-orbit 
interaction
\begin{equation}
\Gamma_{ST_0}=\Gamma_{T_+T_-}=0,
\label{forbiddentransSTTT}
\end{equation}
as well as the reverse rates which follow from the
detailed balance (\ref{detailedbalance}).

\subsection{Singlet-triplet relaxation rates}
\label{RelaxRates}
The Hilbert space of our system is spanned by $N=4$ states: $|\Psi_{S}\rangle$,
$|\Psi_{T_+}\rangle$, $|\Psi_{T_-}\rangle$ and $|\Psi_{T_0}\rangle$.
However, all of these states, except $|\Psi_{T_0}\rangle$, are mixed with
one another by the interaction $H_Z^{SO}$ in Eq.~(\ref{DeltaHZ}).
In Sec.~\ref{ESPSOgfinite}, we found the eigenstates and the corresponding
energies of the quantum dot in the presence of $H_Z^{SO}$.
The eigenstates are given in Eq.~(\ref{Psiabc}), with
the coefficients in Eq.~(\ref{aaacoeff}), and the renormalized energy levels are
given in Eqs.~(\ref{Epmonehalf}) and (\ref{E0ETpm0}).
Here, we calculate the relaxation rates between these eigenstates, 
taking into account only the mixing due to the {\em strong} matrix elements
of $H_Z^{SO}$.

We proceed by letting $U_{\rm int}=U_{\rm ph}$ and focusing on the states
\begin{equation}
|\Psi_{0}\rangle,\; |\Psi_{+}\rangle,\; |\Psi_{-}\rangle\, ,
\end{equation}
which are linear combinations of the singlet $|\Psi_S\rangle$ and two 
triplet $|\Psi_{T_\pm}\rangle$ states, as shown in Eq.~(\ref{Psiabc}).
Then, it is straightforward to obtain the matrix elements of $U_{\rm ph}$,
\begin{eqnarray}
\langle\Psi_\alpha|U_{\rm ph}|\Psi_\beta\rangle&=& 
a_\alpha a_\beta
\left[
\langle\psi_S|U_{\rm ph}|\psi_S\rangle-
\langle\psi_T|U_{\rm ph}|\psi_T\rangle
\right]\nonumber\\
&&+\delta_{\alpha\beta}\langle\psi_T|U_{\rm ph}|\psi_T\rangle,
\label{PsiPsiUint}
\end{eqnarray}
where $\alpha=0,\pm$ labels the considered eigenstates, and $|\psi_{S(T)}\rangle$ is the
singlet (triplet) orbital wave function. 
The coefficients $a_{\alpha}$ read
\begin{eqnarray}
a_\alpha&=&\left[1+\frac{\Delta_+^2}{(E_\alpha-E_{T_+})^2}+
\frac{\Delta_-^2}{(E_\alpha-E_{T_-})^2}\right]^{-1/2},\;\;\;\;\;
\label{aalphaEalpha}
\end{eqnarray}
where $E_\alpha=E_0,E_\pm$ are the exact solutions of Eq.~(\ref{seceqST}).
The last term in Eq.~(\ref{PsiPsiUint}) is irrelevant for our further discussion, since we consider only
$\Gamma_{\alpha\beta}$ with $\alpha\neq\beta$.
Thus, from the first line in Eq.~(\ref{PsiPsiUint}), we see that this mechanism of spin relaxation
relies on the difference between the singlet and triplet charge profiles,
$|\psi_S({\bf r}_1,{\bf r}_2)|^2-|\psi_T({\bf r}_1,{\bf r}_2)|^2$.
We show below that the Coulomb interaction has a strong effect on the magnitude of this difference,
making the singlet and triplet charge distributions look the same in the limit of strong
Coulomb interaction ($\lambda/a^*_B\gg 1$).
Thus, we expect this mechanism to be most efficient in quantum dots with weak and moderate Coulomb interaction
($\lambda/a^*_B\lesssim 1$), and less efficient in quantum dots with strong Coulomb interaction
and in double quantum dots. 

\begin{figure}
 \begin{center}
  \includegraphics[angle=0,width=0.4\textwidth]{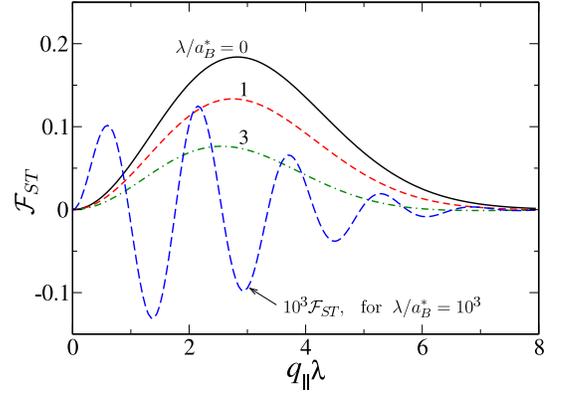}
    \caption{\small
(Color online) The form-factor ${\cal F}_{ST}(q_\parallel)$ for $\lambda/a_B^*=0,1,3$, as well as
 $10^3{\cal F}_{ST}(q_\parallel)$ for $\lambda/a_B^*=10^3$.
}
    \label{calFSTfig}
 \end{center}
\end{figure}

Using the wave functions found in Sec.~\ref{EnWaveFuncs}, we evaluate next the following form-factor
\begin{eqnarray}
{\cal F}_{ST}(\mbox{\boldmath $q$}_\parallel)&=&\langle\psi_S|
e^{i{\bf q}_\parallel{\bf R}}
\cos\left({\bf q}_\parallel{\bf r}/2\right)
|\psi_S\rangle\nonumber\\
&&-\langle\psi_T|
e^{i{\bf q}_\parallel{\bf R}}
\cos\left({\bf q}_\parallel{\bf r}/2\right)
|\psi_T\rangle.
\label{calFSTdeff}
\end{eqnarray}
For the wave functions $|\psi_S\rangle=|0000\rangle$ and $|\psi_T\rangle=|000,-1\rangle$, 
we obtain
\begin{eqnarray}
{\cal F}_{ST}(q_\parallel)&=&e^{-{1\over 4}q_\parallel^2\Lambda^2}
\left[
e^{-{1\over 16}q_\parallel^2\tilde\lambda_0^2}{}_1F_1\left(-\sqrt{\gamma_0},1;{q_\parallel^2\tilde\lambda_0^2\over 16}\right)
\right.
\nonumber\\
&&\left.
-e^{-{1\over 16}q_\parallel^2\tilde\lambda_1^2}{}_1F_1\left(-\sqrt{1+\gamma_1},1;{q_\parallel^2\tilde\lambda_1^2\over 16}\right)
\right],\;\;\;\;\;\;\;\;\;\;
\label{calFST}
\end{eqnarray}
where $\gamma_m$ and $\tilde\lambda_m$, with $m=0,1$, are the variational parameters found in Appendix~\ref{appA}.
Note that ${\cal F}_{ST}$ is independent of the in-plane angle of the emitted phonon, due to the the
circular symmetry of the confining potential.
In the absence of Coulomb interaction ($\gamma_m=0$ and $\tilde\lambda=\lambda$), Eq.~(\ref{calFST}) reduces to
the following expression ($\lambda/a_B^*=0$)
\begin{equation}
{\cal F}_{ST}\left(q_\parallel\right)=\frac{\hbar q_\parallel^2}{8m^*\omega}
e^{
-\frac{\hbar q_\parallel^2}{4m^*\omega}},
\label{calFSTzeroCI}
\end{equation}
where we used that $\lambda=\sqrt{2\hbar/m^*\omega}$ and $\Lambda=\sqrt{\hbar/2m^*\omega}$.
Note that the function ${\cal F}_{ST}\left(q_\parallel\right)$ in Eq.~(\ref{calFSTzeroCI}) has a single
scale, $q_\parallel\sim \sqrt{4m^*\omega/\hbar}$.
For strong Coulomb interaction, however, we find that ${\cal F}_{ST}\left(q_\parallel\right)$ has
two scales; an additional scale arises due to the Coulomb interaction and it is given by
$q_\parallel\sim 1/\langle r\rangle$, where $\langle r\rangle=\lambda(\lambda/2a_B^*)^{1/3}$.
In the limit $\lambda/a_B^*\gg 1$, we find ${\cal F}_{ST}\left(q_\parallel\right)$ from
Eq.~(\ref{calFSTdeff}) for both $q_\parallel\ll 1/\langle r\rangle$ and $q_\parallel\gg 1/\langle r\rangle$,
and then match the two asymptotes into the following crossover function ($\lambda/a_B^*\gg 1$)
\begin{equation}
{\cal F}_{ST}\left(q_\parallel\right)=
\frac{q_\parallel a_B^*}{3}
J_1\left(\frac{q_\parallel\langle r\rangle}{2}\right)
e^{
-\frac{(1+\sqrt{3})\hbar q_\parallel^2}{8\sqrt{3}m^*\omega}},
\label{calFSTlargeCI}
\end{equation}
where $\langle r\rangle=\lambda(\lambda/2a_B^*)^{1/3}$ and $J_{1}(x)$ is a Bessel function of the
first kind.
We checked numerically that Eq.~(\ref{calFSTlargeCI}) coincides with Eq.~(\ref{calFST})
in the whole range of $q_\parallel$ for $\lambda/a_B^*\gg 1$.
By comparing Eq.~(\ref{calFSTlargeCI}) to Eq.~(\ref{calFSTzeroCI}), we find that
the maximal value of ${\cal F}_{ST}\left(q_\parallel\right)$
scales like $\propto(a_B^*/\lambda)^{7/6}$ for strong Coulomb interaction. 
In Fig.~\ref{calFSTfig}, we plot ${\cal F}_{ST}\left(q_\parallel\right)$ as a function of 
$q_\parallel\lambda$ for $\lambda/a_B^*=0,1,3$, which shows that
${\cal F}_{ST}\left(q_\parallel\right)$ gets suppressed by the Coulomb interaction for $\lambda/a_B^*\gtrsim 1$.
In addition, we also plot $10^3{\cal F}_{ST}\left(q_\parallel\right)$ for $\lambda/a_B^*=10^3$, 
which shows that ${\cal F}_{ST}\left(q_\parallel\right)$ oscillates on the scale of
$\Delta q_\parallel=1/\langle r\rangle$, see also Eq.~(\ref{calFSTlargeCI}).

Next, we find the relaxation rates $\Gamma_{\alpha\beta}$ 
by substituting Eq.~(\ref{PsiPsiUint}) into Eq.~(\ref{GabGR})
and averaging over phonons at thermal equilibrium. 
We obtain
\begin{equation}
\Gamma_{\alpha\beta}=a_\alpha^2a_\beta^2
\left[\tilde\Gamma_{\rm\it LA}^{\rm\it DP}(\omega_{\beta\alpha})+\tilde\Gamma_{\rm\it LA}^{\rm\it PE}(\omega_{\beta\alpha})
+\tilde\Gamma_{\rm\it TA}^{\rm\it PE}(\omega_{\beta\alpha})\right],
\label{Gammaabres}
\end{equation}
where $\omega_{\beta\alpha}=(E_\beta-E_\alpha)/\hbar$ is the transition frequency,
the coefficients $a_\alpha$ are given in Eq.~(\ref{aalphaEalpha}), and
\widetext
\begin{eqnarray}\label{eqGammaLADP}
\tilde\Gamma_{\rm\it LA}^{\rm\it DP}(w)&=&\left(1+N_w\right)
\frac{2\Xi_0^2w^3}{\pi\hbar\rho_cs_1^5}
\int_{0}^{\pi/2}
{d\vartheta}
\left|{\cal F}_{ST}\left({w\over s_1}\sin\vartheta\right)
F\left({w\over s_1}\cos\vartheta\right)\right|^2
\sin{\vartheta},\\
\label{eqGammaLAPE}
\tilde\Gamma_{\rm\it LA}^{\rm\it PE}(w)&=&\left(1+N_w\right)
\frac{36\pi e^2h_{14}^2w}{\kappa^2\hbar\rho_cs_1^3}
\int_{0}^{\pi/2}{d\vartheta} 
\left|{\cal F}_{ST}\left({w\over s_1}\sin\vartheta\right)
F\left({w\over s_1}\cos\vartheta\right)\right|^2
\sin^5{\vartheta}\cos^2\vartheta,\\
\label{eqGammaTAPE}
\tilde\Gamma_{\rm\it TA}^{\rm\it PE}(w)&=&\left(1+N_w\right)
\frac{4\pi e^2h_{14}^2w}{\kappa^2\hbar\rho_cs_2^3}
\int_{0}^{\pi/2}{d\vartheta} 
\left|{\cal F}_{ST}\left({w\over s_2}\sin\vartheta\right)
F\left({w\over s_2}\cos\vartheta\right)\right|^2
\left(8\cos^4\vartheta+\sin^4\vartheta\right)\sin^3{\vartheta},
\end{eqnarray}
\endwidetext
\noindent
where $N_w=(e^{\hbar w/T}-1)^{-1}$.
The subscripts ${\rm\it LA}$ and ${\rm\it TA}$ on the left-hand side in Eqs.~(\ref{eqGammaLADP})-(\ref{eqGammaTAPE})
correspond to the dot electrons interacting with either a longitudinal acoustic ({\rm\it LA}) or a
transverse acoustic ({\rm\it TA}) phonon.
The superscripts give the mechanism of the electron-phonon interaction: deformation potential (${\rm\it DP}$)
or piezoelectric (${\rm\it PE}$).
In Eq.~(\ref{Gammaabres}), we have taken into account only the acoustic phonons.
The optical phonons can be neglected due to their large excitation energy 
($\hbar\omega_{\rm\it LO}\approx 37\,{\rm meV}$ and $\hbar\omega_{\rm\it TO}\approx 34\,{\rm meV}$, at $T=0$ in GaAs).
For the acoustic phonons, we used a linear dispersion relation, $\omega_{{\bf q}j}=s_jq$,
which requires that the transition energy $\hbar\omega_{\beta\alpha}$ be much smaller 
than the Debye energy ($\approx 30\,{\rm meV}$).
In practice, the linear dispersion relation is fairly accurate for
$\hbar\omega_{\rm\it LA}<23\,{\rm meV}$ and
$\hbar\omega_{\rm\it TA}<8\,{\rm meV}$, in GaAs.
In this material, the speed of sound is $s_1=4.73\times 10^5\,{\rm cm}/{\rm s}$ for LA phonons,
and $s_2=s_3=3.35\times 10^5\,{\rm cm}/{\rm s}$ for TA phonons.
As mentioned above, we take into account both the deformation-potential ($\Xi_0=6.7\,{\rm eV}$)
and piezoelectric ($h_{14}=-0.16\,{\rm C}/{\rm m}^2$) mechanisms of electron-phonon interaction. 
Finally, we assume a dielectric constant $\kappa=13.1$ and a crystal mass density $\rho_c=5.3\,{\rm g}/{\rm cm}^3$. 

\begin{figure}
 \begin{center}
  \includegraphics[angle=0,width=0.48\textwidth]{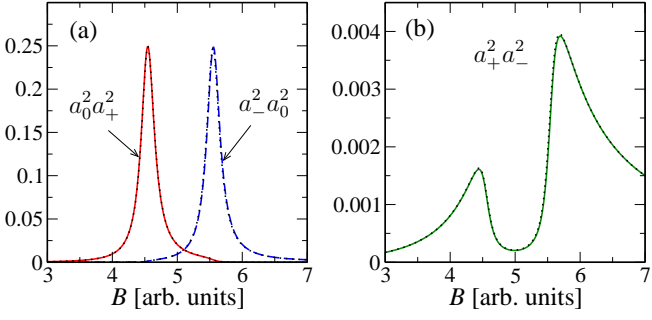}
    \caption{\small
(Color online) (a) The factors $a_0^2a_+^2$ (solid line) and $a_0^2a_-^2$ (dashed line)
in Eq.~(\ref{a02apm2eq}) as functions of the magnetic field $B_z=B$.
(b) Similarly, the factor $a_+^2a_-^2$ in Eq.~(\ref{ap2am2eq}) as a function of $B$.
For the plot we used the following model (arbitrary units):
$\Delta_+=0.15|E_Z|$, $\Delta_-=0.1|E_Z|$, $E_S=0$,
$E_{T_\pm}=10-2B\pm E_Z$, and $E_Z=-0.2B$.
The exact solutions for $a_\alpha^2a_\beta^2$ (dotted line) coincide with the approximated ones
(see Eqs.~(\ref{a02apm2eq}),(\ref{ap2am2eq})) on the scale of the plot.
Note the different scales on the left and right panels.
}
    \label{ai2aj2figure}
 \end{center}
\end{figure}

We consider further the relaxation rates: $\Gamma_{+,0}$, $\Gamma_{0,-}$, and $\Gamma_{+,-}$.
The other relaxation rates are either zero, in this approximation, or can be found from
the detailed-balance principle in Eq.~(\ref{detailedbalance}).
Note that the states $|\Psi_\pm\rangle$ and $|\Psi_0\rangle$ are linear combinations
of the singlet $|\Psi_S\rangle$ and triplet $|\Psi_{T_\pm}\rangle$ states, as given in Eq.~(\ref{Psiabc}).
Also recall that the triplet state $|\Psi_{T_0}\rangle$ decouples for these approximations, and hence it is long lived.
To analyze the relaxation rates $\Gamma_{+,0}$, $\Gamma_{0,-}$, $\Gamma_{+,-}$, 
we first consider the factor $a_\alpha^2a_\beta^2$ in Eq.~(\ref{Gammaabres}) separately.
As before, we assume $\Delta_\pm\ll E_Z$, or equivalently $\langle r\rangle\ll\lambda_{SO}$.
In addition, we also assume that $\Delta_+$ and $\Delta_-$ do not differ much from each other,
{\em i.e.} $\Delta_\pm\gg \Delta_\mp^2/E_Z$.
Under these assumptions and for $E_Z=-|E_Z|$, we obtain the following expressions for $a_\alpha^2a_\beta^2$,
\begin{eqnarray}
2a_0^2a_\pm^2\!\!&=&\!\!\frac{\Delta_\pm^2}{J_\pm^2+4\Delta_\pm^2}\left(1\pm\frac{J_\mp}{\sqrt{J_\mp^2+4\Delta_\mp^2}}\right),
\label{a02apm2eq}
\\
4a_+^2a_-^2\!\!&=&\!\!\!\left(1+\frac{J_+}{\sqrt{J_+^2+4\Delta_+^2}}\right)
\left(1-\frac{J_-}{\sqrt{J_-^2+4\Delta_-^2}}\right),
\label{ap2am2eq}
\;\;\;\;\;\;\;\;\;
\end{eqnarray}
where $J_\pm=E_{T_\pm}-E_S$ (or $J_\pm=E_{TS}\pm E_Z$).
In case of $E_Z=|E_Z|$, one should replace $J_\pm\to J_\mp$ and $\Delta_\pm\to\Delta_\mp$ in Eqs.~(\ref{a02apm2eq}) and (\ref{ap2am2eq}).
In Fig.~\ref{ai2aj2figure},
we plot $a_\alpha^2a_\beta^2$ versus $B$ across the singlet-triplet crossing
in a model specified in the caption of the figure.
\widetext
\begin{center}
\begin{table}[b]
\begin{tabular}{||c||c|c|c|c|c||}
\hline
{} & $E_{TS}>|E_Z|$,  & $E_{TS}\approx |E_Z|$, & $|E_Z|>E_{TS}>-|E_Z|$, & $E_{TS}\approx -|E_Z|$, & $E_{TS}<-|E_Z|$, \\
{} & $E_{TS}-|E_Z|\gg\Delta_+$  & $\left|E_{TS}-|E_Z|\right|\ll\Delta_+$ & $|E_Z|-|E_{TS}|\gg\Delta_\pm$ &
$\left|E_{TS}+|E_Z|\right|\ll\Delta_-$ & $-E_{TS}-|E_Z|\gg\Delta_-$ \\ \hline
$a_0^2a_+^2$ & $\Delta_+^2/(E_{TS}-|E_Z|)^2$ & $1/4$ & $\Delta_+^2/(E_{TS}-|E_Z|)^2$ & $\lesssim\left(\Delta_+/2E_Z\right)^2$ &
$\Delta_+^2\Delta_-^2/(E_{TS}^2-E_Z^2)^2$ \\ \hline
$a_0^2a_-^2$ & $\Delta_+^2\Delta_-^2/(E_{TS}^2-E_Z^2)^2$ & $\lesssim\left(\Delta_-/2E_Z\right)^2$ & 
$\Delta_-^2/(E_{TS}+|E_Z|)^2$ & $1/4$ & $\Delta_-^2/(E_{TS}+|E_Z|)^2$ \\ \hline
$a_+^2a_-^2$ & $\Delta_-^2/(E_{TS}+|E_Z|)^2$ & $\lesssim\left(\Delta_-/2E_Z\right)^2$ 
& $\Delta_+^2\Delta_-^2/(E_{TS}^2-E_Z^2)^2$ & $\lesssim\left(\Delta_+/2E_Z\right)^2$ & $\Delta_+^2/(E_{TS}-|E_Z|)^2$\\ \hline
\end{tabular}
\caption{The factor $a_\alpha^2a_\beta^2$ in different regimes of $E_{TS}$. 
We assume here $E_Z=-|E_Z|$, however, the table can also be used for $E_Z=|E_Z|$ 
after replacing $\Delta_\pm\to\Delta_\mp$.}
\label{Tablea2ia2j}
\end{table}
\end{center}
\endwidetext

From Eq.~(\ref{a02apm2eq}), we find that $a_0^2a_\pm^2$
has a Lorentzian line-shape as a function of $E_{TS}$
in the vicinity of $E_{TS}=\pm|E_Z|$, see also Fig.~\ref{ai2aj2figure}(a).
The maximal value of $a_0^2a_\pm^2$ approaches $1/4$ in the vicinity of $E_{TS}=\pm|E_Z|$.
Around this point ($E_{TS}=\pm|E_Z|$), the spin and orbital degrees of freedom of the two electrons are strongly intermixed
between themselves. 
As a consequence,
the relaxation rate $\Gamma_{+,0}$ ($\Gamma_{0,-}$) is independent of the spin-orbit interaction strength,
and it is given by $(1/4)\tilde\Gamma(2\Delta_\pm/\hbar)$ according to Eq.~(\ref{Gammaabres}).
The quantity $\tilde\Gamma(w)$ depends on the type of bath that is coupled to the electron charge. 
Whether an appreciable relaxation takes place or not depends thus on the value of $\tilde\Gamma(w)$
in the limit $w\to 0$.
As we show below, $\tilde\Gamma(w)$ due to the electron-phonon interaction is generally suppressed for
$w<s/\lambda$ for all mechanisms of electron-phonon interaction.  
Therefore, the resonant enhancement of spin relaxation at $E_{TS}=\pm|E_Z|$ can occur due to phonons only in structures
with $\Delta_\pm\sim\hbar s/\lambda$.
However, $\Delta_\pm\ll\hbar s/\lambda$ in typical GaAs quantum dots used for studying spin relaxation,
and therefore the phonon-induced spin relaxation is suppressed at $E_{TS}=\pm|E_Z|$ in these structures.
Finally, we note that $a_0^2a_-^2$ is strongly suppressed for $E_{TS}>|E_Z|$, and
$a_0^2a_+^2$ is strongly suppressed for $E_{TS}<-|E_Z|$.
In both cases $a_0^2a_\pm^2$ is suppressed
owing to the selection rule of $H_{SO}$,
which allows only a single spin to be flipped in a first order process.
Thus, for example, the state $|\Psi_-\rangle$ at $E_{TS}>|E_Z|$ consists
mostly of $|\Psi_{T_-}\rangle$ and has an extremely small (second order in $H_{SO}$)
admixture of $|\Psi_{T_+}\rangle$.
Similarly, $|\Psi_0\rangle\approx|\Psi_{T_+}\rangle+{\cal O}(H_{SO}^2)|\Psi_{T_-}\rangle$. 
As a result,
a scalar fluctuation, like $U_{\rm ph}(t)$, does not connect $|\Psi_-\rangle$ and $|\Psi_0\rangle$ 
in the first order of $H_{SO}$.

Next, we consider the factor $a_+^2a_-^2$. 
From Eq.~(\ref{ap2am2eq}), we find that $a_+^2a_-^2=\Delta_-^2/J_-^2$ on the left-hand
side ($E_{TS}>|E_Z|$) and  $a_+^2a_-^2=\Delta_+^2/J_+^2$
on the right-hand side ($E_{TS}<-|E_Z|$) of the singlet-triplet crossing.
$a_+^2a_-^2$ as a function of  $E_{TS}$  has two maxima, at $E_{TS}\approx\pm E_Z$, see Fig.~\ref{ai2aj2figure}(b).
The upper bounds on these maxima read: $a_+^2a_-^2\lesssim \Delta_-^2/4E_Z^2$ for
the maximum at $E_{TS}\approx|E_Z|$, and $a_+^2a_-^2\lesssim \Delta_+^2/4E_Z^2$ for
the maximum at $E_{TS}\approx-|E_Z|$.
In the region between the maxima ($-|E_Z|<E_{TS}<|E_Z|$), the factor $a_+^2a_-^2$ is strongly
suppressed owing to the selection rule of $H_{SO}$,
as discussed above.
We summarize our results for $a_0^2a_\pm^2$ and $a_+^2a_-^2$ in Table~\ref{Tablea2ia2j}, for
a number of limiting cases.
Except for the two regions of avoided crossing, at $E_{TS}=\pm E_Z$, one of the three
factors $a_\alpha^2a_\beta^2$ is always suppressed compared to the other two.
As a result, for most values of $E_{TS}$, there are only two relaxation rates of interest,
which correspond to transitions between the triplet states $|\Psi_{T_\pm}\rangle$ 
and the singlet $|\Psi_S\rangle$.

\begin{figure}
 \begin{center}
  \includegraphics[angle=0,width=0.4\textwidth]{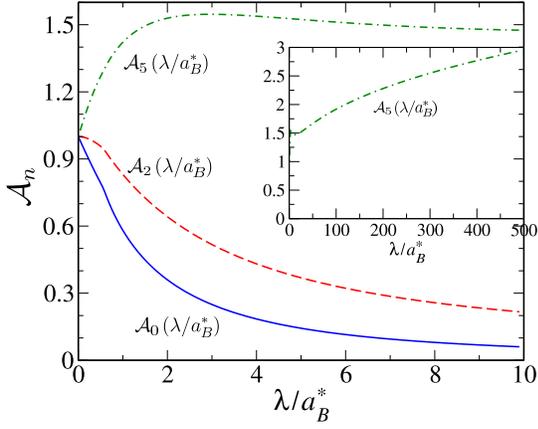}
    \caption{\small
(Color online) The factors ${\cal A}_0(\lambda/a_B^*)$ (solid curve),
${\cal A}_2(\lambda/a_B^*)$ (dashed curve), and
 ${\cal A}_5(\lambda/a_B^*)$ (dot-dashed curve)
as functions of the Coulomb interaction strength $\lambda/a_B^*$.
Inset: the factor ${\cal A}_5(\lambda/a_B^*)$ on a larger scale.
}
    \label{supfacAfigure}
 \end{center}
\end{figure}

We turn now to analyzing the terms in the square brackets in Eq.~(\ref{Gammaabres}).
The spin-orbit interaction enters here indirectly, via the
transition frequency $\omega_{\beta\alpha}$. 
The latter is considerably influenced (renormalized) by spin-orbit coupling only at the points of avoided crossing ($E_{TS}=\pm E_Z$).
Each rate $\tilde\Gamma(w)$ in Eqs.~(\ref{eqGammaLADP})$-$(\ref{eqGammaTAPE})
is thus independent of the spin-orbit coupling and can be studied separately as a function of the transition frequency $w$.
For simplicity, we assume $T=0$, which allows only spontaneous emission of phonons to occur ($N_{w}=0$).
In the low frequency limit, $w\ll s/\langle r\rangle$, we obtain from Eqs.~(\ref{eqGammaLADP})$-$(\ref{eqGammaTAPE})
\begin{eqnarray}
\tilde\Gamma_{\rm\it LA}^{\rm\it DP}(w)&\propto&w^7,\nonumber\\
\tilde\Gamma_{\rm\it LA(TA)}^{\rm\it PE}(w)&\propto&w^5.
\label{eqfrscal75}
\end{eqnarray}
Such a frequency scaling is valid for all quantum dots that have confining potentials
with in-plane inversion symmetry, $U(x,y)=U(-x,-y)$.
For such quantum dots, the form-factor ${\cal F}_{ST}(q_\parallel)$ is real and scales
as ${\cal F}_{ST}(q_\parallel)\propto q_\parallel^2$
at $q_\parallel\ll 1/\langle r\rangle$.
In contrast, for asymmetric quantum dots, 
the terms $\tilde\Gamma_{\rm\it LA}^{\rm\it DP}(w)\propto w^5$
and $\tilde\Gamma_{\rm\it LA(TA)}^{\rm\it PE}(w)\propto w^3$ are possible in the leading order of $w$.
For our quantum dot, with $U(x,y)=m^*\omega_0^2(x^2+y^2)/2$,
we obtain explicitly ($w\ll s/\langle r\rangle$)
\begin{eqnarray}\label{eqGammaLADPwto0}
\tilde\Gamma_{\rm\it LA}^{\rm\it DP}(w)&=&
\frac{\Xi_0^2w^7\lambda^4}{240\pi\hbar\rho_cs_1^9}
{\cal A}_0(\lambda/a_B^*),\\
\label{eqGammaLAPEwto0}
\tilde\Gamma_{\rm\it LA}^{\rm\it PE}(w)&=&
\frac{2\pi e^2h_{14}^2w^5\lambda^4}{385\kappa^2\hbar\rho_cs_1^7}
{\cal A}_0(\lambda/a_B^*),\\
\label{eqGammaTAPEwto0}
\tilde\Gamma_{\rm\it TA}^{\rm\it PE}(w)&=&
\frac{26\pi e^2h_{14}^2w^5\lambda^4}{3465\kappa^2\hbar\rho_cs_2^7}
{\cal A}_0(\lambda/a_B^*),
\end{eqnarray}
where ${\cal A}_0(\lambda/a_B^*)\leq1$ is a suppression factor, due to
the Coulomb interaction, given by
\begin{equation}
{\cal A}_n(\lambda/a_B^*)=\frac{\lambda_{1,0}^n}{\lambda^{n+4}}\left[(1+\sqrt{1+\gamma_1})\tilde\lambda_1^2-
(1+\sqrt{\gamma_0})\tilde\lambda_0^2\right]^2,
\end{equation}
with $\gamma_m$ and $\tilde\lambda_m$, for $m=0,1$, being the variational parameters
found in Appendix~\ref{appA}, and $\lambda_{1,0}$ given in Eq.~(\ref{lambdamm}).
Note that ${\cal A}_n(\lambda/a_B^*)$ depends only on the ratio $\lambda/a_B^*$.
We plot ${\cal A}_0(\lambda/a_B^*)$ as a function of $\lambda/a_B^*$ in Fig.~\ref{supfacAfigure} (solid curve).
In limiting cases, we obtain
\begin{eqnarray}
{\cal A}_0(x)=
\left\{
\begin{array}{ll}
1-\frac{\pi}{4}x,& \;\;\;\; x\ll1,\\
\frac{4}{9}\left(\frac{2}{x}\right)^{4/3},& \;\;\;\; x\gg1.
\end{array}
\right.
\end{eqnarray}
Thus, the Coulomb interaction has a strong effect on the spin relaxation rates.

Assuming $E_Z\ll E_{TS}\ll\hbar s/\langle r\rangle$ and $E_{TS},-E_Z>0$,
we find the triplet-to-singlet relaxation rates ($\Gamma_{ST_+}\equiv\Gamma_{+0}$ and $\Gamma_{ST_-}\equiv\Gamma_{+-}$)
\begin{equation}\label{GamSTpm0eqPETA}
\Gamma_{ST_\pm}=
\frac{13\pi e^2h_{14}^2E_{TS}^3E_Z^2\lambda^6}{13860\kappa^2\hbar^6\rho_cs_2^7\Lambda_\pm^2}{\cal A}_2\left(\lambda/a_B^*\right),
\end{equation}
where we introduced two effective spin-orbit lengths,
\begin{equation}
\frac{1}{\Lambda_\pm}=\sqrt{\frac{1-l_{x'}^2}{\lambda_-^2}+\frac{1-l_{y'}^2}{\lambda_+^2}\mp\frac{2l_z}{\lambda_+\lambda_-}}.
\end{equation}
In Eq.~(\ref{GamSTpm0eqPETA}), we retained only the piezoelectric interaction with transverse phonons 
($\tilde\Gamma_{\rm\it TA}^{\rm\it PE}$), 
because it gives the largest rate for GaAs quantum dots.
The suppression factor ${\cal A}_2\left(\lambda/a_B^*\right)$ is shown in Fig.~\ref{supfacAfigure} (dashed curve).
For weak Coulomb interaction, we have ${\cal A}_2\left(\lambda/a_B^*\right)\approx 1-0.203\left(\lambda/a_B^*\right)^2$,
and, for strong Coulomb interaction, we have ${\cal A}_2\left(\lambda/a_B^*\right)=(4/9)\left(\lambda/2a_B^*\right)^{-2/3}$.
Thus, in this case, the effect of Coulomb interaction is to reduce the singlet-triplet relaxation rates.
This reduction is despite the fact that the Coulomb interaction enhances the spin-orbit effects in the quantum dot,
{\em e.g.}, the splittings $\Delta_\pm$ in Eq.~(\ref{gapsD}), see also  Fig.~\ref{STtranSchetch}.
The enhancement of spin-orbit effects occurs due to an effectively larger quantum-dot size, 
$\lambda_{1,0}\approx \langle r\rangle$.
Note that $\lambda_{1,0}/\lambda=1+(\sqrt{\pi}/8)(\lambda/a_B^*)$, for $\lambda/a_B^*\ll1$,
and  $\lambda_{1,0}/\lambda=(\lambda/2a_B^*)^{1/3}$, for $\lambda/a_B^*\gg1$.
Also note that ${\cal A}_2\left(\lambda/a_B^*\right)=(\lambda_{1,0}/\lambda)^2{\cal A}_0\left(\lambda/a_B^*\right)$,
and therefore $(\lambda_{1,0}/\lambda)^2$ can be viewed as an enhancement factor and
${\cal A}_0\left(\lambda/a_B^*\right)$ as a suppression factor.
The suppression factor ${\cal A}_0\left(\lambda/a_B^*\right)$ originates
from the form-factor ${\cal F}_{ST}(q_\parallel)$ in Eq.~(\ref{calFST}),
which is sensitive to the difference $|\psi_S({\bf r}_1,{\bf r}_2)|^2-|\psi_T({\bf r}_1,{\bf r}_2)|^2$, see
Eq.~(\ref{calFSTdeff}).
The Coulomb interaction makes the singlet and triplet charge profiles look more similar to each other,
and thus, reduces the relaxation rates.
An additional reduction of relaxation rates stems from $E_{TS}$, which enters Eq.~(\ref{GamSTpm0eqPETA}).
The dependence of $E_{TS}$ on the Coulomb interaction strength is studied in Section~\ref{EnWaveFuncs},
see Eqs.~(\ref{EST}) and (\ref{deltafit}).

Assuming now $\left|E_{TS}-|E_Z|\right|\ll\Delta_+$ and $|E_Z|\ll\hbar s/\langle r\rangle$, we
obtain three relaxation rates,
\begin{eqnarray}\label{GammaETSeqEZp0}
&&\Gamma_{+,0}=\frac{13\sqrt{2}\pi e^2h_{14}^2\left|E_Z\right|^5\lambda^9}{55440\kappa^2\hbar^6\rho_cs_2^7\Lambda_+^5}
{\cal A}_5(\lambda/a_B^*),\\
&&\Gamma_{0,-}=\Gamma_{+,-}=\frac{26\pi e^2h_{14}^2|E_Z|^5\lambda^6}{3465\kappa^2\hbar^6\rho_cs_2^7\Lambda_-^2}
{\cal A}_2(\lambda/a_B^*),\;\;\;\;\;\;\;\;\;
\label{GammaETSeqEZ0mpm}
\end{eqnarray}
where the factor ${\cal A}_5(\lambda/a_B^*)\geq1$ is shown in Fig.~\ref{supfacAfigure} (dot-dashed curve).
Note that ${\cal A}_5\left(\lambda/a_B^*\right)= 1+\left(3\sqrt{\pi}/8\right)\left(\lambda/a_B^*\right)$,
for $\lambda/a_B^*\ll1$, and
${\cal A}_5\left(\lambda/a_B^*\right)=(4/9)\left(\lambda/2a_B^*\right)^{1/3}$, for $\lambda/a_B^*\gg1$.
The relaxation rate $\Gamma_{+,0}$ in Eq.~(\ref{GammaETSeqEZp0}) is enhanced
by the Coulomb interaction, via the factor ${\cal A}_5(\lambda/a_B^*)\geq1$, whereas
the relaxation rates $\Gamma_{0,-}$ and $\Gamma_{+,-}$ are suppressed,
due to the factor ${\cal A}_2(\lambda/a_B^*)\leq1$, like in the previous regime.
The enhancement due to the factor ${\cal A}_5(\lambda/a_B^*)$ is not significant
for a realistic structure, however, it confirms our picture that there are
two competing effects of the Coulomb interaction on the singlet-triplet relaxation, as described above.
In the case of ${\cal A}_5(\lambda/a_B^*)$, the increase of the quantum dot size ($\langle r\rangle\approx\lambda_{1,0}$)
wins over the suppression due to $f_{00}^2(r)-f_{01}^2(r)$.

Considering further $E_{TS}\ll E_Z\ll\hbar s/\langle r\rangle$, we
obtain
\begin{eqnarray}\label{GammaESTeq0p0}
&&\Gamma_{+,0}=\frac{13\pi e^2h_{14}^2|E_Z|^5\lambda^6}{13860\kappa^2\hbar^6\rho_cs_2^7\Lambda_+^2}
{\cal A}_2(\lambda/a_B^*),\\
&&\Gamma_{0,-}=
\frac{13\pi e^2h_{14}^2|E_Z|^5\lambda^6}{13860\kappa^2\hbar^6\rho_cs_2^7\Lambda_-^2}
{\cal A}_2(\lambda/a_B^*),\;\;\;\;\;\;\;\;\;
\end{eqnarray}
where ${\cal A}_2(\lambda/a_B^*)$ was analyzed above.
Again, we see that the Coulomb interaction reduces the relaxation rates in this regime.

On the triplet side of the crossing ($E_{TS}<0$), one can use Eqs.~(\ref{GammaETSeqEZp0}) and (\ref{GammaETSeqEZ0mpm}) with 
$\Gamma_{+,0}\leftrightarrow\Gamma_{0,-}$ and $\Lambda_\pm\to\Lambda_\mp$ for the regime 
$E_{TS}\approx -|E_Z|\ll\hbar s/\langle r\rangle$, and 
Eqs.~(\ref{GamSTpm0eqPETA}) with $\Gamma_{S,T_\pm}\to\Gamma_{T_\pm,S}$ and 
$E_{TS}\to |E_{TS}|$ for the regime $E_Z\ll E_{TS}\ll\hbar s/\langle r\rangle$.

\begin{figure}
 \begin{center}
  \includegraphics[angle=0,width=0.35\textwidth]{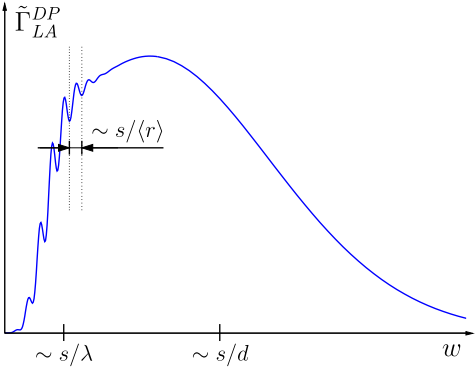}
    \caption{\small
(Color online) Schematic of $\tilde{\Gamma}_{\rm\it LA}^{\rm\it DP}(w)$, showing the relevant scales
of $w$: $s/\lambda$,  $s/d$, and $s/\langle r\rangle$.
The curve was obtained using the form-factor in Eq.~(\ref{calFSTlargeCI}), for a fairly large
value of the Coulomb interaction strength, $\lambda/a_B^*=10^3$.
}
    \label{tGLADP10to3}
 \end{center}
\end{figure}

So far, we have considered only the regimes of small transition frequencies, $w\ll s/\langle r\rangle$.
The quantities $\tilde\Gamma(w)$ in Eqs.~(\ref{eqGammaLADP})$-$(\ref{eqGammaTAPE})
are functions of the transition frequency $w$, and contain several scales of $w$:
$s/\langle r\rangle$, $s/\lambda$ and $s/d$.
In principle, $s/\langle r\rangle$ can be much smaller than $s/\lambda$,
provided the Coulomb interaction is strong enough.
In the regime $s/\langle r\rangle\ll w\ll s/\lambda$,
we find oscillations of $\tilde\Gamma(w)$ as function of $w$ 
with a period $\Delta w\sim s/\langle r\rangle$, see Fig.~\ref{tGLADP10to3}.
These oscillations stem from the oscillations of
${\cal F}_{ST}(q_\parallel)$ in Fig.~\ref{calFSTfig}.
Note, however, that a fairly large strength of the Coulomb interaction is required to 
observe these oscillations.
The reason for this is that $\langle r\rangle=\lambda(\lambda/2a_B^*)^{1/3}$ scales
weakly (power $1/3$) with $\lambda/a_B^*\gg 1$.
In a typical GaAs structure with $\lambda/a_B^*\simeq 5$,
there is not sufficient room between the scales $s/\langle r\rangle$ and $s/\lambda$ for these oscillations to occur.

\begin{figure}
 \begin{center}
  \includegraphics[angle=0,width=0.4\textwidth]{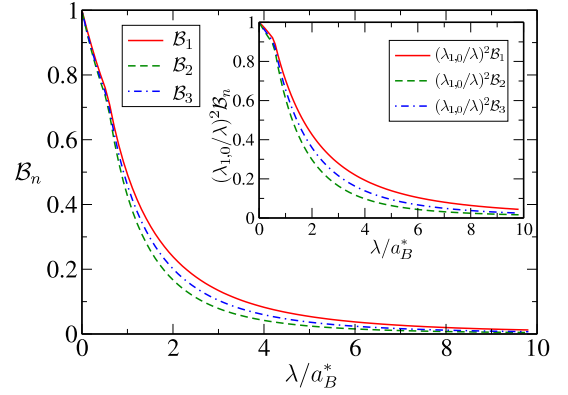}
    \caption{\small
(Color online) The factors ${\cal B}_1$, ${\cal B}_2$, and ${\cal B}_3$ as functions of the Coulomb interaction strength
$\lambda/a_B^*$. 
Inset: the factors in the main figure multiplied by $(\lambda_{1,0}/\lambda)^2$.
}
    \label{calBnfig}
 \end{center}
\end{figure}

In the regime $s/\lambda\ll w\ll s/d$, we find
\begin{eqnarray}
\tilde\Gamma_{\rm\it LA}^{\rm\it DP}(w)&\propto&w,\\
\tilde\Gamma_{\rm\it LA}^{\rm\it PE}(w)&\propto&w^{-5},\\
\tilde\Gamma_{\rm\it TA}^{\rm\it PE}(w)&\propto&w^{-3}.
\end{eqnarray}
These scaling laws are universal in the sense that the power of $w$ does not depend on the dot confining potential,
provided $\lambda$ is understood as the smallest lateral scale.
The phonon of frequency $w$ is emitted at an angle $\vartheta\simeq\arcsin\left(s/\lambda w\right)$ to the $z$-axis.
From Eqs.~(\ref{eqGammaLADP})$-$(\ref{eqGammaTAPE}), we obtain
\begin{eqnarray}\label{eqGammaLADPintermed}
\tilde\Gamma_{\rm\it LA}^{\rm\it DP}(w)&=&
\frac{\Xi_0^2w}{2\pi\hbar\rho_cs_1^3\lambda^2}
{\cal B}_1(\lambda/a_B^*),\\
\label{eqGammaLAPEintermed}
\tilde\Gamma_{\rm\it LA}^{\rm\it PE}(w)&=&
\frac{1728\pi e^2h_{14}^2s_1^3}{\kappa^2\hbar\rho_c\lambda^6w^5}
{\cal B}_2(\lambda/a_B^*),\\
\label{eqGammaTAPEintermed}
\tilde\Gamma_{\rm\it TA}^{\rm\it PE}(w)&=&
\frac{96\pi e^2h_{14}^2s_2}{\kappa^2\hbar\rho_c\lambda^4w^3}
{\cal B}_3(\lambda/a_B^*),
\end{eqnarray}
where the suppression factors ${\cal B}_n(\lambda/a_B^*)$ are shown in Fig.~\ref{calBnfig}.
We note that ${\cal B}_n\sim (\lambda/a_B^*)^{-7/3}$ at  $\lambda/a_B^*\gg 1$.

Next one can consider again different regimes given in Table~\ref{Tablea2ia2j} 
and using Eq.~(\ref{Gammaabres}) obtain relaxation rates for different mechanisms of electron-phonon
interaction.
We skip this part here and consider only the experimentally most relevant regime
$E_{TS}\gg E_Z,\hbar s/\lambda$. 
For GaAs, we estimate $\hbar s_2/\lambda\approx 22\,\mu{\rm eV}$ for $\lambda=100\,{\rm nm}$.
Retaining only the deformation potential mechanism, we obtain using Eq.~(\ref{eqGammaLADPintermed}),
\begin{equation}\label{rateGammaLADPintermed}
\Gamma_{ST_\pm}=
\frac{\Xi_0^2E_Z^2\lambda_{1,0}^2}{16\pi\hbar^2\rho_cs_1^3E_{TS}\Lambda_\pm^2\lambda^2}
{\cal B}_1(\lambda/a_B^*),
\end{equation}
for $E_{TS}>0$.
On the other side of the crossing ($E_{TS}<0$), we have the same expression,
but with $\Gamma_{S,T_\pm}\to\Gamma_{T_\pm,S}$ and 
$E_{TS}\to |E_{TS}|$.
Finally, we note that $(\lambda_{1,0}/\lambda)^2{\cal B}_n$ is a monotonically decaying function
of $\lambda/a_B^*$, see inset of Fig.~\ref{calBnfig}, 
and thus the Coulomb interaction reduces the relaxation rates.

In the regime $w\gg s/d$, the scaling of $\tilde\Gamma(w)$ with $w$ is non-universal, in the
sense that it depends on the quantum dot lateral and transverse confinement potentials.
Considering the quantity
\begin{equation}
\max_\vartheta \left|{\cal F}_{ST}\left({w\over s_2}\sin\vartheta\right)
F\left({w\over s_2}\cos\vartheta\right)\right|,
\label{maxeqcalFF}
\end{equation}
one can find a line $\vartheta(w)$ in the $(w,\vartheta)$-plane 
such that for a given $w$ the product of form-factors is maximal at $\vartheta=\vartheta(w)$ .
Then, except for the special cases: $\vartheta(w)=0,\pi/2$, 
the phonon emission is peaked at $\vartheta(w)$.
The scaling of $\tilde\Gamma(w)$ with $w$ follows the square of the
expression in Eq.~(\ref{maxeqcalFF}), up to some prefactors which are power-law in $w$.
In the case of our harmonic lateral confinement, the form-factor
${\cal F}_{ST}(q_\parallel)$ decays Gaussian-like at $q_\parallel\gg 1/\lambda$.
Provided the transverse form-factor $F(q_z)$ has not a stronger decay law,
then the least-suppression line is $\vartheta(w)=0$.
The phonons are then emitted at an angle close to $0,\pi$
(more precisely at $\vartheta\simeq\arcsin\left(s/\lambda w\right)$).
The scaling of $\tilde\Gamma(w)$ with $w$ follows the scaling of 
$|F(w/s)|^2$, up to prefactors which are power-law in $w$.
In the other extreme case, when $F(q_z)$ has a stronger 
decay law (at $q_z\gg 1/d$) than ${\cal F}_{ST}(q_\parallel)$, then
 $\vartheta(w)\sim \arccos(s/dw)$ and 
$\tilde\Gamma(w)\propto |{\cal F}_{ST}(w/s)|^2$.
Thus, the regime of large transition frequencies ($w\gg s/d$)
allows one to learn about the dot confinement, however the singlet-triplet
relaxation rates are strongly suppressed in this regime.
For GaAs, we estimate that the singlet-triplet energy $E_{TS}$ needs to be
larger than $\hbar s_2/d\approx 0.4\,{\rm meV}$ at $d=5\,{\rm nm}$,
for this regime to be accessible.

\begin{figure}
 \begin{center}
  \includegraphics[angle=0,width=0.42\textwidth]{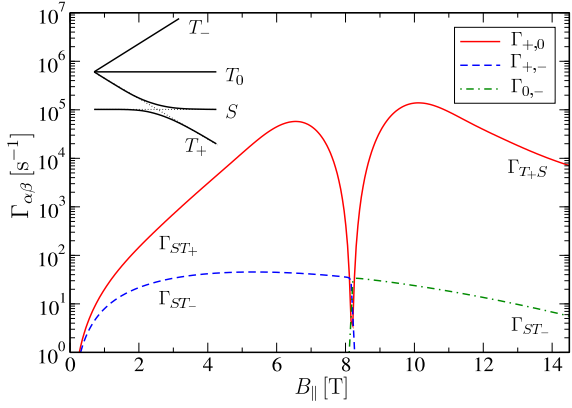}
    \caption{\small
(Color online) Relaxation rates $\Gamma_{+,0}$, $\Gamma_{+,-}$, and  $\Gamma_{0,-}$ as functions of an in-plane magnetic field
$\mbox{\boldmath $B$}=(0,B_\parallel,0)$.
An avoided crossing of the singlet $|S\rangle$ and triplet $|T_+\rangle$ occurs at $E_{TS}=-E_Z\approx 0.75\,{\rm meV}$ 
($B_\parallel\approx 8.2\,{\rm T}$).
We used parameters typical for a GaAs quantum dot: $\hbar\omega_0=2\,{\rm meV}$, $\lambda_\pm=8\,\mu{\rm m}$, $d=5\,{\rm nm}$, $g=-0.44$,
and $\kappa=13$.
Inset: Schematic of level crossing for applying an in-plane magnetic field.
}
    \label{STratesBparfig}
 \end{center}
\end{figure}

To summarize this section, we present the calculated relaxation rates as 
functions of the magnetic field $B$.
For an in-plane magnetic field $\mbox{\boldmath $B$}=(0,B_\parallel,0)$,
we show the relaxation rates in Fig.~\ref{STratesBparfig}.
Only a single avoided crossing (at $E_{TS}=-E_Z$) is present in this case, 
see inset of Fig.~\ref{STratesBparfig}.
The states $\Psi_{T_+}$ and $\Psi_S$ are strongly intermixed with each other,
which results in a large relaxation rate between these states.
For the parameters used in Fig.~\ref{STratesBparfig},
the relaxation between  $\Psi_{T_+}$ and $\Psi_S$
(see $\Gamma_{+,0}$ in Fig.~\ref{STratesBparfig})
is about $10^3$ times stronger than the relaxation
between $\Psi_{T_-}$ and $\Psi_S$.
At $E_{TS}=-E_Z$, the relaxation rate $\Gamma_{+,0}$ has a dip,
because the phonon-emission energy $\hbar w=2\Delta_+$ is very small 
($\Delta_+\ll \hbar  s/\lambda$) at this point.
We note that the value of $\Gamma_{+,0}$ in the dip depends strongly
on $\Delta_+$, as it can be seen after rewriting  Eq.~(\ref{GammaETSeqEZp0}) as follows,
\begin{equation}\label{GammaETSeqEZp0Delta}
\Gamma_{+,0}=\frac{208\pi e^2h_{14}^2\Delta_+^5\lambda^4{\cal A}_0}{3465\kappa^2\hbar^6\rho_cs_2^7}.
\end{equation}
The sensitivity of $\Delta_+$ to the magnetic field direction, illustrated in 
Figs.~\ref{DeltaPlus1} and \ref{DeltaSym1},
can be used here to tune $\Gamma_{+,0}$ by orders of magnitude.
It is also important to note that at this point ($E_{TS}=-E_Z$)
the eigenstates $|\Psi_0\rangle$ and $|\Psi_+\rangle$ 
are approximately given by
\begin{eqnarray}\label{psipatEZeqETS}
|\Psi_{+}\rangle&\approx &\frac{1}{\sqrt{2}}\left(|\Psi_S\rangle
+ |\Psi_{T_+}\rangle\right),\\
|\Psi_{0}\rangle&\approx &\frac{1}{\sqrt{2}}\left(|\Psi_S\rangle
- |\Psi_{T_+}\rangle\right),
\label{psimatEZeqETS}
\end{eqnarray}
where $|\Psi_S\rangle=|S\rangle|\psi_S\rangle$ and $|\Psi_{T_+}\rangle=|T_+\rangle|\psi_T\rangle$,
with $|\psi_{S(T)}\rangle$ being the singlet (triplet) orbital wave function.
In Eqs.~(\ref{psipatEZeqETS}) and (\ref{psimatEZeqETS}), we neglected
terms $\propto{\cal O}(\lambda/\lambda_{SO})|\Psi_{T_-}\rangle$.
Thus, at this degeneracy point (and in its vicinity), there is no spin selectivity
in coupling to potential fluctuations, because the spin is not a quantum number here,
see Eqs.~(\ref{psipatEZeqETS}) and (\ref{psimatEZeqETS}).
While the phonon-induced relaxation is suppressed at small frequencies,
the coupling to some other degrees of freedom in the solid state 
({\em e.g.} $1/f$-noise, noisy gates, $B$-field fluctuations, particle-hole excitations on the Fermi sea, 
{\em etc.}) can dominate the relaxation between $|\Psi_0\rangle$ and $|\Psi_+\rangle$.
The relaxation rate is given by the orbital (charge) relaxation rate at the energy $\hbar w=2\Delta_+$.
Furthermore, since the states (\ref{psipatEZeqETS}) and (\ref{psimatEZeqETS}) are difficult
to discriminate and prepare by the currently used 
techniques,\cite{FujisawaT1Nature,HansonST2el,SingleShotNature,Kroutvar}
it is more convenient to study relaxation at the avoided crossing point by means of the Ramsey fringes.
Then the relaxation rate in Eq.~(\ref{GammaETSeqEZp0Delta}) appears as a decoherence rate.
The required rise (fall) time of pulses for tuning to the degeneracy point should be smaller than 
$\pi\hbar/\Delta_+\approx 200\,{\rm ps}$ for $\Delta=100\,\mu{\rm eV}$, 
which is feasible experimentally (note that $E_{TS}$ can be tuned by electrical gates on sort time-scales,
while keeping $E_Z$ constant).

Far away from the dip in Fig.~\ref{STratesBparfig}, say to the left ($|E_Z|<E_{TS}$), the relaxation rates read
\begin{equation}\label{GammatotheleftSTpm}
\Gamma_{ST_\pm}=\frac{\Xi_0^2\Delta_\pm^2{\cal B}_1}{2\pi\hbar^2\rho_cs_1^3\lambda^2(E_{TS}\mp|E_Z|)},
\end{equation}
where we assumed the regime $s/\lambda\ll w\ll s/d$, with $w=(E_{TS}\mp|E_Z|)/\hbar$,
 and took into account only the deformation
potential mechanism.
The latter approximation is valid only for large transition frequencies $w$,
typically $w^2>10\pi e|h_{14}|s^2/\kappa\Xi_0\lambda$.
At smaller $w$ the piezoelectric TA mechanism gives the largest 
contribution to the relaxation rates.
Thus, the rate $\Gamma_{ST_+}$ turns to
\begin{equation}\label{GammatotheleftSTpPE}
\Gamma_{ST_+}=
\frac{96\pi\hbar^2 e^2h_{14}^2s_2\Delta_+^2{\cal B}_3}{\kappa^2\rho_c\lambda^4(E_{TS}-|E_Z|)^5},
\end{equation}
as $E_{TS}-|E_Z|$ is decreased, but still remaining in the regime 
$s/\lambda\ll (E_{TS}-|E_Z|)/\hbar\ll s/d$.

\begin{figure}
 \begin{center}
  \includegraphics[angle=0,width=0.42\textwidth]{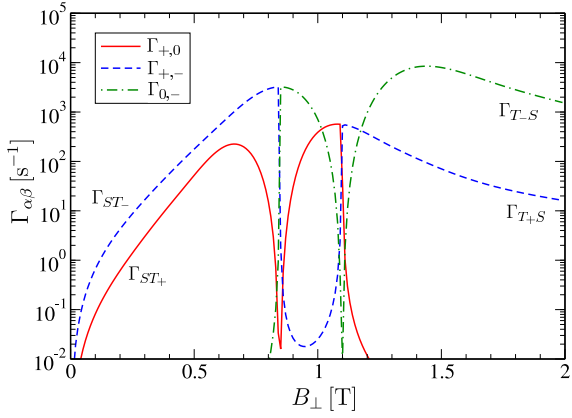}
    \caption{\small
(Color online) Relaxation rates $\Gamma_{+,0}$, $\Gamma_{+,-}$, and  
$\Gamma_{0,-}$ as functions of the transverse magnetic field
$\mbox{\boldmath $B$}=(0,0,B_\perp)$.
Two avoided crossings occur at $E_{TS}=\pm E_Z$, as shown in Fig.~\ref{STtranSchetch}.
We use the same parameters as in Fig.~\ref{STratesBparfig}, except for
$\lambda_+=\lambda_-/2=8\,\mu{\rm m}$.
For these parameters, we obtain $E_{TS}\approx 0.75\,{\rm meV}$ at $B_\perp=0$, 
and $E_{TS}= 0$ at $B_\perp\approx 0.95\,{\rm T}$.
}
    \label{STratesBperpfig}
 \end{center}
\end{figure}

For a magnetic field perpendicular to the 2DEG plane, 
$\mbox{\boldmath $B$}=(0,0,B_\perp)$, the relaxation rates have a richer 
behavior, because more than one avoided crossing can occur in this case,
see for example Fig.~\ref{STtranSchetch}.
Furthermore,  $\Delta_+$ and $\Delta_-$ can strongly differ from each other, 
due to their different dependence on the $B$-field direction,
see Eq.~(\ref{gapsD}).
In Fig.~\ref{STratesBperpfig}, we plot the relaxation rates
as functions of the magnetic field $B_\perp$, which is tuned across 
the singlet-triplet crossing.
Two avoided crossing occur at $B\approx 0.85\,{\rm T}$ 
($E_{TS}=|E_Z|$) and at $B\approx 1.1\,{\rm T}$ ($E_{TS}=-|E_Z|$).
We choose $\lambda_+=2\lambda_-$, which means that
$\Delta_-=3\Delta_+$, see Eq.~(\ref{gapsD}).
The relaxation rates increase strongly as one approaches the neighborhood of the 
singlet-triplet transition from either side (maxima of rates correspond to the
optimal phonon emission condition: transition energy on the order of $\hbar s/\lambda$).
In contrast to the case of in-plane field (see Fig.~\ref{STratesBparfig}),
both relaxation rates are strongly affected here.
Although the overall behavior of the rates in Fig.~\ref{STratesBperpfig} is more complicated
than in Fig.~\ref{STratesBparfig}, the behavior of the rates in the vicinity of an avoided 
crossing is not much different.
For definiteness, let us consider the avoided crossing on the left-hand side,
occurring at $B\approx 0.85\,{\rm T}$ or equivalently at $E_{TS}=-E_Z=|E_Z|$.
There, the rate $\Gamma_{+,0}$ exhibits a dip as a function of $B_\perp$,
whereas the rate $\Gamma_{+,-}$ ($\Gamma_{0,-}$) decreases (increases) abruptly
at the point of avoided crossing.
The occurrence of the dip is explained, as before, by the suppression
of phonon emission at small energies $\hbar w=2\Delta_+ \ll \hbar s/\lambda$,
see also Eq.~(\ref{eqfrscal75}).
This behavior is similar to the behavior of the rate $\Gamma_{+,0}$ in Fig.~\ref{STratesBparfig}.
The abrupt dependence of the rates $\Gamma_{+,-}$ and $\Gamma_{0,-}$ on $B_\perp$ 
at the avoided crossing is due to the fact that the transition occurs between
a distant level $|\Psi_-\rangle\approx|\Psi_{T_-}\rangle$ onto two closely spaced
levels $|\Psi_+\rangle$ and $|\Psi_0\rangle$, which 
change from $|\Psi_+\rangle\approx|\Psi_S\rangle$ and $|\Psi_0\rangle\approx|\Psi_{T_+}\rangle$
on the left-hand side of the avoided crossing to 
$|\Psi_+\rangle\approx|\Psi_{T_+}\rangle$ and $|\Psi_0\rangle\approx|\Psi_S\rangle$
on the right-hand side.
Obviously, the transition between $|\Psi_{T_-}\rangle$ and $|\Psi_{T_+}\rangle$
is forbidden due to the spin selection rules, and therefore, the rate
$\Gamma_{0,-}$ ($\Gamma_{+,-}$) is suppressed on the left-hand side (right-hand side)
of the avoided crossing.
As one sweeps the magnetic field across the avoided crossing
the rate $\Gamma_{+,-}$ ($\Gamma_{0,-}$) decreases (increases)
in the neighborhood $-\Delta_+\lesssim E_{TS}+E_Z\lesssim \Delta_+$.
This increase/decrease appears abrupt on the scale of the plot,
because $\Delta_+$ is very small for GaAs ($\Delta_\pm\ll E_Z$).
Note that the same abrupt-change behavior is present for the
rates $\Gamma_{+,-}$, $\Gamma_{0,-}$ in Fig.~\ref{STratesBparfig}.
The second avoided crossing (at $B\approx 1.1\,{\rm T}$) in
Fig.~\ref{STratesBperpfig} shows an analogous behavior
with the rate $\Gamma_{0,-}$ exhibiting a dip and the rates
$\Gamma_{+,0}$ and $\Gamma_{+,-}$ interchanging their roles.
We thus conclude that the physics of the singlet-triplet relaxation 
for perpendicular and parallel $B$-fields is very much alike, although the 
case of perpendicular field is richer because two avoided crossings take place.
As for tilted magnetic fields, 
we did not find any qualitative difference as to the case of a perpendicular field, 
as long as two avoided crossings were present in the considered $B$-field range.

\subsection{Triplet-triplet relaxation rates}
\label{RelaxRatesTT}
Away from the singlet-triplet crossing, the triplet
consists of three levels symmetrically split by the Zeeman energy,
$E_{T_\pm}=E_T\pm E_Z$ and $E_{T_0}=E_T$.
The transition between $|T_-\rangle$ and $|T_+\rangle$ is forbidden
at the first order of $H_{SO}$, as we mentioned already in 
Eq.~(\ref{forbiddentransSTTT}).
The transition between $|T_0\rangle$ and $|T_\pm\rangle$
is allowed and occurs due to virtual processes with going to states
outside the singlet-triplet subspace (see Fig.~\ref{realVSvitrual}(a) 
and Sec.~\ref{ssecWeakME}).
At the singlet-triplet (avoided) crossing the states $|T_\pm\rangle$
mix with the singlet state (see Sec.~\ref{ESPSOgfinite}).
We are, therefore, entitled
to consider the relaxation of $|T_0\rangle$ onto the eigenstates
\begin{equation}
|\Psi_\alpha\rangle=a_\alpha|\Psi_S\rangle
+b_\alpha|\Psi_{T_+}\rangle
+c_\alpha|\Psi_{T_-}\rangle,
\label{PsiabcTT}
\end{equation}
where $\alpha=0,\pm$ labels the eigenstates and 
the coefficients $a_\alpha$, $b_\alpha$, and
$c_\alpha$ were found in Sec.~\ref{ESPSOgfinite}.
Next, we use $U_{\rm int}$ given in Eq.~(\ref{UintSB}) 
with $S_{\rm w}$ given in Eq.~(\ref{SwHarmonic}). 
In Eq.~(\ref{SwHarmonic}), we retain only the first term 
(see Fig.~\ref{realVSvitrual}
and discussion in Sec.~\ref{ssecWeakME}).
We evaluate first the matrix elements of $U_{\rm int}$
with respect to the unmixed states.
As a result, we obtain
\begin{eqnarray}
\langle\Psi_{T_\pm}|U_{\rm int}(t)|\Psi_{T_0}\rangle=\mp \frac{i}{\sqrt{2}}E_Z
(\mbox{\boldmath $X$}\mp i\mbox{\boldmath $Y$})\cdot\mbox{\boldmath $\Omega$}(t),
\label{matelPsiTpmUPsiT0Om}
\end{eqnarray}
where $\mbox{\boldmath $\Omega$}=(\Omega_{x'},\Omega_{y'},0)$ is a dimensionless quantum fluctuating field.
Its components are given by
\widetext
\begin{eqnarray}
\Omega_{x'}(t)&=&\frac{-i}{m^*\omega_0^2\lambda_-}\sum_{{\bf q}\,j}q_{y'}M_{{\bf q}\,j}
{\cal F}_T(\mbox{\boldmath $q$}_\parallel)F(q_z)
\left[b^\dagger_{-{\bf q}\,j}(t)+b_{{\bf q}\,j}(t)\right],\nonumber\\
\Omega_{y'}(t)&=&\frac{-i}{m^*\omega_0^2\lambda_+}\sum_{{\bf q}\,j}q_{x'}M_{{\bf q}\,j}
{\cal F}_T(\mbox{\boldmath $q$}_\parallel)F(q_z)
\left[b^\dagger_{-{\bf q}\,j}(t)+b_{{\bf q}\,j}(t)\right],
\label{eqOmegaxypp}
\end{eqnarray}
\endwidetext
\noindent
where ${\bf q}=({\bf q}_\parallel,q_z)$ and ${\bf q}_\parallel=(q_{x'},q_{y'})$.
In Eq.~(\ref{eqOmegaxypp}), ${\cal F}_T(q_\parallel)$ denotes the following form-factor,
\begin{equation}
{\cal F}_T(\mbox{\boldmath $q$}_\parallel)=
\left\langle\psi_T\right|e^{i{\bf q}_\parallel{\bf R}}\cos({\bf q}_\parallel{\bf r}/2)\left|\psi_T\right\rangle.
\label{TTformFT}
\end{equation}
We evaluate ${\cal F}_T(\mbox{\boldmath $q$}_\parallel)$ for the triplet state 
$\left|\psi_T\right\rangle=\left|000,-1\right\rangle$ and obtain
\begin{equation}
{\cal F}_T(q_\parallel)=e^{-\frac{1}{16}q_\parallel^2\left(\lambda^2+\tilde\lambda_1^2\right)}
{}_1F_1\left(-\sqrt{1+\gamma_1},1;\frac{q_\parallel^2\tilde\lambda^2_1}{16}\right),
\label{CalFTqparTonly}
\end{equation}
where $\gamma_1$ and $\tilde\lambda_1$ are the variational parameters for $m=1$ studied in Appendix~\ref{appA}.
Note that ${\cal F}_T$ is independent of the in-plane angle of the emitted phonon, 
due to the circular symmetry of the confining potential.

Next we use the states in Eq.~(\ref{PsiabcTT}) and the matrix elements 
(\ref{matelPsiTpmUPsiT0Om}) to evaluate the relaxation rates according to Eq.~(\ref{GabGR}),
\begin{equation}
\Gamma_{\alpha T_0}=(1-a_\alpha^2)\Gamma_{\alpha T_0}^{(0)}+
2{\rm Re}[b_\alpha c^*_\alpha]\Gamma_{\alpha T_0}^{(1)}+
2{\rm Im}[b_\alpha c^*_\alpha]\Gamma_{\alpha T_0}^{(2)}.
\label{Gratesinterfer}
\end{equation}
The first term in Eq.~(\ref{Gratesinterfer}) gives the
triplet-triplet relaxation without accounting for interference
between the two processes in Eq.~(\ref{matelPsiTpmUPsiT0Om}).
The other two terms in Eq.~(\ref{Gratesinterfer}) are
due to the interference, and are therefore relevant only close to 
the singlet-triplet crossing.
Away from the singlet-triplet crossing the first term
alone gives the triplet-triplet relaxation rates.
To shorten notations, we consider further 
$\Gamma_{\alpha \beta}^{(0)}$, where either $\alpha\equiv T_0$ 
or $\beta\equiv T_0$. 
We find
\widetext
\begin{equation}
\Gamma_{\alpha\beta}^{(0)}=\frac{E_Z^2\omega_{\beta\alpha}^3(1+N_{\omega_{\beta\alpha}})}
{2\pi\hbar\rho_c\left(m^*\omega_0^2\Lambda_{SO}\right)^2}
\sum_j\int_0^{\pi/2}d\vartheta\frac{\sin^3\vartheta}{s_j^5}
\left|{\cal F}_T\left(\frac{\omega_{\beta\alpha}}
{s_j}\sin\vartheta\right)F\left(\frac{\omega_{\beta\alpha}}{s_j}\cos\vartheta\right)\right|^2
\left(e^2\bar\beta_{j\vartheta}^2+\frac{\omega_{\beta\alpha}^2}{s_j^2}\bar\Xi_j^2\right),
\label{TtoTGammap0res}
\end{equation}
\endwidetext
\noindent
where 
$\omega_{\beta\alpha}=(E_\beta-E_\alpha)/\hbar$ is the transition frequency.
The effective spin-orbit length $\Lambda_{SO}$ is defined as follows
\begin{equation}
\frac{1}{\Lambda_{SO}}=\sqrt{\frac{1-l_{x'}^2}{\lambda_-^2}+\frac{1-l_{y'}^2}{\lambda_+^2}}.
\end{equation}
In Eq.~(\ref{TtoTGammap0res}), the summation runs over the acoustic phonon branches, with $j=1,2,3$ corresponding,
respectively, to longitudinal and two transverse polarizations.
The quantities $\bar\beta_{j\vartheta}$ and $\bar\Xi_j$ in Eq.~(\ref{TtoTGammap0res}) are introduced as
follows
\begin{equation}
\left(\begin{array}{c}
\bar\beta_{j\vartheta}^2\\
\bar\Xi_j^2
\end{array}
\right)
=\frac{1}{\pi}\int_0^{2\pi}d\phi\cos^2\phi
\left(\begin{array}{c}
|\beta_{{\bm q}j}|^2\\
|\Xi_{{\bm q}j}|^2
\end{array}
\right).
\end{equation}
After integrating over the phonon angle $\phi$ we obtain explicitly
\begin{eqnarray}
\bar\Xi_1&=&\Xi_0,\;\;\;\;\;\;\;\;\bar\Xi_2=\bar\Xi_3=0,\nonumber\\
\bar\beta_{1,\vartheta}&=&\frac{3\sqrt{2}\pi h_{14}}{\kappa}\sin^2\vartheta\cos\vartheta,\nonumber\\
\bar\beta_{2,\vartheta}&=&\frac{\sqrt{2}\pi h_{14}}{\kappa}\sin2\vartheta,\nonumber\\
\bar\beta_{3,\vartheta}&=&\frac{\sqrt{2}\pi h_{14}}{\kappa}(3\cos^2\vartheta-1)\sin\vartheta.
\end{eqnarray}
We note that the explicit form of $\bar\beta_{2,\vartheta}$ and $\bar\beta_{3,\vartheta}$ depends
on the choice of transverse phonon polarizations, whereas 
$\bar\beta_{2,\vartheta}^2+\bar\beta_{3,\vartheta}^2$ is independent of this choice.

The quantities $\Gamma_{\alpha T_0}^{(1)}$ and $\Gamma_{\alpha T_0}^{(2)}$ in Eq.~(\ref{Gratesinterfer})
do not have the meaning of rates, but rather of corrections to the rate.
They are, therefore, not positively defined.
We find for $\Gamma_{\alpha T_0}^{(1)}$ and $\Gamma_{\alpha T_0}^{(2)}$
similar expressions as for $\Gamma_{\alpha T_0}^{(0)}$ in Eq.~(\ref{TtoTGammap0res}),
but with a different geometrical prefactor.
To obtain $\Gamma_{\alpha T_0}^{(1)}$ and $\Gamma_{\alpha T_0}^{(2)}$ from
Eq.~(\ref{TtoTGammap0res}), one has to replace the factor
$1/\Lambda_{SO}^2$ by
\begin{eqnarray}
\left(\Lambda_{SO}^{'}\right)^{-2}
&=&
\left(\frac{1}{\lambda_-^2}-\frac{1}{\lambda_+^2}\right)
\cos 2\varphi'-\frac{l_{x'}^2}{\lambda_-^2}-\frac{l_{y'}^2}{\lambda_+^2},\;\;\;\;\;\;\; \\
\left(\Lambda_{SO}^{''}\right)^{-2}
&=& -\left(\frac{1}{\lambda_-^2}-\frac{1}{\lambda_+^2}\right)
l_z \sin 2\varphi',
\end{eqnarray}
respectively.
Here, $\varphi'$ is the azimuthal angle of ${\bm l}$.

Next, we evaluate the product $b_\alpha c^*_\alpha$ using
Eqs.~(\ref{aaacoeff}) and (\ref{TDHZS}).
The resulting expression (with separated real and imaginary parts) reads
\begin{equation}
b_\alpha c^*_\alpha
=\frac{E_Z^2\lambda_{1,0}^2a_\alpha^2}{8E_{\alpha T_+}E_{\alpha T_-}}
\left[\left(\Lambda_{SO}^{'}\right)^{-2}+ i\left(\Lambda_{SO}^{''}\right)^{-2}\right],
\label{bcsReImpI}
\end{equation}
where $E_{\alpha T_\pm}=E_\alpha-E_{T_\pm}$.
Using Eq.~(\ref{bcsReImpI}) and the proportionality of $\Gamma_{\alpha T_0}^{(1)}$ and $\Gamma_{\alpha T_0}^{(2)}$
to $\Gamma_{\alpha T_0}^{(0)}$, we obtain for the rate in Eq.~(\ref{Gratesinterfer})
\begin{equation}
\frac{\Gamma_{\alpha T_0}}{\Gamma_{\alpha T_0}^{(0)}}
=1-a_\alpha^2
\left[1-\frac{E_Z^2\lambda_{1,0}^2\Lambda_{SO}^2}{4E_{\alpha T_+}E_{\alpha T_-}}
\left(\frac{1}{\Lambda_{SO}^{'4}}+\frac{1}{\Lambda_{SO}^{''4}}\right)\right].
\end{equation}
The latter equation allows one to evaluate the relaxation rate
$\Gamma_{\alpha T_0}$ knowing the ``bare'' rate 
$\Gamma_{\alpha T_0}^{(0)}$, which in its turn can be calculated
from Eq.~(\ref{TtoTGammap0res}).
The physical meaning of $\Gamma_{\alpha T_0}^{(0)}$ becomes
clear if one considers taking the avoided crossing splittings $\Delta_\pm$ to zero,
while keeping the virtual transitions ({\em weak} matrix elements) finite.
Then, there exist only two rates, $\Gamma_{T_-T_0}$ and $\Gamma_{T_0T_+}$.
The reverse rates are trivially given through the detailed balance equation (\ref{detailedbalance}).
Due to spin-symmetry, these rates are equal to each other, $\Gamma_{T_-T_0}=\Gamma_{T_0T_+}$,
and can be considered as a function of the transition frequency.
This function is nothing but the rate $\Gamma_{\alpha T_0}^{(0)}$ in 
Eq.~(\ref{TtoTGammap0res}); note that the only dependence on the indices $\alpha$ and 
$\beta$ in $\Gamma_{\alpha \beta}^{(0)}$ is through the transition frequency $\omega_{\beta\alpha}$.
Therefore, $\Gamma_{\alpha \beta}^{(0)}$ has the meaning of the 
``bare'' triplet-triplet relaxation rate.

The ``bare'' rate $\Gamma_{\alpha \beta}^{(0)}$ is closely related
to the one-electron spin-flip rate.\cite{GKL,Khaetskii}
To illustrate this, we briefly outline the derivation of the
one-electron rate.
It is straightforward to verify that, for
one-electron case, Eq.~(\ref{matelPsiTpmUPsiT0Om}) takes the following form
\begin{equation}
\langle\Psi_{\uparrow}|U_{\rm int}(t)|\Psi_{\downarrow}\rangle=-iE_Z
(\mbox{\boldmath $X$}-i\mbox{\boldmath $Y$})\cdot\mbox{\boldmath $\Omega$}_1(t),
\label{matelPsiupUPsidownOm}
\end{equation}
where $\mbox{\boldmath $\Omega$}_1(t)$ is
the field acting on a single electron.
An expression for $\mbox{\boldmath $\Omega$}_1(t)$ can be obtained from 
Eq.~(\ref{eqOmegaxypp}) by replacing
${\cal F}_T(q_\parallel)\to\frac{1}{2}{\cal F}_0(q_\parallel)$.
Here, ${\cal F}_0(q_\parallel)$ is the single-electron form-factor,
given by
\begin{equation}
{\cal F}_0(q_\parallel)=\langle\psi_0|e^{i{\bm q}_\parallel{\bm r}_1}|\psi_0\rangle=
e^{-{1\over 4}q_\parallel^2\lambda_d^2},
\label{TTcalF0qprl}
\end{equation}
where we assumed that the dot is in its ground state 
$\psi_0({\bm r}_1)=\lambda_d^{-1}\pi^{-1/2}\exp\left(-r_1^2/2\lambda_d^2\right)$,
with $\lambda_d=\sqrt{\hbar/m^*\omega}$.
The spin-flip rate is then given by the following expression 
(assuming $E_Z=-|E_Z|$ and denoting $\omega_Z=|E_Z|/\hbar$) 
\widetext
\begin{equation}
\Gamma_{\uparrow\downarrow}=\frac{\hbar\omega_Z^5(1+N_{\omega_Z})}{4\pi\rho_c(m^*\omega_0^2\Lambda_{SO})^2}
\sum_j\int_0^{\pi/2}d\vartheta\frac{\sin^3\vartheta}{s_j^5}
\left|{\cal F}_0\left(\frac{\omega_Z}{s_j}\sin\vartheta\right)F\left(\frac{\omega_Z}{s_j}\cos\vartheta\right)\right|^2
\left(e^2\bar\beta_{j\vartheta}^2+\frac{\omega_Z^2}{s_j^2}\bar\Xi_j^2\right).
\label{down2upGammares}
\end{equation}
\endwidetext
\noindent
Note that Eqs.~(\ref{TtoTGammap0res}) and (\ref{down2upGammares}) are very similar to each other,
differing only by 
$\left|{\cal F}_T\left(q_\parallel\right)\right|^2\to{1\over2}\left|{\cal F}_0\left(q_\parallel\right)\right|^2$
in going from Eq.~(\ref{TtoTGammap0res}) to (\ref{down2upGammares}).
The form-factors ${\cal F}_T\left(q_\parallel\right)$ and ${\cal F}_0\left(q_\parallel\right)$
differ significantly only at large strengths of the Coulomb interaction.
In the absence of the Coulomb interaction, Eq.~(\ref{CalFTqparTonly}) gives
\begin{equation}
{\cal F}_{T}\left(q_\parallel\right)=
\left(1-\frac{1}{16}q_\parallel^2\lambda^2\right)
e^{-\frac{1}{8}q_\parallel^2\lambda^2},
\end{equation}
which differs from Eq.~(\ref{TTcalF0qprl}) only by the
prefactor in the parentheses. 
Note that $\lambda=\sqrt{2}\lambda_d$ and ${}_1F_1(-1,1;x)=1-x$.
For strong Coulomb interaction, we find the following (approximate) 
crossover expression for 
${\cal F}_{T}\left(q_\parallel\right)$, by matching
the limits of small and large $q_\parallel$,
\begin{equation}
{\cal F}_{T}\left(q_\parallel\right)=
J_0\left(\frac{q_\parallel\langle r\rangle}{2}\right)
e^{
-\frac{(1+\sqrt{3})\hbar q_\parallel^2}{8\sqrt{3}m^*\omega}}.
\label{calFTgreatCI}
\end{equation}
Equation (\ref{calFTgreatCI}) differs from Eq.~(\ref{TTcalF0qprl})
both by the prefactor and the argument of the exponent.
We note that, since the form-factor ${\cal F}_T\left(q_\parallel\right)$
is integrated over its argument in Eq.~(\ref{TtoTGammap0res}),
the resulting rate is, by order of magnitude, equal to the one-electron spin-flip 
rate in Eq.~(\ref{down2upGammares}), except for the case of strong Coulomb interaction.
In the case of strong Coulomb interaction,
a new frequency scale, $\omega \sim s/\langle r \rangle$,
emerges in the triplet-triplet relaxation,
analogously to the singlet-triplet relaxation studied in Sec.~\ref{RelaxRates}.

Finally, we note that for weak spin-orbit coupling the interference
terms in Eq.~(\ref{Gratesinterfer}) can be neglected, since
they are smaller by at least one power of $\lambda/\lambda_{SO}$ 
as compared to the first term.
In this case, the triplet-triplet relaxation is governed by
the transitions $T_-\leftrightarrow T_0$ and
$T_0\leftrightarrow T_+$, both of which occurring at the same rate.
For weak spin-orbit coupling, it is also hard to
discriminate between eigenstates and bare states
at the points of avoided crossings in usual
spin relaxation experiments.
Therefore, for the accessible regions, the
factor $1-a_\alpha$ is either zero or unity.
Thus, we arrive at a single non-zero rate 
$\Gamma_{T_-T_0}=\Gamma_{T_0T_+}$, 
which is obtained by evaluating $\Gamma_{\alpha\beta}^{(0)}$ 
in Eq.~(\ref{TtoTGammap0res}) at the transition frequency
$\omega_{\beta\alpha}=|E_Z|/\hbar$, for $E_Z=-|E_Z|$.
The reverse rate $\Gamma_{T_+T_0}=\Gamma_{T_0T_-}$ 
is obtained by multiplying the result by
$\exp\left(-|E_Z|/T\right)$.

                      \section{Conclusions}                      %
\label{secConclusions}
In this paper, we studied spin relaxation in a quantum dot with two electrons.
The spin relaxation occurs, as in the case of single-electron quantum dots,
due to a combined effect of the spin-orbit, Zeeman, and electron-phonon interactions.
We focused on the vicinity of a singlet-triplet crossing, which occurs with 
applying an orbital magnetic field, considering thus relaxation between states
of the singlet-triplet subspace.
We found that the spin-orbit interaction induces avoided crossings
between the singlet $|S\rangle$ and the triplet states $|T_\pm\rangle$ 
as the quantum dot is tuned across the singlet-triplet crossing.
At the same time, the degeneracy at the crossing of the singlet $|S\rangle$ and the triplet 
state $|T_0\rangle$ is not lifted at the first order of the spin-orbit interaction.
This result has the implication that the spin relaxation occurs efficiently 
only between the following states:
\begin{eqnarray}
&& S \leftrightarrow T_+,\;\;\;\;\;\; S \leftrightarrow T_-,\\
&& T_0 \leftrightarrow T_+,\;\;\;\;\;\; T_0 \leftrightarrow T_-.
\end{eqnarray}
Close to an avoided crossing point, the singlet and triplet 
($T_\pm$) states are superimposed with each other and form,
thus, states of no definite spin. 
We find, nevertheless, that the phonon emission is inefficient
in this case for typical GaAs quantum dots, because the transition energy
is too small for the spin-phonon coupling to be relevant.
As usual, the strongest coupling to the phonons occurs at the 
``optimal phonon'' emission energy.\cite{Bockelmann}
This energy is on the order of $\hbar s/\lambda_d$, which is
the energy of an acoustic phonon with the wavelength equal to the
dot size $\lambda_d$.

The Coulomb interaction between the electrons has a number of effects on the
spin relaxation.
Firstly, the Coulomb interaction affects the relaxation rates
via the singlet-triplet energy $E_{TS}$, because the relaxation rates 
depend strongly on $E_{TS}$.
The singlet-triplet energy $E_{TS}$ is sensitive to the Coulomb interaction
and it is generally smaller than in the non-interacting case.
Moreover, the singlet-triplet crossing ($E_{TS}=0$) is possible in finite
orbital magnetic fields only if the Coulomb interaction is present.
Secondly, the Coulomb interaction modifies the form-factor of the
spin-phonon interaction and introduces a new frequency scale 
(for strong Coulomb interaction) in the relaxation rates, 
$\omega\sim s/\langle r\rangle$, where $\langle r\rangle$
is the average distance between the electrons and $s$ is the speed of sound.
Finally, the spin admixture mechanism that allows for the 
singlet-triplet relaxation to occur is essentially different 
in the limits of weak and strong Coulomb interaction.
For weak Coulomb interaction, superimposing states inside 
the singlet-triplet subspace is sufficient for obtaining
efficient singlet-triplet relaxation.
For strong Coulomb interaction, this mechanism becomes 
inefficient.
Despite the fact the states in the singlet-triplet subspace
are strongly admixed to each other, the orbital parts
of the wave functions of the singlet and triplet become
similar to each other.
More precisely, the electron charge distributions in the
singlet and triplet states converge onto each other
in the limit of strong Coulomb interaction.
Therefore, a different admixture mechanism, which was 
neglected in this paper, is responsible for 
singlet-triplet relaxation in the limit of strong 
Coulomb interaction.

We modeled the quantum dot by a circularly symmetric harmonic 
confining potential
and took into account the Coulomb interaction 
with a high accuracy using a variational method.
The quantum dots used in experiment might, of course,
not have a harmonic confining potential and
not even have circular or inversion symmetry.
We, therefore, indicate here which of our results change
if the quantum dot confinement assumes a different form.
Our results about the spin relaxation rates
remain qualitatively the same for any
confining potential. This refers to the
non-monotonic dependence of the spin relaxation rates
on the magnetic field shown in 
Figs.~\ref{STratesBparfig}~and~\ref{STratesBperpfig}.
The expressions in 
Eqs.~(\ref{eqGammaLADP}), (\ref{eqGammaLAPE}) and (\ref{eqGammaTAPE}),
as well as in Eq.~(\ref{TtoTGammap0res}), are
valid only for circularly symmetric confining
potentials. If the confining potential has
no circular symmetry, then an additional
integration over the in-plane angle $\phi$ of the
emitted phonon is present in each of these expressions.
The dependence on the form of the dot confinement
enters in the relaxation rates through the form-factors 
${\cal F}_{ST}(\mbox{\boldmath $q$}_\|)$ and 
${\cal F}_T(\mbox{\boldmath $q$}_\|)$. 
In Eqs.~(\ref{calFST}) and (\ref{CalFTqparTonly}),
the form-factors were evaluated for the harmonic
confinement and in the presence of the Coulomb interaction.
An important difference between confining potentials
with and without center of inversion for the
singlet-triplet relaxation was mentioned 
below Eq.~(\ref{eqfrscal75}).
In the limit $q_\|\to 0$, one obtains
${\cal F}_{ST}(q_\|)\sim q_\|^2$ for
confining potentials with a center of inversion.
If no center of inversion is present, then
${\cal F}_{ST}(q_{x'},q_{y'})\sim q_{x',y'}$,
i.e. the singlet-triplet form-factor is proportional to the first
power of $q_\|$.
As a result, in the limit of vanishing phonon transition energy,
$w\to 0$, the singlet triplet relaxation rates for the case of a confinement 
without inversion symmetry have a weaker $w$-dependence 
(two powers of $w$ less) than for the case of a confinement 
with inversion symmetry. The scale of $w$ where the
crossover takes place is determined by the degree of asymmetry
of the potential.
The dependence of the avoided splitting gaps $\Delta_\pm$
on the direction of the magnetic field is also sensitive
to the form of the confining potential. 
Two limiting cases, namely of a circularly symmetric dot
and of an elongated dot (or a double dot), have been
considered in Sec.~\ref{ESPSOgfinite} 
(cf. Fig.~\ref{DeltaPlus1} and Fig.~\ref{DeltaSym1}).
The matrix elements of the spin-orbit interaction have 
general forms given in Eqs.~(\ref{belowconv0}),
(\ref{belowconv}), (\ref{belowconvST0}), and
(\ref{belowconvST}), which are valid for an arbitrary
shape of the dot confinement.
They can be used to express $\Delta_\pm$ in terms
of the matrix elements 
$\langle{\bf n}|\mbox{\boldmath $\xi$}^r|{\bf n}'\rangle$
of a dot with an arbitrary confining potential,
as long as the characteristic extension of the orbital
wave function is much smaller than $\lambda_{SO}$.
In Sec.~\ref{ESPSOgeq0}, we have 
not assumed any specific form of the quantum dot potential,
however, we restricted our analysis to quantum dots that
have no orbital degeneracies of the two-particle levels.

Another important assumption made in this paper is that
the spin orbit interaction is linear in the electron momentum.
This type of the spin-orbit interaction is most commonly
used for GaAs quantum dots, since the confinement in the
normal to the 2DEG direction ($z$-axis) is strongest.
In Sec.~\ref{ESPSOgeq0}, we showed that,
for this type of spin-orbit interaction at the leading order,
no spin relaxation occurs within the singlet-triplet subspace
without the presence of a Zeeman splitting.
As a result, the lowest-order singlet-triplet relaxation rates are 
proportional to the second power of the Zeeman energy, in agreement 
with the experiments on triplet-to-singlet relaxation in quantum 
dots.\cite{HansonST2el}

In conclusion, we found strong variations of the spin-relaxation rates 
(by many orders of magnitude) across the singlet-triplet crossing,
with extremely large values ($1-100\,{\rm s}$) of spin lifetimes
at the avoided crossing points.

\acknowledgments						 %

We acknowledge support from the Swiss NSF, NCCR Nanoscience,  ONR, and JST ICORP.

  \appendix\section{Variational parameters $\tilde\omega$ and $\gamma$}                      %
\label{appA}
Here, we study the variational parameters $\tilde\omega$ and $\gamma$ that 
minimize the energy $\varepsilon_{0m}$ in Eq.~(\ref{eg0m}).
We derive asymptotic expressions for $\tilde\omega$ and $\gamma$
in the limits of strong ($\lambda/a_B^*\gg 1$) and weak ($\lambda/a_B^*\ll 1$) 
Coulomb interaction.
Equating to zero the derivatives of $\varepsilon_{0m}$ in Eq.~(\ref{eg0m}) with
respect to $\tilde\omega$ and $\gamma$, we obtain two equations
\begin{equation}
(1+x)y^4-1-\frac{m^2}{x}-
y\frac{\lambda}{2a_B^*}\frac{\Gamma(1/2+x)}{x\Gamma(x)}=0,
\label{eq4tildeomega}
\end{equation}
\begin{equation}
x^2y^4-m^2-
xy\frac{\lambda}{a_B^*}
\frac{\Gamma(1/2+x)}{\Gamma(x)}
\left[\Psi(1+x)-\Psi(1/2+x)\right]=0,
\label{eq4gamma}
\end{equation}
where $x=\sqrt{m^2+\gamma}$, 
$y=\tilde\lambda/\lambda=\sqrt{\omega/\tilde\omega}$, and
$\Psi(x)$ is the digamma function. 
Solving Eqs.~(\ref{eq4tildeomega}) and (\ref{eq4gamma})
with respect to $x$ and $y$ gives the variational parameters
$\gamma$ and $\tilde\omega$ as functions of the Coulomb
interaction strength $\lambda/a_B^*$.

We consider first the case $m=0$.
For weak Coulomb interaction ($\lambda/a_B^*\ll1$),
we expand Eq.~(\ref{eq4gamma}) in terms of $x\ll1$ and
obtain
\begin{equation}\label{xeq}
x=\frac{2\ln2-\pi^{-1/2}y^3a_B^*/\lambda}{4\ln^22+\pi^2/3}.
\end{equation}
Note that, for $\lambda/a_B^*< y^3/2\sqrt\pi\ln2$,
Eq.~(\ref{xeq}) gives $x<0$,
whereas by definition $x=\sqrt{\gamma}\geq 0$.
In this case,
 the minimum of energy in Eq.~(\ref{eg0m})
 is achieved at $\gamma=0$ and thus, 
Eq.~(\ref{eq4gamma}) should  be replaced by $x=0$ for  
$\lambda/a_B^*\leq y^3/2\sqrt\pi\ln2$.
Setting $x=0$ in Eq.~(\ref{eq4tildeomega}), we obtain 
 an equation for $y$,
\begin{equation}
y^4-\frac{\sqrt\pi}{2}\frac{\lambda}{a_B^*}y-1=0,
\label{yeq}
\end{equation}
 which is valid for $\lambda/a_B^*\leq y^3/2\sqrt\pi\ln2$.
Considering simultaneously the equality 
 $\lambda/a_B^*=y^3(2\sqrt\pi\ln2)^{-1}$ and
 Eq.~(\ref{yeq}), we find 
 that $\lambda/a_B^*$ has a critical value, 
$(\lambda/a_B^*)_c\equiv\xi_c$, given by
\begin{equation}\label{critpoint}
\xi_c=\frac{1}{2\sqrt{\pi}\ln2}
\left(\frac{4\ln2}{4\ln2-1}\right)^{3/4}\approx 0.57,
\end{equation}
 at which $\tilde\omega$ and $\gamma$ are non-analytic functions 
 of $\lambda/a_B^*$. Clearly, such a critical point 
is not present in the exact eigenstates of the Hamiltonian (\ref{Hm}),
and is an artifact of the variational ansatz we use.
In the interval $0\leq\lambda/a_B^*\leq(\lambda/a_B^*)_c$, we
have $\gamma=0$ and $\tilde\omega=\omega/y^2$, where $y$ is
the positive solution of Eq.~(\ref{yeq}).
Although we can solve Eq.~(\ref{yeq}) analytically  for $y$,
it is more convenient to present here an expansion for 
$\tilde\omega$ in terms of $\lambda/a_B^*$,
\begin{equation}
\frac{\tilde\omega}{\omega}=1-\frac{\sqrt\pi}{4}\frac{\lambda}{a_B^*}+
\frac{\pi}{16}\frac{\lambda^2}{{a_B^*}^2}-
\frac{7\pi^{3/2}}{512}\frac{\lambda^3}{{a_B^*}^3}+...\;.
\label{tildeommegaseries}
\end{equation}
Note that, since $\lambda/a_B^*\leq(\lambda/a_B^*)_c\approx0.57$,
Eq.~(\ref{tildeommegaseries}) converges with a good accuracy;
e.g., at $\lambda/a_B^*=(\lambda/a_B^*)_c$ the four terms in the series
(\ref{tildeommegaseries}) give $\tilde\omega/\omega\approx0.797$, whereas
the exact solution reads $\tilde\omega/\omega=\sqrt{1-1/4\ln2}\approx0.799$.

To the right of the critical point (\ref{critpoint}), both
 Eqs.~(\ref{eq4tildeomega}) and (\ref{eq4gamma}) are valid for $m=0$,
 and we can use them to study two limiting cases:
 (i) the neighborhood of the critical point (\ref{critpoint}) 
 for $\lambda/a_B^*\geq(\lambda/a_B^*)_c$ and (ii)
 the limit of strong Coulomb interaction ($\lambda/a_B^*\gg 1$).
Excluding $\lambda/a_B^*$ from Eqs.~(\ref{eq4tildeomega}) and 
(\ref{eq4gamma}), we express $y$ as a function of $x$ only,
\begin{equation}\label{yviax}
\frac{1}{y^4}=1+x-\frac{1}{2\left[\Psi(1+x)-\Psi(1/2+x)\right]}.
\end{equation}
Next we substitute $y$ from Eq.~(\ref{yviax}) into Eq.~(\ref{eq4gamma})
and obtain 
\begin{eqnarray}
\frac{a_B^*}{\lambda}&=&\frac{\Gamma(1/2+x)}{\Gamma(1+x)}
\left[\Psi(1+x)-\Psi(1/2+x)\right]^{1/4}
\nonumber\\
&&\times\left\{(1+x)\left[\Psi(1+x)-\Psi(1/2+x)\right]-1/2\right\}^{3/4}.
\;\;\;\;\;\;\;\;\;
\label{alviax}
\end{eqnarray}
Equation (\ref{alviax}) gives $a_B^*/\lambda$ as a function of $x$.
The inverse function allows one to find $x$ as a function of $a_B^*/\lambda$,
and then, from Eq.~(\ref{yviax}), also $y$ as a function of $a_B^*/\lambda$.
Expanding Eqs.~(\ref{alviax}) and (\ref{yviax}) for $x\ll1$, which 
corresponds to the case (i), we find for $\sqrt{\gamma}=x$ and 
$\tilde\omega/\omega=1/y^2$ in leading order,
\begin{eqnarray}\label{sqrtgammaxi}
&&\sqrt{\gamma}
=C_0\left(\frac{1}{\xi_c}-\frac{a_B^*}{\lambda}\right)+...,
\\
&&\frac{\tilde\omega}{\omega}
=\left(\frac{\tilde\omega}{\omega}\right)_c
+C_1
\left(\frac{1}{\xi_c}-\frac{a_B^*}{\lambda}\right)+...,
\label{tildeomegatxi}
\end{eqnarray}
 where
 $\xi_c$ is given in Eq.~(\ref{critpoint}), 
 $(\tilde\omega/\omega)_c=\sqrt{1-1/4\ln2}$, 
 and the coefficients $C_0$ and $C_1$ read
\begin{eqnarray}
&&C_0=
\frac{(2/\pi)^{1/2}(\ln2)^{3/4}(4\ln2-1)^{1/4}}
{(8\ln2-5)\ln^22+{2\over3}\pi^2(\ln2-{1\over16})}\approx0.16,
\;\;\;\;\;\;\;\;\;\;
\\
&&C_1=\left(1-\frac{\pi^2}{24\ln^22}\right)
\frac{C_0}{\sqrt{4-1/\ln2}}\approx0.014.
\label{Cs}
\end{eqnarray}
Note that the value $\tilde\omega/\omega=(\tilde\omega/\omega)_c$ at
$\lambda/a_B^*=\xi_c$ is the minimum of $\tilde\omega/\omega$
as a function of $\lambda/a_B^*$ 
[see also Eq.~(\ref{tildeommegaseries}) and the text below it].

Next we turn to the case of strong Coulomb interaction ($\lambda/a_B^*\gg1$).
Here, we have $x\gg1$ and by expanding Eq.~(\ref{yviax}) in terms of
$1/x$ we obtain
\begin{equation}
\frac{\tilde\omega}{\omega}\equiv\frac{1}{y^2}=\frac{\sqrt{3}}{2}
-\frac{1}{16\sqrt{3}x}+\frac{11}{768\sqrt{3}x^2}+...\;.
\label{tooox}
\end{equation}
Expanding Eq.~(\ref{alviax}) for large $x$, and inverting the obtained series 
by means of iteration, we find for $\gamma=x^2$,
\begin{equation}
\gamma=\frac{3}{4}\left(\frac{\lambda}{2a_B^*}\right)^{4/3}-
\frac{7}{8\sqrt{3}}\left(\frac{\lambda}{2a_B^*}\right)^{2/3}-
\frac{7}{288}+...\;.
\label{gamm0asympt}
\end{equation} 
Then, Eq.~(\ref{tooox}) can be rewritten as follows
\begin{equation}
\frac{\tilde\omega}{\omega}=\frac{\sqrt{3}}{2}
-\frac{1}{24}
\left(\frac{2a_B^*}{\lambda}\right)^{2/3}
-\frac{1}{192\sqrt{3}}
\left(\frac{2a_B^*}{\lambda}\right)^{4/3}+...\;.
\label{tooola}
\end{equation}
To summarize for the case $m=0$, the variational parameters
$\gamma$ and $\tilde\omega$ as functions of the Coulomb interaction strength
$\lambda/a_B^*$ are given, respectively, by $\gamma=0$ and 
Eq.~(\ref{tildeommegaseries}) in the interval $0\leq\lambda/a_B^*\leq\xi_c$, 
by Eqs.~(\ref{sqrtgammaxi}) and (\ref{tildeomegatxi}) in the neighborhood 
of $\lambda/a_B^*=\xi_c$ (for $\lambda/a_B^*\geq\xi_c$), 
and by Eqs.(\ref{gamm0asympt}) and (\ref{tooola}) 
in the limit $\lambda/a_B^*\gg1$.

We consider now the case $|m|\geq1$, and without loss of generality 
we assume $m=|m|$. 
Here, we find that $\tilde\omega$ and $\gamma$ are analytic functions 
of $\lambda/a_B^*$.
We thus consider
only the limit of weak and strong Coulomb interaction.
Proceeding similarly to the previous case,
we find 
from Eqs.~(\ref{eq4tildeomega}) and (\ref{eq4gamma})
the following 
relations [cf. Eqs.~(\ref{yviax}) and (\ref{alviax})],
\begin{eqnarray}
y^4&=&\frac{2\left(1+m^2/x\right)
\Delta\Psi(x)-m^2/x^2}{2(1+x)\Delta\Psi(x)-1},
\label{y4m1}
\\
\frac{a_B^*}{\lambda}&=&
\frac{x^2\Gamma(1/2+x)}{(x^2-m^2)\Gamma(1+x)}
\left[\left(1+\frac{m^2}{x}\right)\Delta\Psi(x)
-\frac{m^2}{2x^2}\right]^{1/4}
\nonumber\\
&&
\times
\left[(1+x)\Delta\Psi(x)-\frac{1}{2}\right]^{3/4},
\label{alm1}
\end{eqnarray}
where $\Delta\Psi(x)=\Psi(1+x)-\Psi(1/2+x)$ and $x$ takes values 
in the interval $m\leq x<\infty$.

In the limit of weak Coulomb interaction ($\lambda/a_B^*\ll1$), 
we expand the right-hand side of Eqs.(\ref{y4m1}) and (\ref{alm1}) 
in terms of $x-m\approx \gamma/2m\ll1$, and obtain in leading order
\begin{eqnarray}
\frac{\gamma}{m}&=&\frac{\sqrt{\pi}(2m-1)!!}{2^m(m-1)!}
\left[(1+m)\Delta\Psi(m)-\frac{1}{2}\right]
\frac{\lambda}{a_B^*}+...,
\;\;\;\;\;\;\;\;\;
\\
\frac{\tilde\omega}{\omega}&=&1-
\frac{\sqrt{\pi}(2m-1)!!}{2^{m+1}m!}
\left[\frac{1}{2}-m\Delta\Psi(m)\right]
\frac{\lambda}{a_B^*}+...,
\;\;\;\;\;\;\;\;\;
\end{eqnarray}
where $\Delta\Psi(m)=2\ln2-\sum_{k=1}^m\frac{1}{(2k-1)k}$.
In particular for the case $|m|=1$, we obtain
$\gamma=\sqrt{\pi}(2\ln2-5/4)(\lambda/a_B^*)$ and 
$\tilde\omega/\omega=1-\sqrt{\pi}(3/2-2\ln2)(\lambda/4a_B^*)$.

In the limit of of strong Coulomb interaction ($\lambda/a_B^*\gg1$), we expand
the right-hand side of Eqs.(\ref{y4m1}) and (\ref{alm1}) 
in terms of $x\gg1$, and as a result we obtain
\begin{eqnarray}
\gamma&=&\frac{3}{4}\left(\frac{\lambda}{2a_B^*}\right)^{4/3}
-\frac{7}{8\sqrt{3}}\left(\frac{\lambda}{2a_B^*}\right)^{2/3}
+\frac{72m^2-7}{288}+...,
\;\;\;\;\;\;\;\;\;
\label{gamm12345}
\\
\frac{\tilde\omega}{\omega}&=&\frac{\sqrt{3}}{2}
-\frac{1}{24}
\left(\frac{2a_B^*}{\lambda}\right)^{2/3}
+\frac{48m^2-1}{192\sqrt{3}}
\left(\frac{2a_B^*}{\lambda}\right)^{4/3}+...\;.
\;\;\;\;\;\;\;\;\;
\label{tooom123}
\end{eqnarray}
Note that in Eqs.~(\ref{gamm12345}) and (\ref{tooom123}) the dependence
on $m$ arises only in the last terms. Thus, for strong Coulomb 
interaction, the radial part of the wave function in Eq.~(\ref{nm}) 
[see also Eq.~(\ref{fgnm})] is weakly depending on the quantum number $m$.

\begin{figure}
 \begin{center}
  \includegraphics[angle=0,width=0.48\textwidth]{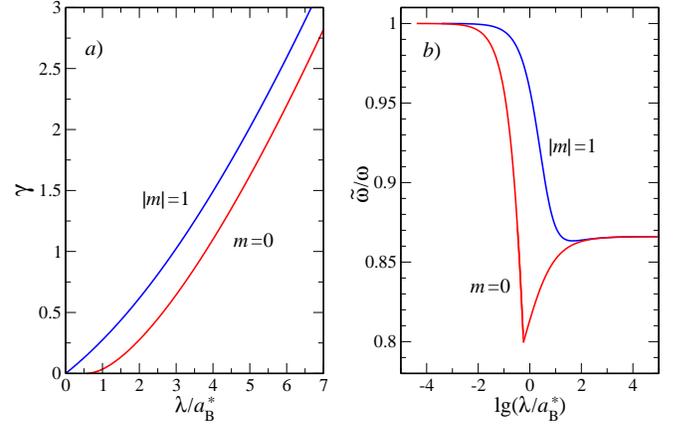}
    \caption{\small
(Color online) (a) The variational parameters $\gamma$ 
as a functions of the Coulomb interaction strength $\lambda/a_B^*$ 
for the singlet $m=0$ and the triplet $|m|=1$ states. 
(b) The ratio $\tilde\omega/\omega$ as a function of the decimal 
logarithm of $\lambda/a_B^*$ for $m=0$ and $|m|=1$.
To find $\gamma$ and $\tilde\omega/\omega$ we minimize the ground state 
energy in Eq.(\ref{eg0m}), see Appendix~\ref{appA} for details.
}
    \label{gammapolygamma}
 \end{center}
\end{figure}

In Fig.~\ref{gammapolygamma}, 
 we plot $\gamma$ and $\tilde\omega/\omega$ 
 as functions of $\lambda/a_B^*$  and $\log_{10}(\lambda/a_B^*)$, respectively,
  both for the singlet state $m=0$ and the triplet states $m=\pm1$.
Note that, for the singlet state $m=0$, the parameters $\gamma$ and 
$\tilde\omega$ behave non-analytically at the point 
$\lambda/a_B^*=\xi_c\approx0.57$, where $\xi_c$ is given in 
 Eq.~(\ref{critpoint}).
Nonetheless, the energy of the $m=0$ state, calculated according 
to Eq.~(\ref{eg0m}), is a smooth function of $\lambda/a_B^*$,
i.e. it has a continuous first derivative with respect to $\lambda/a_B^*$.

\section{Crossover formulae for $\delta$ and $\omega_c^*$}                         %
\label{appdeltaFormula}

\begin{figure}
 \begin{center}
  \includegraphics[angle=0,width=0.48\textwidth]{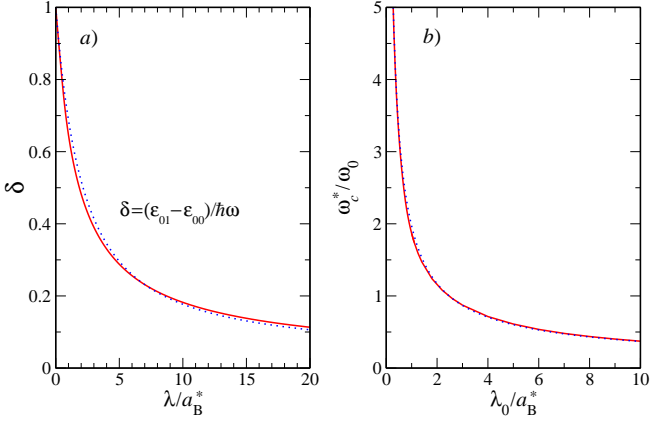}
    \caption{\small
(Color online) (a) The parameter $\delta$ as a function of the Coulomb interaction strength
$\lambda/a_B^*$. 
The dotted curve shows $\delta$ calculated according to 
Eq.~(\ref{deltafit}).
(b) The singlet-triplet degeneracy 
occurs at a megnetic field $B_z^*$ ($E_{TS}=0$).
This figure shows the corresponding $\omega_c^*/\omega_0$ 
as a function of $\lambda_0/a_B^*$,
where $\omega_c^*=eB_z^*/m^*c$ and $\lambda_0=\sqrt{2\hbar/m^*\omega_0}$.
The dotted curve corresponds to the fitting function in Eq.~(\ref{wcstarow0}).
}
    \label{en0en1deJ}
 \end{center}
\end{figure}

The singlet-triplet splitting $E_{TS}$ defined in Eq.~(\ref{EST})
contains an interaction parameter
\begin{equation}
\delta=\frac{\varepsilon_{0,-1}-\varepsilon_{00}}{\hbar\omega},
\end{equation}
where the energies $\varepsilon_{nm}$ are the eigenvalues of 
${\cal H}_m$ in Eq.~(\ref{Hm}).
The parameter $\delta$ shows by how much the energy levels are
renormalized by the Coulomb interaction.
In Fig.~\ref{en0en1deJ}$(a)$,
we plot the parameter $\delta$ as a function of $\lambda/a_B^*$,
calculated with the help of Eq.~(\ref{eg0m}) (solid curve).
For analytic calculations, it is convenient to use the following 
formula
\begin{equation}
\delta(\lambda/a_B^*)=
\frac{\sqrt{1+c(\lambda/a_B^*)^{3/2}}}{(1+b\lambda/a_B^*)^{17/12}},
\label{deltafit}
\end{equation}
where $b={3\over17}\sqrt{\pi}\approx0.31$ and $c=2^{-2/3}b^{17/6}\approx0.023$.
We obtained Eq.~(\ref{deltafit}) by matching the asymptotes of
$\delta$ for the cases of strong and weak Coulomb interaction.
Therefore,
Eq.~(\ref{deltafit}) is exact in the limits of weak and strong Coulomb
interaction, and is accurate within $7\%$ in the crossover region 
($b^{-1}\lesssim\lambda/a_B^*\lesssim c^{-2/3}$).
We plot $\delta(\lambda/a_B^*)$ given by Eq.~(\ref{deltafit}) 
in Fig.~\ref{en0en1deJ}$(a)$ (dotted curve).

The singlet-triplet splitting $E_{TS}$ in Eq.~(\ref{EST}) goes to zero at
a value of cyclotron frequency $\omega_c^*$.
In Fig.~\ref{en0en1deJ}$(b)$, we plot the ratio $\omega_c^*/\omega_0$
as a function of $a_B^*/\lambda_0$; the dotted curve corresponds
to using the following crossover function
\begin{equation}\label{wcstarow0}
\frac{\omega_c^*}{\omega_0}=
\left[1+\left(\frac{2}{\pi^{1/3}}-1\right)
{\cal F}(\lambda_0/a_B^*)
\right]
\left(\frac{\lambda_0}{2a_B^*}\right)^{-2/3},
\end{equation}
where
${\cal F}(x)=\left(1+0.11x^2\right)/\left(1+0.3x^{4/3}\right)^2$.
The crossover function (\ref{wcstarow0}) was obtained, similarly to
Eq.~(\ref{deltafit}), with the help of
asymptotic expressions for $\tilde\omega$ and
$\gamma$ from Appendix~\ref{appA} for the limits of weak and strong Coulomb
interaction.

\section{Fidelity of variational method}                      %
\label{appFidelity}

\begin{figure}
 \begin{center}
  \includegraphics[angle=0,width=0.38\textwidth]{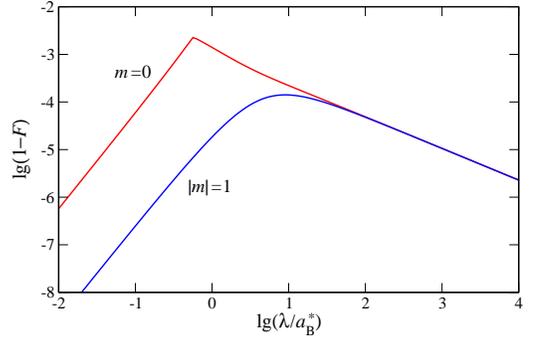}
    \caption{\small
(Color online) Infidelity $1-F$ of the ground state 
of ${\cal H}_m$ for $|m|=0,1$. 
We used Eq.~(\ref{fidFFF}) with the matrix elements
of $V$ evaluated in appendix~\ref{appMEofV}.
Note that both axes are logarithmic
(with decimal base).
}
    \label{fidelityF}
 \end{center}
\end{figure}

We discuss now the accuracy of our variational method.
We rewrite Eq.~(\ref{Hm}) in the following form
\begin{equation}\label{HmV}
{\cal H}_{m}=\tilde{\cal H}_{m}+V,
\end{equation}
where $\tilde{\cal H}_{m}$ has the eigenvalues
$\tilde{\varepsilon}_{nm}=\hbar\tilde\omega(2n+1+\sqrt{m^2+\gamma})$
and the eigenfunctions given in Eq.~(\ref{fgnm}).
The perturbation
\begin{equation}\label{pertV}
V=\frac{\hbar^2}{m^*}\left(\frac{1}{a_B^*r}-\frac{\gamma}{r^2}\right)
+\frac{m^*}{4}\left(\omega^2-\tilde\omega^2\right)r^2,
\end{equation}
is expected to be ``small'' when the variational method works,
with ``small'' meaning that the off-diagonal matrix elements of
$V$ are much smaller than the level spacing (the diagonal part of $V$
needs not be small).
We can show that indeed $V\ll\tilde{\cal H}_{m}$
in the case $\lambda/a_B^*\ll1$, and also in the case $\lambda/a_B^*\gg1$
for transitions involving the ground state.
As for intermediate strengths of the Coulomb interaction 
($\lambda/a_B^*\sim1$), there is no small parameter in the problem,
which can justify the assumption $V\ll\tilde{\cal H}_{m}$.
Nonetheless, we calculate the matrix elements
of $V$ (see Appendix~\ref{appMEofV}) between the ground 
and excited states, and find that they are numerically small, as
compared to the level spacing $2\hbar\tilde\omega$.
In particular, we find that the matrix element 
$\langle \tilde f_{1m}|V|\tilde f_{0m}\rangle$ is identically
zero, if the variational parameter $\tilde\omega$
minimizes the energy of the ground state in Eq.~(\ref{eg0m}) 
[see also Eq.~(\ref{eq4tildeomega})].
Furthermore, we determine an upper bound on the absolute value
of any of the remaining matrix elements ($n\geq2$).
We find that, in
both $m=0$ and $|m|=1$ cases, the matrix element
$\langle \tilde f_{2,m}|V|\tilde f_{0m}\rangle$ 
has the largest value in the crossover region $\lambda/a_B^*\sim1$.
If measured in terms of the level spacing $2\hbar\tilde\omega$, 
this upper bound is approximately $0.1$ and $0.03$,
for the $m=0$ and $|m|=1$ cases, respectively.
Next, we characterize the accuracy of our variational method
in terms of the fidelity
\begin{equation}
F=\left|\langle\Psi|\tilde\Psi\rangle\right|^2,
\end{equation}
where $\Psi(r)\equiv f_{0m}(r)$ is the exact ground state of $H_m$, and
$\tilde\Psi(r)\equiv \tilde f_{0m}(r)$ is the corresponding wave 
function calculated with the help of the variational method.
We estimate the fidelity, using the perturbation theory
expansion\cite{LandauLifshitz}
\begin{equation}
f_{0m}(r)=\tilde f_{0m}(r)+\sum_{n=1}^{\infty}
\frac{\langle\tilde f_{nm}|V|\tilde f_{0m}\rangle}
{\tilde{\varepsilon}_{0m}-\tilde{\varepsilon}_{nm}}
\tilde f_{nm}(r)+...\;,
\label{f0mpertexp}
\end{equation}
where we retain terms up to the second order in $V$ (not shown).
Thus, we obtain the infidelity
\begin{equation}\label{fidFFF}
1-F=\frac{1}{2}\sum_{n=2}^{\infty}
\left|\frac{\langle\tilde f_{nm}|V|\tilde f_{0m}\rangle}
{2n\hbar\tilde\omega}\right|^2,
\end{equation}
where we dropped the $n=1$ term, due to its zero contribution.
The matrix elements 
$V_{nn'}\equiv\langle\tilde f_{nm}|V|\tilde f_{n'm}\rangle$ 
are calculated in Appendix~\ref{appMEofV}.
In Fig.~\ref{fidelityF}, we plot
$\lg(1-F)$ as a function of $\lg(\lambda/a_B^*)$
for $|m|=0,1$, with $\lg(x)$ being the decimal logarithm of $x$.
Figure~\ref{fidelityF} shows that our variational
method is fairly accurate, even in the crossover
region $\lambda/a_B^*\sim1$.
Finally, we note that in Eqs.~(\ref{f0mpertexp}) and (\ref{fidFFF}) we did 
not include the diagonal part of $V$ in the energy denominators.
More accurately, one has to take this part into account as well, since 
$V_{nn}$ can be comparable to the level spacing $\hbar\tilde\omega$.
However, it turns out that the difference $V_{nn}-V_{00}\geq0$
monotonically increases with $n$ in the whole range of $\lambda/a_B^*$.
Therefore, when we replaced 
$2n\hbar\tilde\omega\to2n\hbar\tilde\omega+V_{nn}-V_{00}$
in Eq.~(\ref{fidFFF}), we obtained only an insignificant reduction of 
$1-F$ for $\lambda/a_B^*\sim1$ on the scale of Fig.~\ref{fidelityF} 
(not shown).

\section{Matrix elements of $V$}                      %
\label{appMEofV}
We calculate first the matrix elements 
$\langle\tilde f_{nm}|V|\tilde f_{0m}\rangle$ ($n\geq0$), with
$V$ given in Eq.~(\ref{pertV}) and the wave functions
$\tilde f_{nm}(r)$ in Eq.~(\ref{fgnm}).
We divide the perturbation into three terms,
\begin{equation}
V_{n0}\equiv 
\langle\tilde f_{nm}|V|\tilde f_{0m}\rangle=
V'_{n0}-V''_{n0}+V'''_{n0},
\label{VVpVppVpppn0}
\end{equation}
corresponding, respectively, to the terms proportional to 
$1/r$, $1/r^2$ and $r^2$ in Eq.~(\ref{pertV}).
For the term 
$V'=\hbar^2/m^*a_B^*r$, 
we obtain
\begin{equation}
V'_{n0}=\frac{\hbar^2}{m^*a_B^*\tilde\lambda}
\frac{(2n-1)!!2^{-n}\Gamma(t-1/2)}
{\sqrt{n!\Gamma(t)\Gamma(n+t)}}.
\label{Vpn0}
\end{equation}
Here and below we use the notation $t=1+\sqrt{m^2+\gamma}$.
For the term 
$V''=\hbar^2\gamma/m^*r^2$, 
we obtain
\begin{equation}
V''_{n0}=\frac{\hbar^2}{m^*\tilde\lambda^2}
\frac{\gamma}{(t-1)}
\sqrt{\frac{n!\Gamma(t)}
{\Gamma(n+t)}}.
\label{Vppn0}
\end{equation}
Finally, for the term 
$V'''={1\over4}m^*(\omega^2-\tilde\omega^2)r^2$,
we obtain
\begin{equation}
V'''_{n0}=\frac{m^*\tilde\lambda^2}{4}(\omega^2-\tilde\omega^2)
\left[t\delta_{n,0}-\sqrt{t}\delta_{n,1}
\right],
\label{Vpppn0}
\end{equation}
where $\delta_{nn'}$ is the Kronecker $\delta$-symbol.
We note that in the case $n=1$, we have 
$\langle\tilde f_{1m}|V|\tilde f_{0m}\rangle=0$, 
due to Eq.~(\ref{eq4tildeomega}).

Next, we consider the general case,
\begin{equation}
V_{nn'}\equiv 
\langle\tilde f_{nm}|V|\tilde f_{n'm}\rangle=
V'_{nn'}-V''_{nn'}+V'''_{nn'},
\end{equation}
with the same division of the perturbation into three as 
in Eq.~(\ref{VVpVppVpppn0}).
For the first part, we obtain
\begin{equation}
V'_{nn'}=\frac{\hbar^2}{m^*a_B^*\tilde\lambda}
\sqrt{\frac{(t)_n(t)_{n'}}{n!n'!}}
\sum_{k=0}^{n}\sum_{l=0}^{n'}
\frac{(t)_{k+l-{1\over2}}(-n)_k(-n')_l}
{(t)_k(t)_lk!l!},
\label{Vpnnp}
\end{equation}
where $(x)_n=\Gamma(x+n)/\Gamma(x)$ is the Pochhammer symbol.
For the second part, we obtain
\begin{equation}
V''_{nn'}=\frac{\hbar^2\gamma}{m^*\tilde\lambda^2}
\sqrt{\frac{(t)_n(t)_{n'}}{n!n'!}}
\sum_{k=0}^{n}\sum_{l=0}^{n'}
\frac{(t)_{k+l-1}(-n)_k(-n')_l}
{(t)_k(t)_lk!l!}.
\label{Vppnnp}
\end{equation}
We note that for $n'=n$ we have 
$V''_{nn}=V''_{00}$ for all values of $n$.
Finally, for the last part, we obtain
\begin{eqnarray}
V'''_{nn'}&=&\frac{m^*\tilde\lambda^2}{4}(\omega^2-\tilde\omega^2)
\left[(2n+t)\delta_{nn'}
-\sqrt{n'(n+t)}\,\delta_{n,n'-1}
\right.\nonumber\\
&&\left.
-\sqrt{n(n'+t)}\,\delta_{n,n'+1}
\right].
\label{Vpppnnp}
\end{eqnarray}
We note that for $n'=0$ Eqs.~(\ref{Vpnnp}), (\ref{Vppnnp}) and
(\ref{Vpppnnp}) coincide  with 
Eqs.~(\ref{Vpn0}), (\ref{Vppn0}) and (\ref{Vpppn0}), respectively.



\begin{thebibliography}{99}

\bibitem{Spintronics} 
{\em Semiconductor Spintronics and Quantum Computing}, D. D. Awschalom, D. Loss, and N. Samarth, eds. 
(Springer, New York, 2002). 

\bibitem{Wolf} 
S.~A. Wolf  {\it et al.}, 
Science {\bf 294}, 1488 (2001).

\bibitem{Loss97}
D. Loss  and D.P. DiVincenzo,
Phys.\ Rev.\ A {\bf 57}, 120 (1998).

\bibitem{Cerletti} For a  review see, V. Cerletti, W. A. Coish, O. Gywat, and D. Loss, Nanotechnology {\bf 16}, R27 (2005).

\bibitem{SingleShotNature}
J.M. Elzerman, R. Hanson, L.H. Willems van Beveren, B. Witkamp, L.M.K. Vandersypen, and L.P. Kouwenhoven,
Nature {\bf 430}, 431 (2004).

\bibitem{AmashaZumbuhl}
S. Amasha, K. MacLean, Iu. Radu, D.M. Zumbuhl, M.A. Kastner, M.P. Hanson, and A.C. Gossard,
cond-mat/0607110.

\bibitem{Kroutvar}
M. Kroutvar, Y. Ducommun, D. Heiss, M. Bichler, D. Schuh, G. Abstreiter, and J.J. Finley,
Nature {\bf 432}, 81 (2004).

\bibitem{Golovach04} 
V.N. Golovach and D. Loss,
Phys. Rev. B {\bf 69}, 245327 (2004).

\bibitem{Zumbuel}
D.M. Zumb\"uhl, C.M. Marcus, M.P. Hanson, and A.C. Gossard,
Phys. Rev. Lett. {\bf 93}, 256801 (2004).

\bibitem{PettaT2Science}
J.R. Petta, A.C. Johnson, J.M. Taylor, E.A. Laird, A. Yacoby, M.D. Lukin, C.M. Marcus, M.P. Hanson, and A.C. Gossard
Science {\bf 309}, 2180 (2005).

\bibitem{Laird06} 
E.A. Laird, J.R. Petta, A.C. Johnson, C.M. Marcus, A. Yacoby, M.P. Hanson, and A.C. Gossard,
Phys. Rev. Lett. {\bf 97}, 056801 (2006).

\bibitem{JohnsonPettaT2Nature}
A.C. Johnson, J.R. Petta, J.M. Taylor, A. Yacoby, M.D. Lukin, C.M. Marcus, M.P. Hanson, and A.C. Gossard,
Nature {\bf 435}, 925 (2005). 

\bibitem{KopensFolkScience}
F.H.L. Koppens, J.A. Folk, J.M. Elzerman, R. Hanson, L.H. Willems van Beveren, I.T. Vink, H.P. Tranitz, 
W. Wegscheider, L.P. Kouwenhoven, and L.M.K. Vandersypen,
Science {\bf 309}, 1346 (2005).

\bibitem{Coish05} W. A. Coish and D. Loss,
Phys. Rev. B {\bf 72}, 125337 (2005).

\bibitem{fasth}
C. Fasth, A. Fuhrer, L. Samuelson, V.N. Golovach, and D. Loss,
Phys. Rev. Lett. {\bf 98}, 266801 (2007).

\bibitem{KAT}
L.P. Kouwenhoven, D.G. Austing, and S. Tarucha,
Rep. Prog. Phys. {\bf 64} 701 (2001).

\bibitem{KhaetskiiLossGlazman}
A.V. Khaetskii, D. Loss, and L. Glazman,
Phys. Rev. Lett. {\bf 88} 186802 (2002);
Phys. Rev. B {\bf 67} 195329 (2003).

\bibitem{MerculovEfrosRosen}
I.A. Merkulov,  Al.L. Efros, and M. Rosen,
Phys. Rev. B {\bf 65}, 205309 (2002).

\bibitem{Dyak}
M.I. D'yakonov and V.Yu. Kachorovskii, 
Sov.\ Phys.\ Semicond.\ {\bf 20}, 110 (1986).        

\bibitem{Rashba}
Yu.A. Bychkov and E.I. Rashba, 
JETP Lett. {\bf 39}, 78 (1984).   

\bibitem{Khaetskii} 
A.V. Khaetskii and Yu.V. Nazarov, Phys.\ Rev.\ B {\bf 61}, 12639 (2000);
{\bf 64}, 125316 (2001).

\bibitem{GKL}
V.N. Golovach, A. Khaetskii, and D. Loss, Phys. Rev. Lett. 
{\bf 93}, 016601 (2004).

\bibitem{Halperin}
B.I. Halperin, A. Stern, Y. Oreg, J.N.H.J. Cremers, J.A. Folk, and
C.M. Marcus, Phys.\ Rev.\ Lett.\ {\bf 86}, 2106 (2001).  

\bibitem{Aleiner}
I.L. Aleiner and V.I. Fal'ko,  Phys.\ Rev.\ Lett.\ {\bf 87}, 256801 (2001). 

\bibitem{Khaetski1}
A. Khaetskii, Physica E {\bf 10}, 27 (2001).

\bibitem{BGL}
M. Borhani, V.N. Golovach, and D. Loss,
Phys. Rev. B {\bf 73}, 155311 (2006).

\bibitem{EDSR}
V.N. Golovach, M. Borhani, and D. Loss,
Phys. Rev. B {\bf 74}, 165319 (2006).

\bibitem{TGL}
M. Trif, V.N. Golovach, and D. Loss,
Phys. Rev. B {\bf 75}, 085307 (2007).

\bibitem{Kyriakidis}
J. Kyriakidis, M. Pioro-Ladriere, M. Ciorga, A.S. Sachrajda, and P. Hawrylak,
Phys. Rev. B {\bf 66}, 035320 (2002).

\bibitem{FujisawaT1Nature}
T. Fujisawa, D.G. Austing, Y. Tokura, Y. Hirayama, and S. Tarucha,
Nature {\bf 419}, 278 (2002).

\bibitem{HansonST2el}
R. Hanson, L.H. Willems van Beveren, I.T. Vink, J.M. Elzerman, W.J.M. Naber, F.H.L. Koppens, 
L.P. Kouwenhoven, and L.M.K. Vandersypen,
Phys. Rev. Lett. {\bf 94}, 196802 (2005).

\bibitem{Sasaki05}
S. Sasaki, T. Fujisawa, T. Hayashi, and Y. Hirayama,
Phys. Rev. Lett. 95, 056803 (2005).

\bibitem{Meunier}
T. Meunier, I.T. Vink, L.H. van Beveren, K-J. Tielrooij, R. Hanson, F.H. Koppens, 
H.P. Tranitz, W. Wegscheider, L.P. Kouwenhoven, and L.M. Vandersypen,
Phys. Rev. Lett. {\bf 98}, 126601 (2007).

\bibitem{Fujisawa}
 T. Fujisawa, Y. Tokura and Y. Hirayama, Phys. Rev. B {\bf 63}, R081304 (2001); 
T. Fujisawa, D. G. Austing, Y. Tokura, Y. Hirayama, and S. Tarucha,
Phys.\ Rev.\ Lett.\ {\bf 88}, 236802 (2002).  


\bibitem{Dickmann}
S. Dickmann and P. Hawrylak, Journal of Superconductivity: Incorporating Novel Magnetism {\bf 16}, 387 (2003).

\bibitem{Florescu}
M. Florescu and P. Hawrylak,
Phys. Rev. B {\bf 73}, 045304 (2006).

\bibitem{Chaney}
D. Chaney and P.A. Maksym,
Phys. Rev. B {\bf 75}, 035323 (2007).

\bibitem{Climente}
J.I. Climente, A. Bertoni, G. Goldoni, M. Rontani, and E. Molinari,
Phys. Rev. B {\bf 75}, 081303 (2007);
Phys. Rev. B {\bf 76}, 085305 (2007).

\bibitem{GantmakherLevinson}
V.F. Gantmakher and Y.B. Levinson, {\em Carrier Scattering in Metals 
and Semiconductors} (North-Holland, Amsterdam, 1987).

\bibitem{ErlingssonNazarovFalko}
S.I. Erlingsson, Yu.V. Nazarov, and V.I. Fal'ko,
Phys. Rev. B {\bf 64}, 195306 (2001).

\bibitem{ErlingssonNazarov}
S.I. Erlingsson and Yu.V. Nazarov,
Phys. Rev. B {\bf 66}, 155327 (2002).

\bibitem{Jacak}
L.~Jacak, P.~Hawrylak, and A.~W\'ojs, 
{\em Quantum Dots}, 
(Springer 1998).

\bibitem{LandauLifshitz}
L.D.~Landau and E.M.~Lifshitz, {\it Quantum Mechanics:
non-relativistic theory}, 
(Pergamon Press Ltd. 1977).

\bibitem{note2}
In the presence of Zeeman interaction,
this should be understood as the degeneracy 
of the states $|S\rangle$ and $|T_0\rangle$.

\bibitem{GradsteinRyzhik}
J.S.~Gradstein and I.M.~Ryzhik, 
{\em Tables of Integrals, Series and Products},
(Academic, New York, 1965).

\bibitem{BirPikus}
G.L.~Bir and G.E.~Pikus, 
{\em Symmetry and Strain-Induced Effects in Semiconductors}, 
(Wiley, New York, 1974).

\bibitem{Miller}
J. B. Miller, D. M. Zumb\"uhl, 
C. M. Marcus, Y. B. Lyanda-Geller, 
D. Goldhaber-Gordon, K. Campman, and A. C. Gossard,
Phys. Rev. Lett. {\bf 90}, 076807 (2003). 

\bibitem{Fasth}
C. Fasth, A. Fuhrer, L. Samuelson, V.N. Golovach, and D. Loss,
Phys. Rev. Lett. {\bf 98}, 266801 (2007).

\bibitem{Slichter}  
C.P.~Slichter, {\em Principles of Magnetic Resonance},  
(Springer-Verlag, Berlin, 1980).  

\bibitem{Bockelmann}
U.~Bockelmann,
Phys. Rev. B {\bf 50}, 17271 (1994).


\end{thebibliography}
\end{document}